\documentclass[journal=jacsat,manuscript=article]{achemso}

\usepackage[version=3]{mhchem} % Formula subscripts using \ce{}
\usepackage{graphicx}
\graphicspath{ {images/} }
\usepackage[export]{adjustbox}
\usepackage{wrapfig}
\usepackage{subcaption}
\usepackage{mathrsfs,amsmath,amssymb}
\usepackage{float}
\usepackage{chemformula}
\author{Hanna-Friederike Poggemann}
\affiliation[UULM] {Institute of Electrochemistry, Ulm University, Germany}
\author{Sabrina Klopsch}
\affiliation[RUB] {Applied Microbiology, Faculty of Biology and Biotechnology, Ruhr University Bochum, Germany}
\author{Simon Homann}
\affiliation[UULM] {Institute of Electrochemistry, Ulm University, Germany}
\author{Tim Dirks}
\affiliation[RUB] {Applied Microbiology, Faculty of Biology and Biotechnology, Ruhr University Bochum, Germany}
\author{Sina Schäkermann}
\affiliation[RUB] {Applied Microbiology, Faculty of Biology and Biotechnology, Ruhr University Bochum, Germany}
\author{Julia E. Bandow}
\affiliation[RUB] {Applied Microbiology, Faculty of Biology and Biotechnology, Ruhr University Bochum, Germany}
\author{Timo Jacob}
\affiliation[UULM] {Institute of Electrochemistry, Ulm University, Germany}
\alsoaffiliation[KIT] {Karlsruhe Institute of Technology (KIT), Germany}
\alsoaffiliation[HIU] {Helmholtz Institute Ulm (HIU) Electrochemical Energy Storage, Germany}
\author{Christoph Jung}
\email{christoph.jung@kit.edu}
\affiliation[UULM] {Institute of Electrochemistry, Ulm University, Germany}
\alsoaffiliation[KIT] {Karlsruhe Institute of Technology (KIT), Germany}
\alsoaffiliation[HIU] {Helmholtz Institute Ulm (HIU) Electrochemical Energy Storage, Germany}

\title[An \textsf{achemso} demo]
  {Phenylalanine modification in plasma-driven biocatalysis revealed by solvent accessibility and reactive dynamics in combination with protein mass spectrometry}
\abbreviations{MD}

\begin{document}

\begin{tocentry}

%\begin{figure}
    \includegraphics[width=5cm]{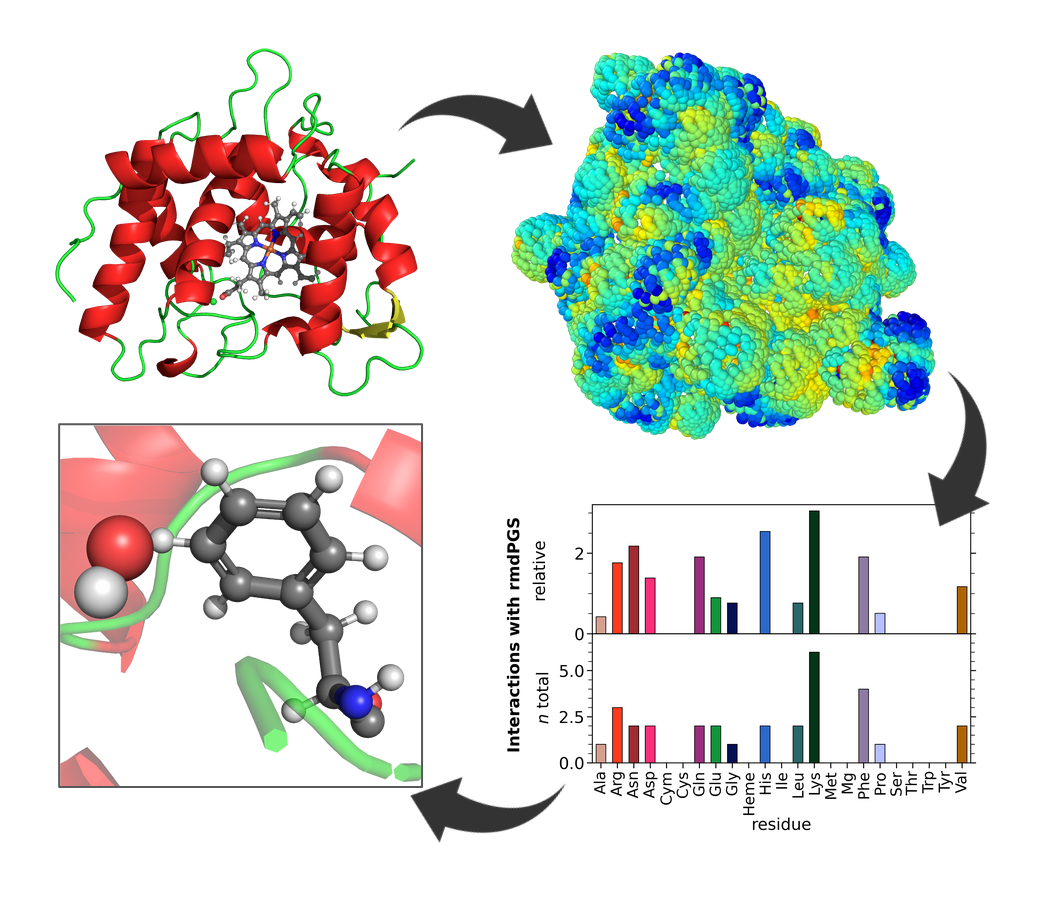}
   % \label{TOC}
%\end{figure}

\end{tocentry}

\begin{abstract}
Biocatalysis is an emerging field that provides an environmentally friendly alternative to conventional catalysis, but still it faces some challenges.
One of the major difficulties for biocatalysts that require reactive species like \ch{H2O2} as co-substrates lies in the concentration of these reactive species. On the one hand, they are used as reactants, but on the other hand, they inactivate the enzymes at high concentrations.
When utilizing non-thermal plasma to deliver \ch{H2O2} for biocatalysis, it is essential to understand the potential interactions between plasma-generated species (PGS) and enzymes. This is particularly important because, alongside \ch{H2O2}, other reactive species such as hydroxyl radicals, atomic oxygen, superoxide, and nitric oxide are also produced. The investigation of the localized reactivity of the solvent accessible surface area (SASA) of an enzyme, with certain species, is an important tool for predicting these interactions. In combination with reactive molecular dynamics (MD) simulations this enabled us to identify amino acid residues that are likely targets for modifications by the PGS. A subset of the theoretical predictions made in the present study were confirmed experimentally by mass spectrometry, underlining the utility of the SASA and MD based screening approach to direct time-consuming experiments and assist their interpretation.
\end{abstract}

%%%%%%%%%%%%%%%%%%%%%%%%%%%%%%%%%%%%%%%%%%%%%%%%%%%%%%%%%%%%%%%%%%%%%
%% Introduction
%%%%%%%%%%%%%%%%%%%%%%%%%%%%%%%%%%%%%%%%%%%%%%%%%%%%%%%%%%%%%%%%%%%%%
\section{Introduction} \label{introduction}
Catalysis plays a pivotal role in our industrial world as most chemical syntheses only became feasible due to the development of appropriate catalysts. It contributes to the production of about 90 percent of all industrial chemicals~\cite{Rothenberg2017CatalysisConceptsGreen}. However, conventional catalysis often has drawbacks, such as the use of harsh chemicals and the generation of heavy metal waste, for example.

Here, the emerging field of biocatalysis offers several advantages over classical catalysis. Enzymes as catalysts not only exhibit high efficiency, but they also can show remarkable specificity towards their substrates \cite{Berg2018Biochemie}. 
This reduces the need for downstream processing and minimizes waste production. The use of enzymes as catalysts not only benefits the environment, but also enables the production of specific chemicals on a smaller scale \cite{Bell2021Biocatalysis, Wu2021BiocatalysisEnzymaticSynthesis}.

When using enzymes that work with co-substrates such as \ch{H2O2}, ensuring a constant supply of reactive species at the right concentration remains a challenge in biocatalysis. These species are essential for fueling the catalytic reaction, but excessive amounts of reactive species can lead to enzyme inactivation limiting turnover rates \cite{Vasudevan1990Deactivationcatalasehydrogen, Valderrama2002SuicideInactivationPeroxidases, Karich2016Exploringcatalaseactivity}.

Several approaches have been tested to make biocatalysis a more competitive process compared to conventional catalysis. A recent review from H. L. Wapshott-Stehli and A. M. Grunde emphasises significant turnover numbers in different biocatalysis approaches highlighting the potential of biocatalysis for various applications \cite{WapshottStehli2021situH2O2generation}.
One innovative approach to address this challenge for \ch{H2O2} dependent enzymes is plasma-driven biocatalysis \cite{Yayci2020PlasmaDrivenSitu}, which uses plasma-generated species (PGS) as reactants for the catalytic reaction. A plasma source, such as a capillary plasma jet (CPJ), a microscale-atmospheric pressure plasma jet (µAPPJ), or a dielectric barrier discharge (DBD) device, generates the desired concentration of reactants. The reactants are then transported into the liquid phase and therefore to the enzymes via the effluent of the plasma source. However, using a plasma source for in-line production is challenging because the reactive plasma species can alter and deactivate the enzymes. Therefore, it is crucial to understand the PGS-protein interactions. 
Recent research by Yacyi \textit{et al.} provides valuable insights into the benefits of plasma species in biocatalysis. The authors came to the conclusion that a plasma-driven production of \ch{H2O2} can prove beneficial for the catalytic reactions, as it enables controlled inline production at the desired concentration \cite{Yayci2020PlasmaDrivenSitu, Yayci2020MicroscaleAtmosphericPressure}.

\begin{figure}[H]
    \centering
    \begin{subfigure}{0.3\textwidth}
    \includegraphics[width=1\textwidth]{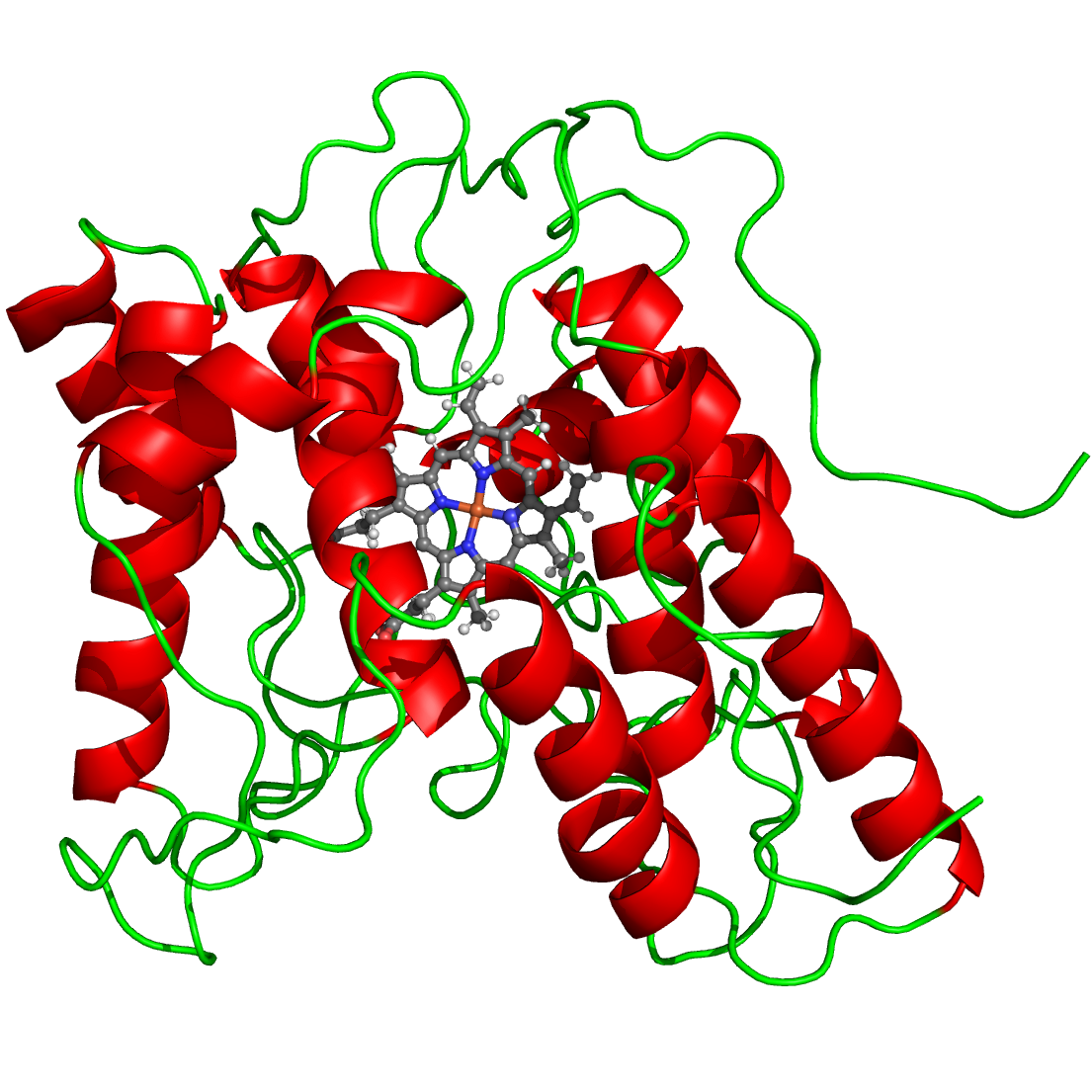}
    \caption{}
    \end{subfigure}
    \begin{subfigure}{0.3\textwidth}
    \includegraphics[width=1\textwidth]{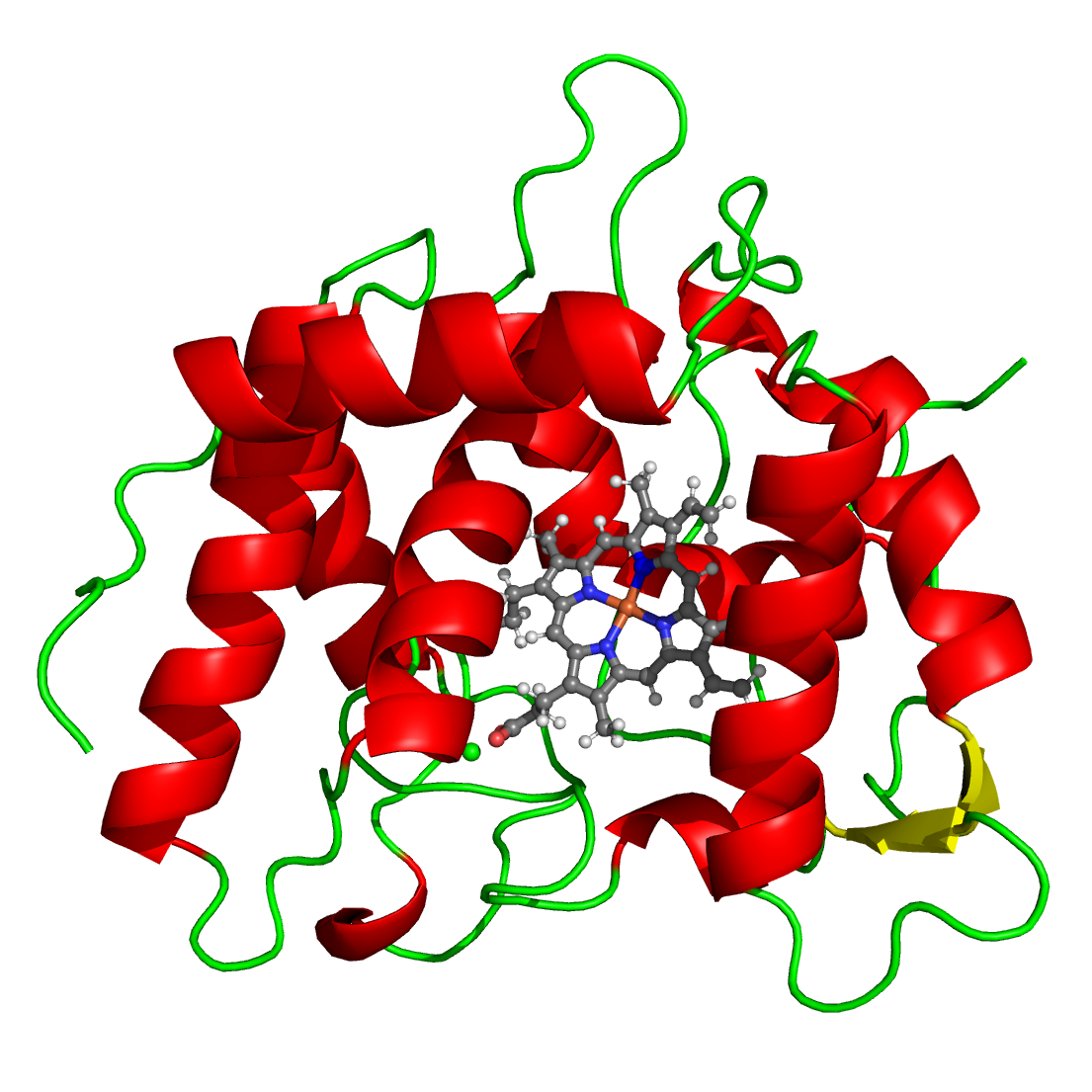}
    \caption{}
    \end{subfigure}
    \begin{subfigure}{0.3\textwidth}
    \includegraphics[width=1\textwidth]{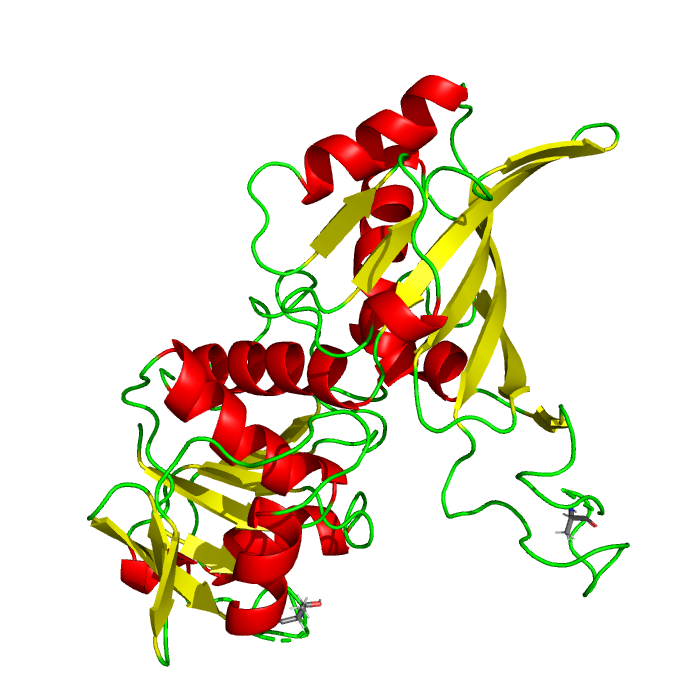}
    \caption{}
    \end{subfigure}
    \caption[]{3D structural representations of the enzymes \textit{Aae}UPO (PDB code: 5OXU) (a), \textit{Cvi}UPO (PDB code: 7ZCL) (b) and GapA (PDB code: 7C5H) (c). Helices are shown in red, sheets in yellow, and turns and coils in green. The heme cofactor and the coordinating Mg ions in \textit{Aae}UPO and \textit{Cvi}UPO are shown in atomic representation using the standard atomic color coding (H: white, C: gray, N: blue, O: red, Fe: orange, Mg: green). The 3D representations were created with PyMOL \cite{PyMOL}.}
    \label{Enzyme_Strcutres}
\end{figure}

The present study seeks to elucidate the interactions between the plasma-generated species \ch{H2O2}, \ch{OH}, \ch{O2}, \ch{O} and \ch{NO} and three enzymes of interest (see Figure \ref{Enzyme_Strcutres}) from an atomistic point of view, using mainly reactive molecular dynamics simulations. The PGS listed have been selected, since \ch{H2O2}, \ch{OH}, \ch{O2}, \ch{O} and \ch{NO} occur at high densities in the plasma-effluent of the used dielectric barrier discharge plasma \cite{Dickenson2018generationtransportreactive}. In addition to the other PGS, we included \ch{H} which is not known to be generated by the plasma effluent but it could emerge from secondary reactions with the solvent. 
At this point it must be emphasized that the species modeled with reactive molecular dynamics are not real plasma-generated species. Due to the limitation of the method, these \textit{rmd}PGS may differ slightly in their chemical behavior from the real species, a detailed explanation can be found in the SI (Section "Explanation to the species nature in ReaxFF"\ref{SI_Species}).
We decided to investigate the enzymes \textit{Aae}UPO, \textit{Cvi}UPO and GapA. \textit{Aae}UPO is a well studied unspecific peroxygenase and has been the model enzyme for plasma-driven biocatalysis. \textit{Cvi}UPO is a promising alternative candidate, therefore the focus of this study lies on \textit{Cvi}UPO as the main model enzyme.
GapA is neither a nonspecific peroxygenase nor does it use \ch{H2O2} as a co-substrate. Instead, it is an enzyme involved in central metabolism that is known to be reversibly inactivated by \ch{H2O2}. This, in turn, enables a comparison between the interaction behavior of an enzyme that is inhibited by \ch{H2O2} and those that actively utilize it. In our investigation, the interaction profiles of each species with the enzymes were screened using a solvent accessible surface analysis (SASA) approach. Subsequently, reactive molecular dynamics simulations were performed in order to verify possible PGS-induced modifications of the predicted residues. Further, the role of the surrounding solvent was addressed by performing these MD simulations both without and with the surrounding solvent. 
Finally, the simulation results were compared to mass spectrometry measurements of plasma-treated \textit{Cvi}UPO, in order to validate the approach against experimental findings. The experimental analysis concentrates exclusively on \textit{Cvi}UPO, as the glycosylation of \textit{Aae}UPO complicates spectrometric analysis of this enzyme.
%%%%%%%%%%%%%%%%%%%%%%%%%%%%%%%%%%%%%%%%%%%%%%%%%%%%%%%%%%%%%%%%%%%%%
%% Computational details
%%%%%%%%%%%%%%%%%%%%%%%%%%%%%%%%%%%%%%%%%%%%%%%%%%%%%%%%%%%%%%%%%%%%%
\section{Computational details}\label{com_methods}
All non-reactive molecular dynamics simulations in this work have been performed with GROMACS (version 2022.2 released June 16th, 2022) \cite{2022GROMACS2022.2Source, Abraham2015GROMACSHighperformance, Berendsen1995GROMACSmessagepassing, Hess2008GROMACS4Algorithms, Lindahl2001GROMACS3.0package, Pall2015TacklingExascaleSoftware, Pall2020Heterogeneousparallelizationacceleration, Pronk2013GROMACS4.5high, VanDerSpoel2005GROMACSFastflexible} and the latest version of the CHARMM force field, CHARMM36, for GROMACS (Updated July 2022) \cite{SoterasGutierrez2016Parametrizationhalogenbonds, Vanommeslaeghe2010CHARMMgeneralforce, Vanommeslaeghe2012AutomationCHARMMGeneral, Vanommeslaeghe2012AutomationCHARMMGenerala, Yu2012ExtensionCHARMMgeneral}.
The non-reactive simulations were used for solvation and pre-equilibration of the enzymes. Prior to the simulations, hydrogen atoms were added to the X-ray structures of the enzymes to emulate neutral pH conditions. Afterwards, all proteins were solvated in cubic simulations boxes (80x80x80 \AA) with periodic boundary conditions using the GROMACS solvation procedure, energy minimized and equilibrated by first a short run in an $NVT$ ensemble (100~ps), followed by an 100~ps $NPT$ run. Both, the $NVT$ and the $NPT$ runs had a position restrain on all enzyme atoms but hydrogen. Afterwards, a longer MD production run was performed for 1~ns at 300\,K or any other desired temperature. The procedure follows the suggestions of J. A. Lemkul \cite{Lemkul2019ProteinsPerturbedHamiltonians}. The hence prepared structures were then used as input for the reactive molecular dynamics simulations. \\
All reactive molecular dynamics simulations have been performed with the latest stable release of the LAMMPS simulation package (version August 2nd, 2023), \cite{2023LAMMPSLargescale, Aktulga2012Parallelreactivemoleculara, Plimpton1995FastParallelAlgorithms, Thompson2022LAMMPSflexiblesimulationa} employing a ReaxFF potential as originally developed by van Duin \textit{et al.} \cite{Duin2001ReaxFFReactiveForce}. The ReaxFF force field used in this work is the biomolecule force field trained by Monti \textit{et al.} in 2013 \cite{Monti2013Exploringconformationalreactive}. ReaxFF potentials allow dynamic bond breaking and formation by working with a concept of interatomic distances to determine bond orders, which are then used to compute the potential energy function.
The time-integrated molecular dynamics simulations were conducted using a time step of 0.15\,fs.
\\
The SASA package for the interaction analysis is built on VMD \cite{HUMP96,VARSH1994} for the determination of the solvent accessible surface area, and on the ReaxFF implementation of LAMMPS for the calculation of the interaction energies. 
The program first automatically converts the GROMCAS output file of an equilibrated protein into a LAMMPS data file, deleting the solvent molecule in the process, and then performs an energy minimization with the ReaxFF potential. Afterwards, all points on the protein surface that are accessible for the solvent (SASA points) are determined  for the minimized structure by an interface to VMD \cite{HUMP96,VARSH1994}.
The distance of the SASA points to the protein surface is set to 1.4~\AA~per default, because its approximately the radius of water, the most commonly used solvent \cite{VARSH1994}. For the calculation of the interaction energy between the protein and a probe molecule, in our case the PGS, one probe molecule at a time is then placed at each of the SASA points. All probe molecules are initially rotated so that they are oriented horizontally to the nearest neighbor atoms of the protein. To find the optimal orientation of the probe molecule to the protein, a constrained energy minimization is performed, allowing the probe molecule to rotate, keeping the center of mass at its original position. At last, the program will run a final single point calculation.  
The resulting total energy $E_\mathrm{SASA}$ is used to calculate the local interaction energy at a specific SASA point $E_\mathrm{int}$ by subtracting the total energy of the protein $E_\mathrm{macro\_mol}$ and the total energy of the probe molecule $E_\mathrm{probe\_mol}$ (compare equation \ref{eq:1}). The latter ones are calculated separately beforehand. All SASA calculations are performed in vacuum. 
\begin{equation} \label{eq:1}
    E_\mathrm{int} = E_\mathrm{SASA} - E_\mathrm{macro\_mol} - E_\mathrm{probe\_mol} 
\end{equation}
The method gives one interaction energy per SASA point, resulting in an interaction map around the enzyme showing regions of strong interaction (most negative values) and regions of weak interactions (less negative values). 
The package further provides two post processing methods, generating plots showing the protein residues and the atom types that are strongest interacted with. The residue analysis includes a calculation of the total interactions per residue and the relative interactions per residue. The relative interactions are calculated as follows:
\begin{equation} \label{eq:2.1}
        Expected\, Interactions = \frac{Residue\, Count}{Total\,Residue\,Count} \cdot Interactions_{\mathrm{Total\,Number}}
\end{equation}
\begin{equation} \label{eq:2.2}
    Relative\, Interactions = \frac{Interactions_{\mathrm{Residue}}}{Expected\, Interactions}
\end{equation}
For our analysis we defined the relative interactions as the quotient of the observed and and expected interactions. 
Per default the post processing functions only take into account the thirty strongest interaction energies. This cutoff value was determined empirically by analyzing the average number of data points required to cover the interaction energies of interest. The post processing functions strongly rely on the python API of the Ovito Open Visualization Tool \cite{Stukowski2009Visualizationanalysisatomistic}. The SASA package for the interaction analysis is available on GitHub ( \url{https://github.com/hpoggemann/SASA-Analysis}).

\section{Experimental workflow} \label{exp_methods}
\subsection{Enzyme preparation} \label{enzyme_prep}
The gene \textit{cviUPO} was overexpressed in ZYM-5052 autoinduction medium using a pET21a(+) expression plasmid containing ampicillin resistance and an N-terminal His-tag. \cite{Studier.2005} The plasmid was transformed into \textit{E. coli} BL21 competent cells and cells were grown in LB medium containing 75 µg ml\textsuperscript{-1} ampicillin at 37°C overnight. For overexpression, 10~ml of overnight culture were added into 1~L autoinduction medium with 75~µg ml\textsuperscript{-1} ampicillin, 500~µM $\alpha$-aminolevulinic acid (Roth), and 200~µM hemin (Roth). Overexpression was performed at 16°C and 120~rpm for 5~days. \\
Cells were harvested by centrifugation and lysed by a homogenizer (pressure cell homogenizer SPCH-EP, Stansted) in lysis buffer containing 20~mM sodium phosphate, 500~mM NaCl, 10\% glycerol, 0.2~mg~ml\textsuperscript{-1} DNase (Sigma), 0.2~mg~ml\textsuperscript{-1} RNase (Sigma), 0.35~mg ml\textsuperscript{-1} lysozyme (Roth), and 2~mM complete protease inhibitor (Roche). After cell lysis, the lysate was centrifugated at 4°C and 21.000~$\times$~g for 60~min and the supernatant was filtered with 45~µm and 20~µm cut-off filters. Purification was carried out using the ÄKTA pure 25 system and a 5~ml HisTrap FF crude column (GE healthcare). Protein elution was performed using 4 column volumes of elution buffer B (20~mM sodium phosphate, 500~mM NaCl, 500~mM imidazole, pH 7.4). Proteins were eluted stepwise at 100 mM imidazole (20\%), 200~mM imidazole (40\%), and 500~mM (100\%) imidazole in buffer B and fractions were collected in 1~ml steps, while fractions showing an absorption signal at 420~nm were united (heme-loaded protein). \\
After protein purification, dialysis was performed for buffer exchange in 5~L buffer A (20~mM sodium phosphate, 500 mM~NaCl, 10\% glycerol pH 7.4) overnight. Finally, a Bradford assay was used to determine protein concentration and heme loading of the enzyme was measured with UV-VIS spectroscopy (V-750 Spectrophotometer, Jasco), calculating the Reinheitszahl ($r/z$-value; A\textsubscript{420}/A\textsubscript{280}) of purified enzyme.\\

\subsection{Plasma treatment}
To assess the influence of plasma on \textit{Cvi}UPO, a dielectric barrier discharge (DBD, Cinogy,) source was used for plasma treatment. To this end, 40~\textmu l enzyme (1~mg~ml\textsuperscript{-1} \textit{Cvi}UPO) were placed onto a metal plate and treated with the DBD for 5~min (electrode diameter 20~mm; 13.5~kV pulse amplitude; 300~Hz trigger frequency; 1~mm distance to sample). The treated samples were transferred to a reaction tube by centrifugation (2,000~$\times$~g for 1~min). The samples were stored at $-70$°C until mass spectrometrical analysis. 
Untreated protein used as control underwent the same procedure but was not exposed to plasma.\\

\subsection{Mass spectrometry of \textit{Cvi}UPO}
For mass spectrometry (MS) analysis, the protein samples were reduced, alkylated, and then digested with trypsin. For this purpose, 0.1\% RapiGest (Waters) and 2.5~mM tris-(2-carboxyethyl) phosphine hydrochloride were added to 1~mg~ml\textsuperscript{-1} enzyme sample and made up to 60~\textmu l total volume with \textit{A. dest.}. The samples were then reduced at 60°C for 45~min. After addition of 5~mM iodoacetamide, the samples were incubated again at 25°C for 15~min in the dark for alkylation. For tryptic digestion, 0.5~\textmu g trypsin (0.5~\textmu g~\textmu l\textsuperscript{-1} stock solution) were added prior to incubation at 37°C and 300~rpm for 5~h. To precipitate RapiGest, 1~\textmu l trifluoroacetic acid (TFA; concentrated) was added and the samples were centrifuged at 10,000~$\times$~g and 4°C for 10~min. The supernatant was transferred to a new reaction tube and the procedure was repeated until no more pellets were visible. For the MS measurement, samples were diluted to a concentration of 0.25~\textmu g \textmu l\textsuperscript{--1} in 5\% acetonitrile with 0.1\% TFA (total volume 20 \textmu l) and 2 \textmu l of sample (0.25 \textmu g/\textmu l) were injected to an ACQUITY UPLC M-Class System (Waters), equipped with an nanoEase m/z peptide CSH column (Waters, particle size 1.7 \textmu m, column size 0.3 x 100 mm) and eluted online to a Synapt XS (Waters) mass spectrometer equipped with an ESI source and a low flow probe (Waters). Peptides were eluted using a gradient of 0.1\% formic acid (FA) in MS grade \textit{A. dest.} (solvent A) or acetonitrile (solvent B) with a flow rate of 7 \textmu l/min: 0-3 min, 1\% B; 3-100 min, 35\% B; 109 min, 90\% B; 110 min, 90\% B; 115 min, 1\% B; 120 min, 1\% B.  The column temperature was set to 40 °C and MS\textsuperscript{E} spectra were recorded from 50-2000 m/z in positive resolution mode with a scan time of 0.7 s. Argon served as collision gas with a collision ramp of 17-60 V. The following parameters were used: capillary voltage 2.5 kV, cone voltage 40 V, source offset, 4 V; cone gas flow 50 L/h, desolvation gas flow 500 L/h, source temperature 80 °C, desolvation temperature 250 °C. Glu-1-fibrinopeptide B was recorded as lock mass. To analyze the spectra of \textit{Cvi}UPO peptides, ProteinLynx Global Server (Version 3.0.3, Waters) was used to detect protein modifications using a \textit{Escherichia coli} BL21 database (Uniprot UP000002032) containing the sequence for r\textit{Cvi}UPO. The amino acid sequence of the \textit{rCvi}UPO used in this study is provided in the supplementary information (Figure \ref{sequence_cviUPO}). The His-tag attached to the protein was not considered in the numbering of amino acid positions, resulting in a 21 position shift when compared to the full-length sequence that includes the His-tag. The chromatographic peak width and MS TOF resolution for MS spectra analysis was set to automatic, lock mass for charge 2 was set to 785.8426 Da/e and the lock mass window was set to 0.25 Da. The low energy threshold and elevated energy threshold were specified as 50.0 counts and 25.0 counts, respectively. Peptide tolerance and fragment tolerance were selected as automatic, minimal fragment ion matches per peptide was 3, minimal fragment matches per protein was 7, and minimal peptide matches per protein was 1. The maximal protein mass was set to 250 kDa and primary digest reagent was trypsin. For modification analysis, up to one missed cleavage per peptide was allowed and the false discovery rate was set to 1. Spectra were processed and searched for the following modifications: carbamidomethylation of cysteine (delta mass +57.0215 Da), which was defined as fixed modifier reagents; oxidation of methionine (delta mass +14.0157 Da), double and triple hydroxylation of cysteine (delta mass +31.9988 Da and +47.9982 Da), single hydroxylation of phenylalanine (delta mass +15.9994 Da), dehydrogenation of arginine (delta mass -4.0316 Da), dehydrogenation of lysine (delta mass -4.0316 Da), hydroxylation and dehydrogenation of lysine (delta mass -3.0468 Da). All modifications were defined as variable modifications.
\\

%%%%%%%%%%%%%%%%%%%%%%%%%%%%%%%%%%%%%%%%%%%%%%%%%%%%%%%%%%%%%%%%%%%%%
%% Results
%%%%%%%%%%%%%%%%%%%%%%%%%%%%%%%%%%%%%%%%%%%%%%%%%%%%%%%%%%%%%%%%%%%%%
\section{Results and Discussion} \label{results}
\subsection{SASA interaction analysis} \label{SASA}
\subsubsection{\textit{Cvi}UPO} \label{SASA-Cvi}
\begin{figure}[h]
    \centering
    \includegraphics[width=1\textwidth]{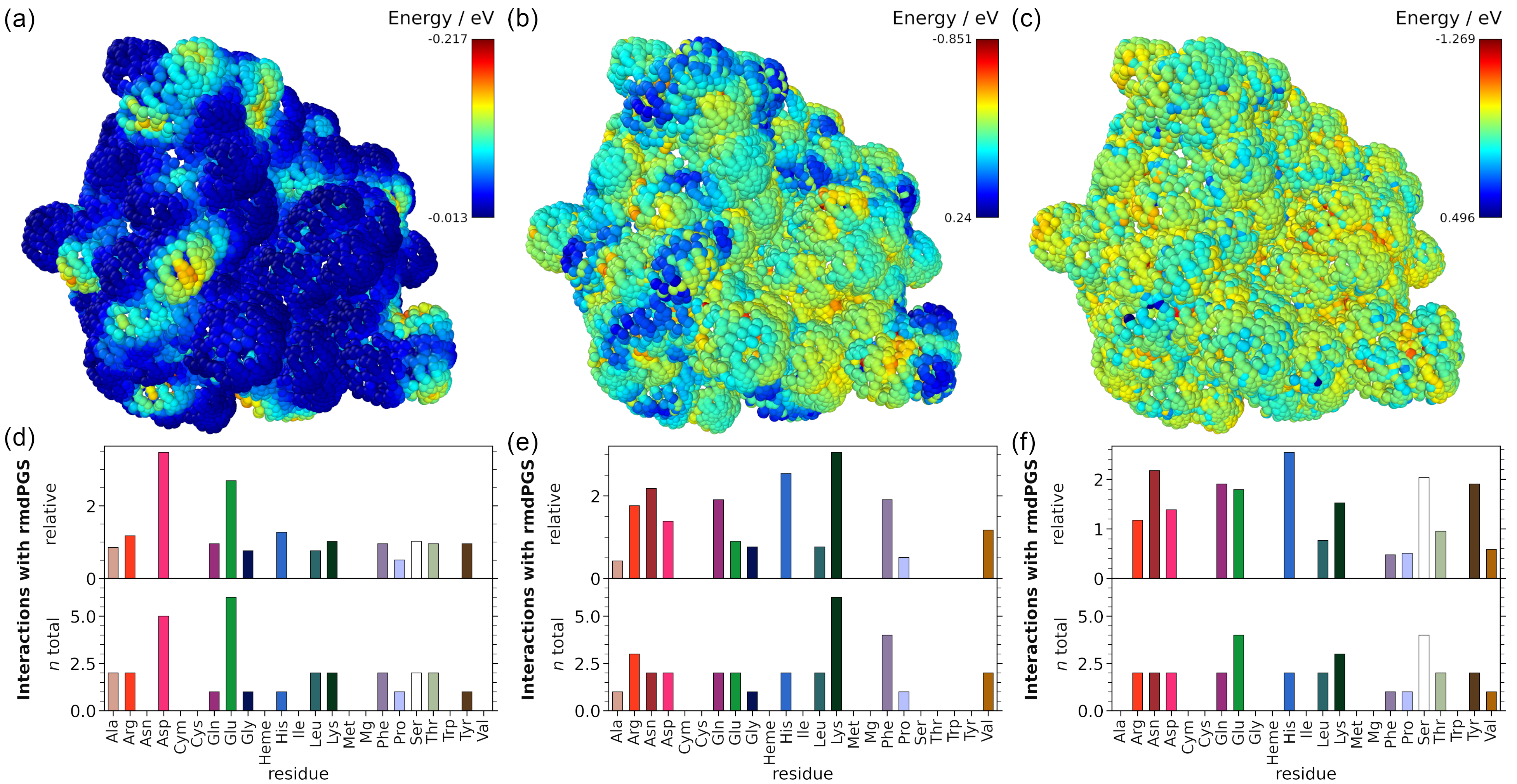}
    \caption{Interaction profiles from SASA for the enzyme \textit{Cvi}UPO at 300\,K. (a), (b) and (c) show the interaction maps with \ch{H}, \ch{OH} and \ch{H2O2}, respectively.  The bar plots below show the relative and total interactions per amino acid, heme cofactor (Heme), the coordinated Mg ion (Mg), and Cym cysteine group bound to the heme center, which differs chemically from a Cys with a free thiol, and is therefore highlighted separately.} Panel (d) corresponds to the interactions with \ch{H}, (e) and (f) show the interactions with \ch{OH} and \ch{H2O2} respectively. The computational details provide a detailed explanation of the SASA procedure.
    \label{SASA_Cvi}
\end{figure}

The SASA interaction analysis was performed for the \textit{rmd}PGS \ch{H2O2}, \ch{OH}, \ch{O2}, \ch{O}, \ch{NO} and \ch{H}. Figure \ref{SASA_Cvi} shows only the results for \ch{H}, \ch{OH} and \ch{H2O2} interactions, data for the other species can be found in the supplementary information (Figure \ref{SI-SASA_Cvi} -- \ref{SI-SASA-3}).

Comparing Figures \ref{SASA_Cvi} (a), (b) and (c) it is apparent that \ch{H} has the least negative interaction energy, which means that the interactions with \ch{H} are weaker than with \ch{OH} or \ch{H2O2}, for example. \ch{H} shows the strongest interaction with glutamic acid (Glu) and aspartic acid (Asp) residues both in absolute and relative terms (Figure \ref{SASA_Cvi} (d)). Both of these residues are negatively charged at neutral pH and in the simulations and can act as proton acceptor residues \cite{Berg2018Biochemie}.
OH shows the strongest absolute interactions with lysine (Lys), followed by phenylalanine (Phe) and arginine (Arg). When taking into account the frequency of each amino acid occurring in textit{Cvi}UPO (relative interaction plot in \ref{SASA_Cvi} (e)), Lys interactions are the most frequent followed by histidine (His), Asp, Glu and Phe. Lys and Arg are known to be proton donor residues, which makes these amino acids attractive for \ch{OH} and other negatively charged PGS. Phe is one of the most frequent amino acids in \textit{Cvi}UPO and also exposed on the outside of the protein, which could explain strong interactions. Also OH radicals are known to form adducts on the aromatic ring \cite{Solar1985ReactionOHphenylalanine}.
OH also shows a nearly inverse local interaction profile to H, which is expected since H is positively charged and OH occurs either as a radical species or as a negatively charged ion. 
\ch{H2O2} on the other hand does display a less locally pronounced interaction profile. Most of the protein surface seems to be equally attractive for this species, except for some sampling points inside the reactive channel. In absolute terms, \ch{H2O2} interacts strongly with serine (Ser) and also with Glu, followed by Lys. When accounting for amino acid frequency (relative values) His, ASn and Ser interact most frequently with \ch{H2O2}. Ser is a polar but uncharged amino acid and it has proton donor as well as acceptor atoms, which could favor an interaction with an ambivalent molecule, such as \ch{H2O2}. 
\ch{O} exhibits a similar interaction profile to \ch{OH}, with the highest number of interactions occurring with Lys, followed by Arg (Supplementary Figure \ref{SI-SASA-3} (d)).
\ch{NO} interacts strongly with aspartic acid (Asp), which is negatively charged and often acts as hydrogen acceptor residue (Supplementary figure \ref{SI-SASA-2} (d)). Furthermore, \ch{NO} interacts often with Lys, as well as Glu and leucin (Leu). Leu similar to Phe has no tendency to act as a hydrogen
acceptor or donor, nevertheless there are frequent interactions with this residue. That can
be explained by the fact that Leu is the most frequent amino acid in \textit{Cvi}UPO.
The most dominant interaction partner of \ch{O2} is Lys, followed by Arg, Leu, isoleucine(Ile), tyrosine (Tyr) and valine (Val), all with an equal amount of total interactions in the SASA analysis (see Figure \ref{SI-SASA-3} (a) in the SI).
All tested \textit{rmd}PGS interact strongest with charged or polar residues of the enzyme, which is the expected behavior. But charge is not the only criterium for interaction. The position of the amino acid is also of importance, as is the total frequency of occurrence of the amino acid in the enzyme. Residues such as Leu or Lys, which appear often and are exposed on the outside of the protein, are more likely to have many interactions with the \textit{rmd}PGS and are therefore also more likely to be modified under experimental conditions.
The analysis shows no strong tendency of neither the \textit{rmd}PGS interact with cysteine (Cys) nor methionine (Met), even though other studies found sulfur-containing amino acids to be strongly modified by the plasma \cite{Takai2014Chemicalmodificationamino, Guo2023Reactivemoleculardynamics, Lackmann2018Chemicalfingerprintscold}. One possible explanation for this behavior might be that neither Met nor Cys appear frequently in \textit{Cvi}UPO (Met only five times and Cys only two times). Also all seven residues are buried in the protein structure and not at the solvent accessible surface area. Furthermore, there is one free thiol available, because one of the two Cys is bound covalent to the heme center (Cym).
In addition, not all species seem to be equally prone to interacting with the active pocket. Only \ch{H2O2} and \ch{O2} have strong interactions with residues close to the heme center. \ch{H2O2} interacts with a Lys (Lys165) and a Glu (Glu162) residue while \ch{O2} interacts with Ile (Ile61), the same Lys residue (Lys165), as well as an hydrogen atom from the heme center. Both the potential removal of a hydrogen atom from the heme center and alterations in Glu162 or Lys165 could be detrimental to the catalytic process. Given that catalysis primarily occurs at the heme center and both of these amino acids play an active role in the catalytic mechanism of \textit{Cvi}UPO, their integrity is crucial for enzymatic function \cite{GonzalezBenjumea.2020}. In the \textit{Cvi}UPO reaction cycle Glu162 serves as a hydrogen acceptor and Lys165 supports the stabilization of the intermediates. 

\begin{figure}[H]
    \centering
    \includegraphics[width=0.8\textwidth]{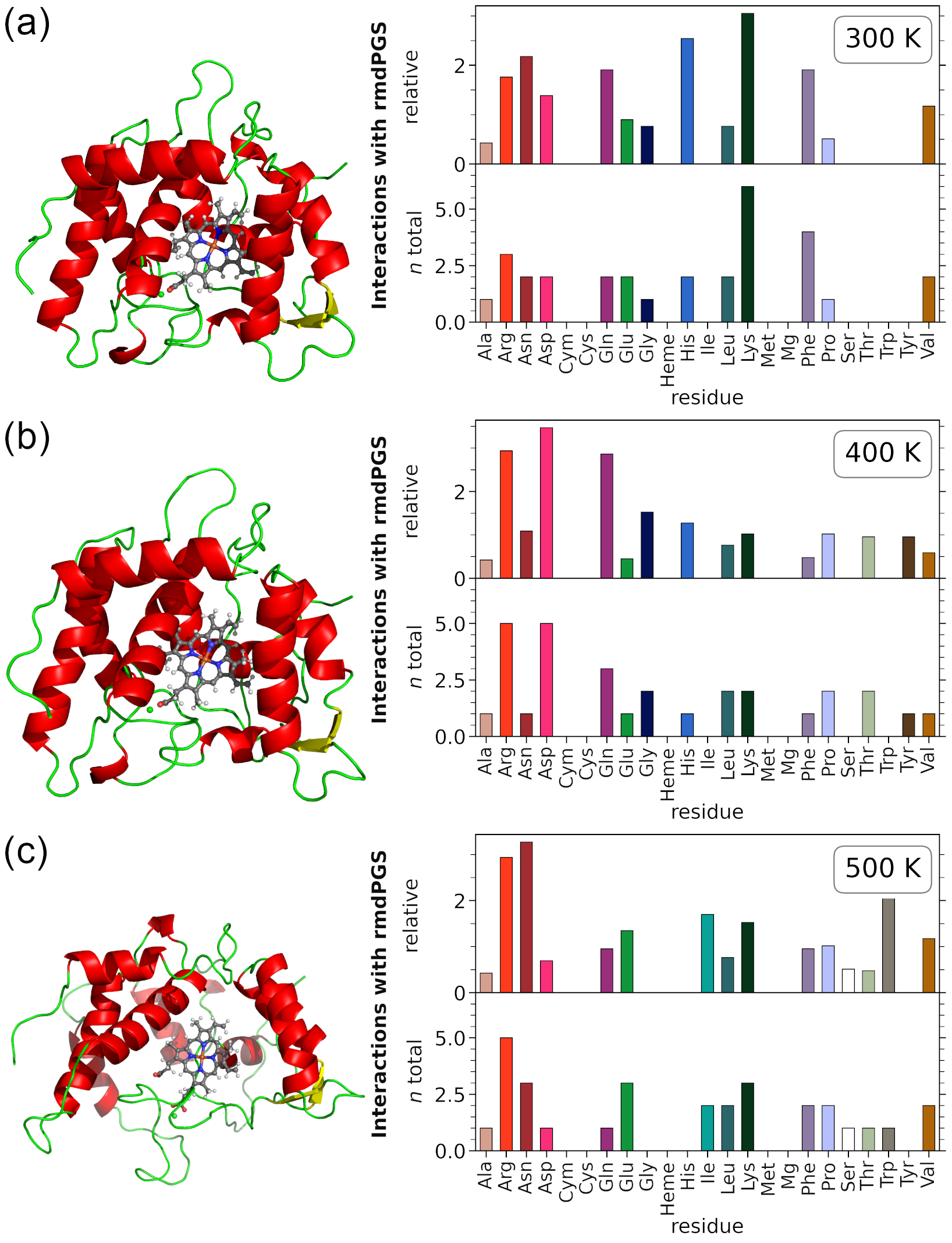}
    \caption{3D representations of \textit{Cvi}UPO (left) and SASA interaction profile (right) for \ch{OH} at different temperatures: (a) at 300~K, (b) 400~K, and (c) at 500~K. The SASA procedure and analysis are consistent with those, used to generate the previous figure. The 3D representations were generated using PyMOL \cite{PyMOL}, with input structures derived from the output of the equilibration simulations at the specified temperatures. For a detailed description of the SASA methodology, refer to the computational details in the methods section.}
    \label{Cvi_temp}
\end{figure}

The interaction profiles change when the temperature rises and the protein begins to unfold. The exact denaturation temperatures are different from enzyme to enzyme, but most of them are inactivated at temperatures above 60\,°C (338,15\,K). Theoretically, higher test temperatures up to 350\,K could also be sufficient. However, proteins in MD simulations do not always behave identically to those in experiments. Additionally, fluctuations in thermostats during simulations often range between 10\,K and 20\,K. Therefore, alongside 300\,K, two higher temperatures — 400\,K and 500\,K — were tested. These temperatures are extremely high by experimental standards, ensuring that the protein is at least partially unfolded at 400\,K and fully denatured at 500\,K. The results for \ch{OH} at 300\,K, 400\,K and 500\,K can be found in Figure \ref{Cvi_temp}. For the other species the results can be found in the supplementary information (Supplementary Figure \ref{SI-SASA-1} to \ref{SI-SASA-3}).
Figure \ref{Cvi_temp} already indicates some differences in behavior at 300\,K and 400\,K.  The most notable observation is the decrease in interactions with Lys and Phe at 400\,K, accompanied by an increase in interactions with Arg and Asp. This shift may be attributed to the initial stages of enzyme unfolding, which likely exposes charged amino acids more to \ch{OH}, thereby enhancing the likelihood of interaction.
The unfolded protein at 500\,K has a greater surface area, exposing even more amino acid residues for potential interactions. But the overall interaction profile of \ch{OH} is consistent with the lower temperatures.
For the other tested \textit{rmd}PGS this is not necessarily the case. With increasing temperature most of the \textit{rmd}PGS display a shift in their interactions towards protein residues which have been protected by the surrounding structure before, like the heme center (Supplementary Figure \ref{SI-SASA-1} to \ref{SI-SASA-3} in the SI). Overall, higher temperatures appear to increase the chance of fatal modification on crucial parts of the enzyme, as key residues become more exposed to reactive species.

To validate whether the \textit{rmd}PGS really react with the protein in the predicted position those ten SASA points, that have the lowest interaction energies, were investigated in detail (the precise coordinates and the corresponding amino acids for each point are listed in supplementary Table \ref{tab:SI-Cvi_SASA_MD}). After an initial energy minimization, a short MD simulation of 75~fs was performed in order to check whether chemical reactions occur. This brief time period was selected to prevent the probe molecule from diffusing to locations other than the predicted position, ensuring that interactions occur specifically at the predicted sites. The simulations were conducted for all \textit{rmd}PGS at 300, 400, and 500\,K in vacuum as well as in the solvent phase.
None of the vacuum simulations showed an immediate bond formation between the \textit{rmd}PGS and the protein during the energy minimization step. Apart from a few exceptions, after 75~fs only \ch{H}, \ch{O} and \ch{OH} form bonds to the enzyme at the predicted positions with the high interaction energies. Of the 10 positions tested, hydrogen binds in eight positions at 300\,K, while oxygen binds in seven and \ch{OH} and \ch{NO} in only five and one of the tested positions, respectively (compare Supplementary Table \ref{tab:SI-Cvi_SASA_MD}). Naturally, \ch{H} binds to oxygen, while \ch{O} and \ch{OH} only bind to hydrogen atoms. 
The detailed results of this analysis are provided together with the simulation data in addition to the paper.
This result is consistent at all three tested temperatures. The other \textit{rmd}PGS tend to associate at the protein surface but they do not form covalent bonds to one of the protein atoms. 
In the solvent simulations, only few bonds are forming between the protein and the \textit{rmd}PGS in this short time period (compare Supplementary Table \ref{tab:SI-Cvi_SASA_MD_solv} in the SI). Most of the species do not react with the protein in the predicted position but rather interact with the solvent molecules. This is a trend that also continues with longer simulations that allow diffusion (see Section Interactions with high \textit{rmd}PGS concentrations\ref{Concentration}). 

This validation shows that the algorithm is indeed able to predict the positions of the protein surface that are most likely to be modified by \textit{rmd}PGS binding, but a high interaction energy at a particular position does not automatically mean that a \textit{rmd}PGS will bind there. A subsequent verification is always recommended. 

%%%%%%%%%%%%%%%%%%%%%%%%%%%%%%%%%%%%%%%%%%%%%%%%%%%%%%%%%%%%%%%%%%%%%%%%%%
\subsubsection{Comparison to \textit{Aae}UPO and GapA} \label{SASA-Comp}
\begin{figure}[H]
    \centering
    \includegraphics[width=1\textwidth]{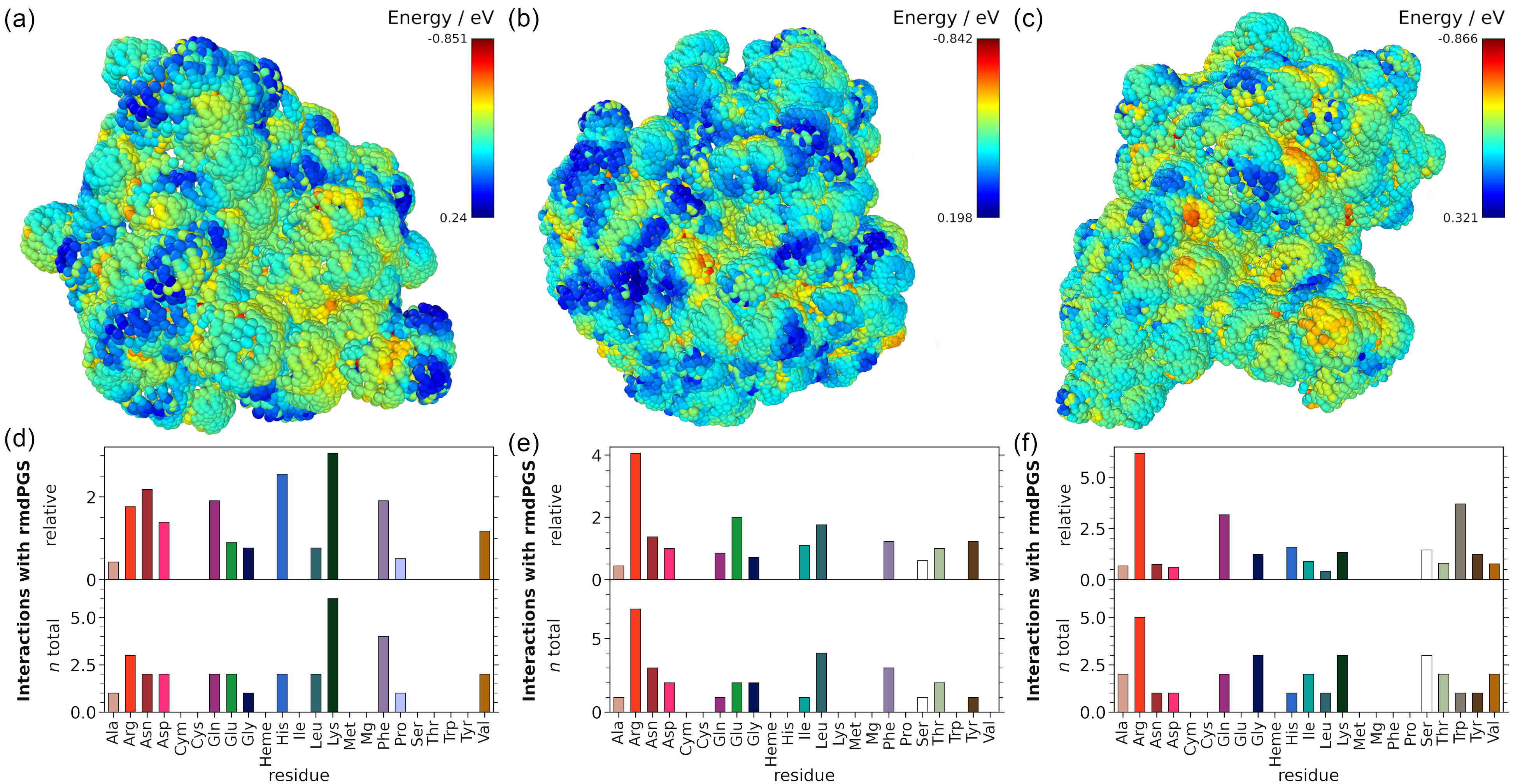}
    \caption{Comparison of the SASA interaction map of \ch{OH} at 300\,K between (a) \textit{Cvi}UPO, (b) \textit{Aae}UPO, and (c) GapA. The diagrams underneath show the respective residue interactions. The SASA procedure and analysis were performed as described above, and methodological details are provided in the computational Details section.}
    \label{SASA_compair}
\end{figure}

To broaden the scope of our investigation, the interaction analysis was also performed for \textit{Aae}UPO, which is a model enzyme for \ch{H2O2}-dependent biocatalysis, and for GapA. GapA is a glycolytic enzyme, that does not have a heme center, but uses a cysteine thiol group as its catalytic center instead. While \ch{H2O2}-based heme poisoning is a common problem of peroxidases, GapA is inactivated reversibly in an \ch{H2O2}-dependent fashion by formation of an intramolecular disulfide bond which engages the catalytic cysteine.
Figure \ref{SASA_compair} shows the comparison of the \ch{OH} interactions with the three enzymes. Similar to Figure \ref{SASA_Cvi} the interaction maps are displayed in the upper panels and the interactions with amino acids in the lower panels. The interactions of the other tested species can be found in the supporting information (Supplementary Figure \ref{SI-SASA_Aae} to \ref{SI-SASA-GapA-2}).   
Even though the three enzymes are differently-shaped, the interactions with the \textit{rmd}PGS follow similar trends. For all three enzymes \ch{OH} shows strong interactions with Arg. The main difference is the strong interaction of \ch{OH} with Lys and Phe in \textit{Cvi}UPO. In \textit{Aae}UPO \ch{OH} has moderate interactions with Phe but not with Lys, whereas the opposite pattern is observed for GapA, where \ch{OH} interacts to some extend with Lys but not with Phe. However, \textit{Aae}UPO shows few strong interactions at Lys also with the other reactive species, such as \ch{O2} (Supplementary Figure \ref{SI-SASA-2} (e)), likely because Lys occurs less frequently in \textit{Aae}UPO than in \textit{Cvi}UPO, which can be clearly seen in the relative data. The same argument applies to GapA, Phe less abundant this enzyme compared to \textit{Cvi}UPO.
While the interaction partners for \ch{H} are almost identical for all three enzymes, the interactions with \ch{H2O2} show some variation between the proteins. In \textit{Aae}UPO, \ch{H2O2} interacts most strongly with Arg and Asp in absolute and Trp in relative terms, whereas in GapA, the strongest interactions are observed with Arg, threonine (Thr) and His. The His interaction is even more pronounced in the relative representation. These differences are probably related to the different shapes and compositions of the proteins. GapA has far more beta sheets in its protein structure, of which Thr is a central component, making it easier to access.
The \ch{NO} and \ch{O} results are again similar for all three proteins, with Glu being targeted by \ch{NO} and Arg and Lys interacting with \ch{O}. Furthermore, NO interacts with only few residues in GapA when compared to \textit{Aae}UPO and \textit{Cvi}UPO. This might suggest that GapA has a low affinity for NO and that it is well protected from modifications by these species, but this would need to be confirmed experimentally.
\ch{O2} interacts most strongly with Arg and Lys in \textit{Cvi}UPO and GapA. In contrast, for \textit{Aae}UPO, the strongest interactions are with Arg, Leu, Phe, and Glu. \ch{H} and \ch{NO} have the least overall interaction with the surface of the different enzymes. Of course, there are certain sites that interact strongly with these species, but overall there are fewer interactions than with the other species tested. \\
Similar to \textit{Cvi}UPO, for \textit{Aae}UPO only two interactions with low energy are detected near the heme center. Both are interactions with \ch{H2O2}, one with Glu196 and the other with Phe199.  For GapA \ch{H2O2}, \ch{O2}, \ch{OH} and \ch{O} are the species most prone to interact with the reactive center, indicated by the low energy interactions in the active pocket. They all interact with Thr152 and Gly210, but only \ch{O2} has a high interaction energy with a hydrogen atom ($\mathrm{HN_{2295}}$) from the reactive cysteine group Cys150.

The result emphasizes that the species important for catalysis, \textit{e.g.} \ch{H2O2} and \ch{O2}, are pulled towards the active site in all three enzymes, but in comparison, the number of interactions at the active site is lower than at other positions on the protein surface. This is partly due to the fact that there is only one reactive center in each of the enzymes studied, so the probability of interaction with this center is lower than with the more abundant residues of the enzyme. Moreover, the reactive center of the protein is less accessible than the amino acids on the protein surface.

With increasing temperature, the interactions of the \textit{rmd}PGS with \textit{Aae}UPO and GapA change in a similar way as for \textit{Cvi}UPO. The residues with the most interactions change slightly because new interaction spots are available when the protein starts unfolding. Therefore, the change between 300\,K and 400\,K is small, because the proteins are still quite stable at these temperatures. However, at around 500\,K, the proteins lose their structure and shape, resulting in a more pronounced change in the interaction profile. Interestingly, \textit{Cvi}UPO and GapA have the most positions with a favorable interaction energy in the active pocket at 400\,K. At 500\,K the active pocket has fully collapsed due to the unfolding and access pathways to the active center are blocked. \textit{Aae}UPO on the other hand shows the most interactions within the active pocket at 500\,K.  
The short validation MD simulations of the predicted positions in \textit{Aae}UPO and GapA also showed similar results to \textit{Cvi}UPO. In vacuum, mostly \ch{H}, \ch{O} and \ch{OH} instantly react with the enzymes at the predicted positions (compare Supplementary Tables \ref{tab:SI-Aae_SASA_MD} -- \ref{tab:SI-GapA_SASA_MD} in the SI). The hydrogen atoms bind to an unsaturated oxygen atom, while \ch{O} and \ch{OH} abduct hydrogen atoms from the protein. This behavior is independent of the tested temperature. In the solvent we also observe identical behavior as for \textit{Cvi}UPO. Most \textit{rmd}PGS rather interact with the solvent than with the proteins. 

%%%%%%%%%%%%%%%%%%%%%%%%%%%%%%%%%%%%%%%%%%%%%%%%%%%%%%%%%%%%%%%%%%%%%%%%%%
\subsection{Interactions of \textit{Cvi}UPO with high \textit{rmd}PGS concentrations} \label{Concentration}
\begin{figure}[H]
    \renewcommand\thesubfigure{\roman{subfigure}}
    \centering
    \begin{subfigure}{0.49\textwidth}
    \includegraphics[width=1\textwidth]{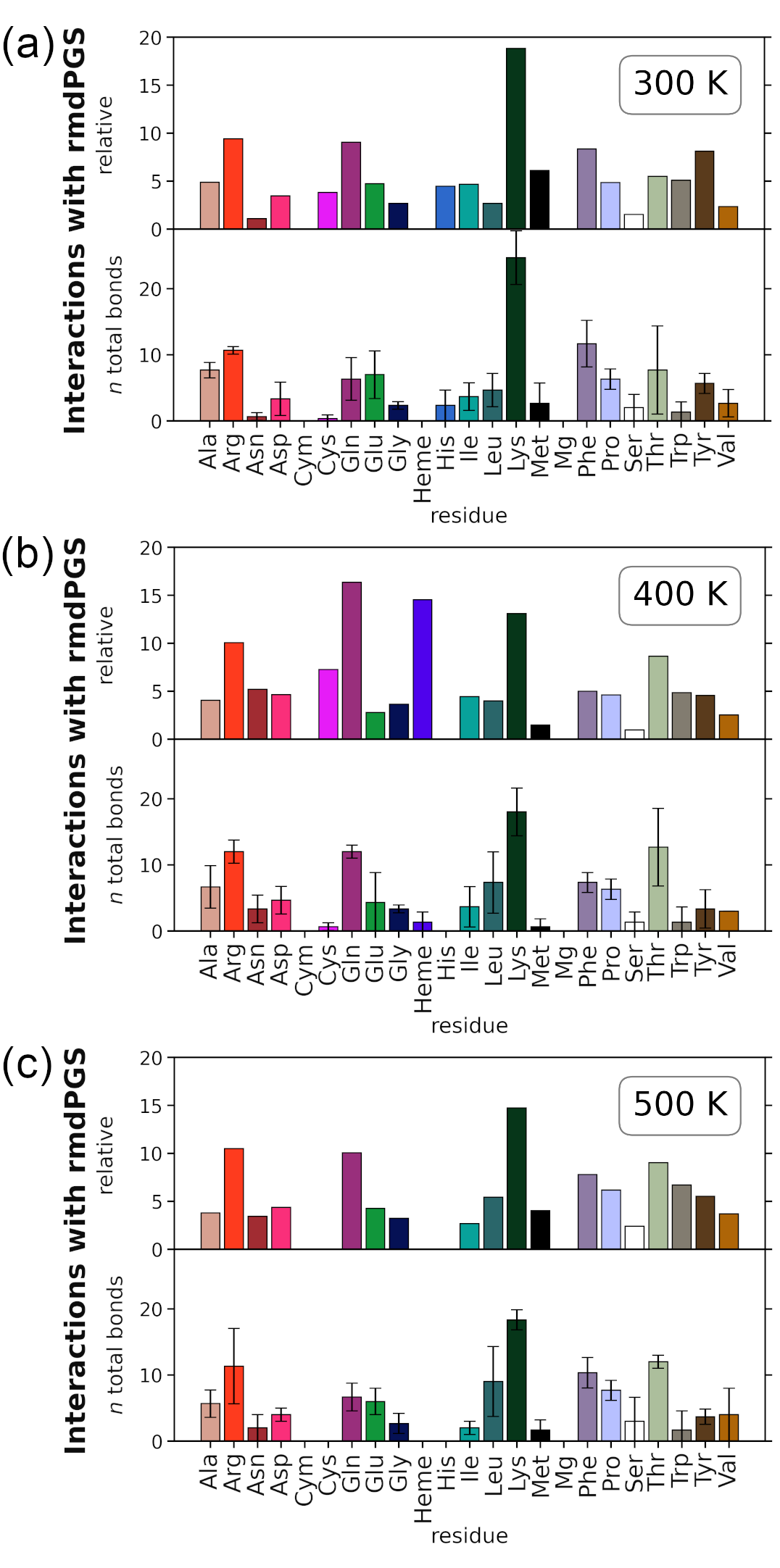}
    \caption{Without solvent}
    \end{subfigure}
    \begin{subfigure}{0.49\textwidth}
    \includegraphics[width=1\textwidth]{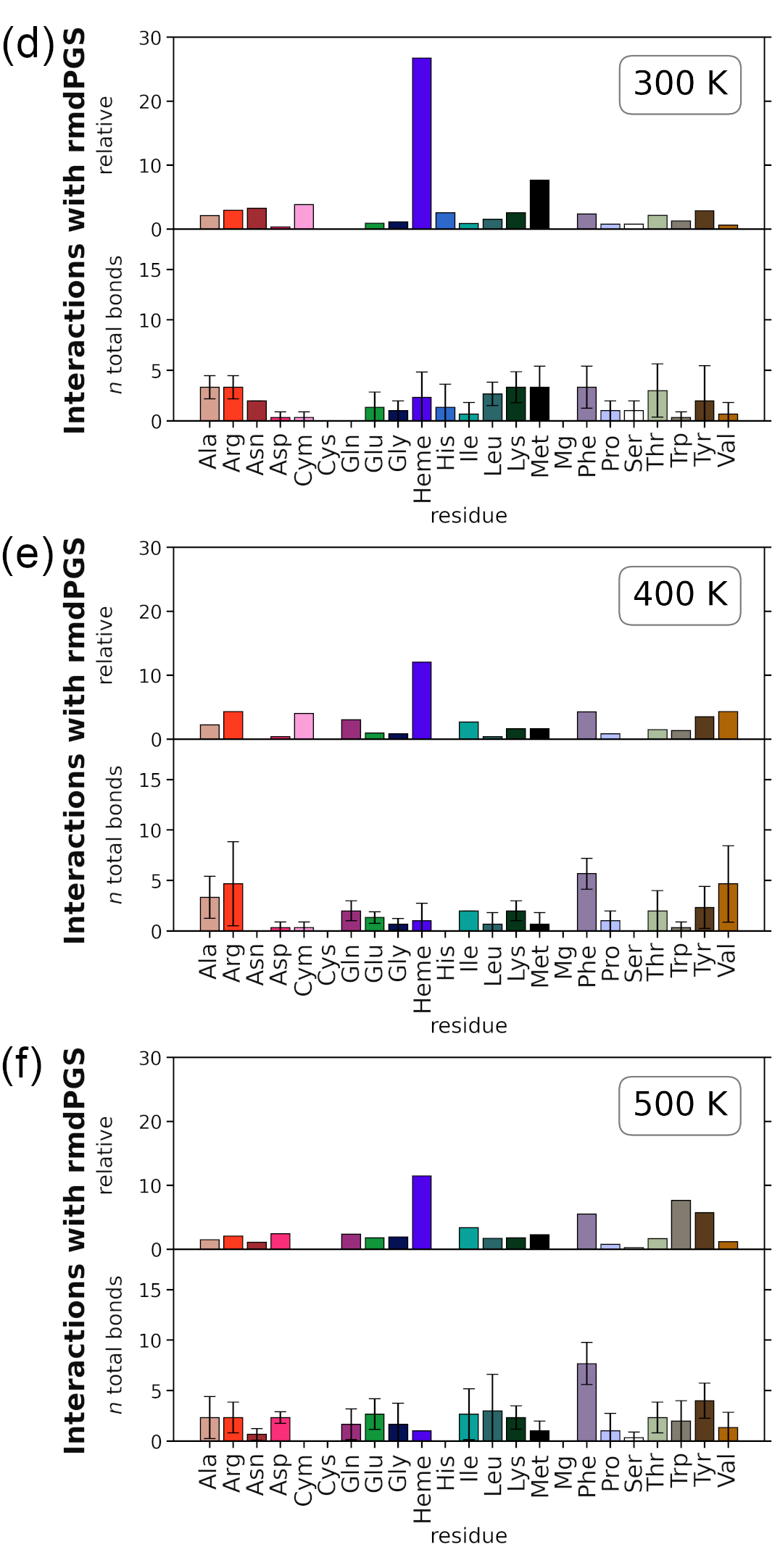}
    \caption{With solvent}
    \end{subfigure}
    \caption{Bond analysis for high concentrations of hydrogen with \textit{Cvi}UPO. All panels show the total number of bonds to an additional \ch{OH} atom per residue and the relative interactions calculated following equation \ref{eq:2.2}. The legend is equivalent to that in the previous graphs. The panels (a) and (d) show the interactions at 300\,K, without and with solvent, respectively. The panels (b) and (e) show the interactions at 400\,K and the panels (c) and (f) show the interactions at 500\,K. All simulations were conducted as described at the beginning of this section as well as in the Computational Details section.}
    \label{Conc_MD_Cvi}
\end{figure}

Under experimental conditions the \textit{rmd}PGS will not diffuse towards the enzyme systematically, but rather via random diffusion from the interface of the plasma (DBD) or the effluent (\textmu APPJ, CPJ) with the liquid through the solvent to the protein. To approach more realistic conditions we performed MD simulations with a high concentration of \textit{rmd}PGS in a random configuration around the enzyme. These simulations are intended to serve as an extended proof of concept for SASA fast screening approach. The simulations were performed at different temperatures (300\,K, 400\,K and 500\,K) and repeated 3 times with different random orientations of the \textit{rmd}PGS. The total simulation time for these simulations was 75\,ps with a 0.15\,fs time step. Additionally, they were conducted in vacuum and in solvent. The results for the interactions of 100 \ch{OH} atoms with \textit{Cvi}UPO are display in Figure \ref{Conc_MD_Cvi}, the other species can be found in the SI (Supplementary Figure \ref{SI-Conc_MD_Cvi-h} to \ref{SI-Conc_MD_Cvi-o}).

Figure \ref{Conc_MD_Cvi} (a) shows that in absence of solvent \ch{OH} is binding frequently to Lys, Phe and Arg at 300\,K, which aligns with the SASA predictions (see Figure\,\ref{Cvi_temp}. At higher temperatures this trend continues and as predicted by the SASA analysis the interactions with Gln increase (Figure \ref{Conc_MD_Cvi} (b) and (c)).
For the other \textit{rmd}PGS, the predictions also agree well with the MD simulation results. The favored interaction partners for \ch{H} are still Glu and Lys (Supplementary Figure \ref{SI-Conc_MD_Cvi-h}) and \ch{O} has nearly identical interactions in the simulations and the SASA prediction (Supplementary Figure \ref{SI-Conc_MD_Cvi-o}) . \ch{H2O2} has less interactions with Ser and Glu than expected based on the SASA analysis, in contrast to that Arg has more interactions especially at 500\,K (Supplementary Figure \ref{SI-Conc_MD_Cvi-h2o2}). Similar to the other \textit{rmd}PGS NO has the most interactions with Lys, but Asp and Glu show less interactions compared to the prediction from the SASA analysis (Supplementary Figure \ref{SI-Conc_MD_Cvi-no}). Interestingly, with rising temperature this deviation decreases. For \ch{NO} the total interactions per residue are again similar, but Tyr and Val do not have as many interactions in the MD simulation results. The only outlier appears to be \ch{O2}, but \ch{O2} has only few interactions with the enzyme in the MD simulations that it is difficult to compare with the prediction (Supplementary Figure \ref{SI-Conc_MD_Cvi-o2}). Overall, the ratio and trend of the favored amino acids stay highly similar within the observed temperature range, which also emphasizes the great correspondence to the SASA prediction. \\
The trend from the simulations without solvent is also visible in the simulations with solvent (Figure \ref{Conc_MD_Cvi} (d)-(f)), despite the high errorbars. A detailed analysis of the newly formed bonds involving the additional \ch{OH} revealed that, in the solvent simulations, the majority react with the solvent rather than with the enzyme surface. This is in accordance with the short validation simulations from Section "SASA interaction analysis \textit{Cvi}UPO"\ref{SASA}. The surrounding water molecules protect the enzyme from being modified by the \textit{rmd}PGS. \\
The solvent simulations with the other \textit{rmd}PGS also follow the trend from the SASA predictions (compare Supplementary Figures \ref{SI-Conc_MD_Cvi-h} to  \ref{SI-Conc_MD_Cvi-o}). However, this observation should be taken with care, because of the relatively large error bars.  The analysis of the high concentration simulations emphasizes that not only properties of the protein surface play a role but also the initial position and the diffusion pathways of the \textit{rmd}PGS towards the protein.\\
Some differences between the SASA predictions and the MD simulations are expected, as the SASA method works with predetermined points and only considers the most favorable interactions, while the MD simulations work with a fixed concentration of \textit{rmd}PGS in a random distribution. However, the reliability of the SASA prediction is encouraging.\\
To achieve more reliable statistics, this type of simulation would need to be repeated more often with varying initial conditions. While this approach would improve statistical accuracy, it is also highly time-intensive. Given that the trends observed in these simulations align with the SASA results, the SASA method is certainly an interesting rapid approach for a fast screening of protein-plasma interactions.

The same behavior can also be observed with the enzymes \textit{Aae}UPO and GapA, a detailed description of their interaction can be found in the SI.

%%%%%%%%%%%%%%%%%%%%%%%%%%%%%%%%%%%%%%%%%%%%%%%%%%%%%%%%%%%%%%%%%%%%%%%%%%
\iffalse
- Experimental tests of plasma treatment of AS showed that TYR prone to different oxidation products
- The aromatic ring of Tyr has been reported to be easily hydroxylated  and nitrated, thus oxidation likely occur in the aromatic ring of Tyr
- Similar to TYR also the aromatic ring of PHE is easily oxidized
- Main modification is introduction of O into structure \cite{Guo2023Reactivemoleculardynamics}
- Aromatic rings are stable and remain uncharged even after H abstraction \cite{Guo2023Reactivemoleculardynamics}
- 5 rings can be oxidized to ring open products \cite{Guo2023Reactivemoleculardynamics}
\fi
%%%%%%%%%%%%%%%%%%%%%%%%%%%%%%%%%%%%%%%%%%%%%%%%%%%%%%%%%%%%%%%%%%%%%%%%%%

\subsection{Mass spectrometry of plasma-treated \textit{Cvi}UPO}
In order to investigate the interactions of amino acids of the \textit{Cvi}UPO with PGS, the r\textit{Cvi}UPO was treated with the DBD for 5~min and was then digested with trypsin for MS measurements.  
Beside Met and Cys, the amino acids Phe, Arg, and Lys were chosen for the modification analysis because of their predicted strong interaction with ROS such as OH (SASA interaction analysis). Initially, only reactions with ROS and not with RNS were investigated experimentally. The mass spectrometrical analysis showed highly significant oxidation of Phe51, Phe67, Phe208 (Phe+O), Met42, and Met220 (Met+O) in all three replicates after 5~min of plasma treatment (Table 1).

\begin{table}[H]
\caption{\textbf{Amino acid residues of \textit{Cvi}UPO oxidized by plasma treatment.} For plasma treatment, 40~\textmu l of r\textit{Cvi}UPO (1~mg~ml\textsuperscript{-1}) were applied to a metal plate and treated with DBD plasma for 5 min (electrode diameter 20 mm; 13.5 kV pulse amplitude; 300 Hz trigger frequency; 1 mm distance to sample). Protein samples were collected, reduced with TCEP, alkylated with iodoacetamide, and digested with trypsin for LC-MS/MS measurement. Five amino acid residues were found to be oxidized by plasma treatment.  The average relative intensities of the peptides containing these five amino acids in unmodified form (bold, black) or oxidized (bold, red) in untreated controls and plasma-treated samples are shown sorted by the starting position of the peptide (N=3). Modifications were detected in all three replicate experiments with relative intensities $>$100. n.d.: not detected.}
    \resizebox{\textwidth}{!}{%
    \begin{tabular}{|c|c|c|r|c|c|}
    \hline
        \textbf{Amino acid} & \textbf{Starting pos.} & \textbf{Peptide nr.} &  \multicolumn{1}{|c|}{\textbf{Peptide sequence}}   & \textbf{\parbox{3cm}{Average relative intensity (\textbf{Control})}} & \textbf{\parbox{3cm}{Average relative intensity (\textbf{Plasma})}} \\ \hline
        \textbf{Met42} & 33 & 1 & DGRNITVA\textbf{\textcolor{red}{M}}LVPVLQEVFHLSPELAQTISTLGLFTAQDPSK & 21 & 2254 \\
        ~ & 33 & 2 & DGRNITVA\textbf{M}LVPVLQEVFHLSPELAQTISTLGLFTAQDPSK & 552 & 24 \\
        ~ & 36 & 3 & NITVA\textbf{\textcolor{red}{M}}LVPVLQEVFHLSPELAQTISTLGLFTAQDPSK & 32 & 417 \\ \hline
        \textbf{Phe51} & 36 & 1 & DGRNITVAMLVPVLQEV\textbf{\textcolor{red}{F}}HLSPELAQTISTLGLFTAQDPSK & 19 & 590 \\ 
        ~ & 33 & 2 & DGRNITVAMLVPVLQEV\textbf{F}HLSPELAQTISTLGLFTAQDPSK & n.d. & 24 \\ \hline
        \textbf{Phe67} & 33 & 1 & DGRNITVAMLVPVLQEVFHLSPELAQTISTLGL\textbf{\textcolor{red}{F}}TAQDPSK & 3 & 265 \\ 
        ~ & 33 & 2 & DGRNITVAMLVPVLQEVFHLSPELAQTISTLGL\textbf{F}TAQDPSK & 552 & 24 \\
        ~ & 51 & 3 & HLSPELAQTISTLGL\textbf{F}TAQDPSK & n.d. & 2 \\
        ~ & 61 & 4 & STLGL\textbf{F}TAQDPSK & 2 & 3 \\ 
        ~ & 63 & 5 & LGL\textbf{F}TAQDPSK & 4 & 2 \\ 
        ~ & 64 & 6 & GL\textbf{F}TAQDPSK & 6 & 6 \\ 
        ~ & 66 & 7 & \textbf{F}TAQDPSK & 4 & 5 \\ \hline
        \textbf{Phe208} & 202 & 1 & AELSG\textbf{\textcolor{red}{F}}SMASDVLELALVTPEK & 37 & 727 \\ 
        ~ & 202 & 2 & AELSG\textbf{F}SMASDVLELALVTPEK & 253 & 3 \\ \hline
        \textbf{Met220} & 202 & 1 & AELSGFS\textbf{\textcolor{red}{M}}ASDVLELALVTPEK & 34 & 731 \\ 
        ~ & 202 & 2 & AELSGFS\textbf{M}ASDVLELALVTPEK & 253 & 3 \\ \hline
    \end{tabular}}
\end{table}
The strong interaction with PGS and Phe may be explained by the fact that many Phe residues in \textit{Cvi}UPO are exposed on the outside of the protein. The aromatic ring of Phe seems to react with OH or other highly reactive oxygen species. Two of the modified phenylalanines, Phe51 and Phe67, are located near the substrate channel and their modification could influence the enzyme activity. The chemical reactions of free amino acids with plasma-derived RONS have been extensively studied and it has been shown that Phe is rapidly oxidized in the aromatic ring and hydroxylation or nitration could be detected after longer plasma treatment of 10~min \cite{Takai.2014}.
Although experimental investigations have confirmed specific modifications, including the oxidation of aromatic amino acids such as Phe in DBD plasma-treated glycinin \cite{Liu.2021}, plasma-induced amino acid modifications in proteins have not been extensively studied to date. Hence, SASA-based interaction predictions are highly relevant, as the observed modifications at Phe were significant and in agreement with our experimental data.\\
Since the active site of an enzyme is responsible for substrate and co-substrate binding and thus also for enzymatic activity, the investigation after plasma treatment is of particular importance. The amino acids His90, Thr158, Phe88, Glu162, Tyr166 and Cys19 are primarily involved in the catalytic cycle. \cite{GonzalezBenjumea.2020} The residues His90, Glu162 and Tyr166, are conserved and located distal to the catalytic center, while the proximal Cys19 is involved in the reaction. Residues Phe88 and Thr158 are also involved in substrate-heme interactions close to the heme access channel. Although interactions with amino acids in the active center of the \textit{Cvi}UPO like Phe88 were not observed in the experimental study, this could occur with longer plasma treatment times.\\
While Phe can undergo hydroxylation reactions, sulfur-containing amino acid residues in proteins like Cys or Met exhibit greater susceptibility to oxidative modifications and rapidly react with ROS \cite{Lackmann.2015}. The thiol group of cysteine readily reacts with ROS, leading to the formation of disulfide bonds, sulfenic (\ch{-SOH}), sulfinic (\ch{-SO2H}), or sulfonic (\ch{-SO3H}) acids, significantly altering protein structure and function \cite{Lackmann.2015}. Methionine, with its thioether group, is also a prime target for oxidation, resulting in the formation of methionine sulfoxide and, under prolonged oxidative stress, methionine sulfone \cite{Takai.2014}. Interaction profiles from SASA showed no interactions between PGS and Met or Cys, which we explain by the fact that these are buried in the protein structure and not on the solvent-accessible surface or are already engaged in covalent bonds (Cys19-Heme). However, mass spectrometry analysis showed modified Met42 and Met220 in all replicates, resulting in the majority of Met to be oxidized after 5 min treatment (Met+O) .\\
SASA interaction profiles predicted strong interactions between PGS such as OH, \ch{H2O2}, \ch{O2} and the basic amino acids Lys and Arg, which are both known to be proton donor residues. However, amino acid modifications such as dehydrogenation of arginine, hydroxylation and/or dehyrogenation of lysine were detected only in some replicates and at low relative intensity. Due to the positive charge, these amino acids could be modified by reactive nitrogen species, which has not yet been investigated experimentally. For more detailed analyses, further possible modifications and plasma treatment times should be included. In addition, modifications to other amino acids should be investigated and the sequence coverage should be increased from 70\% in the current analysis. If amino acids in the active center of the enzyme are modified, this could lead to enzyme inactivation. Depending on the residues affected, enzyme engineering could be a viable approach to selectively exchange amino acids, thereby improving protein plasma stability. \\

%%%%%%%%%%%%%%%%%%%%%%%%%%%%%%%%%%%%%%%%%%%%%%%%%%%%%%%%%%%%%%%%%%%%%
%% Conclusion
%%%%%%%%%%%%%%%%%%%%%%%%%%%%%%%%%%%%%%%%%%%%%%%%%%%%%%%%%%%%%%%%%%%%%
\section{Conclusion} \label{conclusion}
The SASA analysis method presented here allows to identify regions of interest on a protein surface with respect to interaction with a small reactive molecule. This method is not only applicable to protein--plasma interactions but also to other problems, such as the prediction of ligand binding sites on a protein. This approach could be used for a first and rather rough scan of the protein surface for attractive binding positions. 
As discussed in detail at the end of section \ref{SASA_Cvi}.1  ("SASA interaction analyis -- \textit{Cvi}UPO"),  a subsequent validation of any predicted positions is always recommended because a high interaction energy at a certain position does not necessarily mean that a species will bind there. Especially under real experimental condition the presence of the solvent plays an important role. Here, the simulation indicated that most reactive species rather react with water molecules than with the protein surface. Nevertheless, the simulations in the section "Interactions of \textit{Cvi}UPO with high \textit{rmd}PGS concentrations"\ref{Concentration} showed that even in random configuration with enough time some of the reactive species will bind to the enzymes on similar residues as predicted by the SASA analysis method. The results of this study highlight that any experimental setup that prolongs the diffusion path of reactive species in the solvent leads to a reduction in modifications to the enzymes. In the case studied here, prolonged diffusion through the liquid phase increases the probability that the PGS will react with the solvent rather than modifying the enzymes, which is consistent with the experimental results from Yayci \textit{et~al.}\cite{Yayci2020Protectionstrategiesbiocatalytic}\,. In their study the binding of the enzymes to beads placed at the bottom of a beaker prolonged the diffusion path of the PGS towards the enzyme and also promoted steric interference for the PGS to reach potentially attractive binding sites.
Additionally, the experimental results from the mass spectrometry measurements, presented in this work, are in good agreement with the predictions from the SASA method. In particular, the predicted modification of phenylalanine was confirmed by the MS measurements, suggesting that ROS \ch{OH} and \ch{O} play an important role in the \textit{Cvi}UPO inactivation. The Phe modification by plasma treatment has thus been brought into focus by the SASA analysis, which prompted the inclusion of Phe modifications in the MS database search. This suggests that perhaps some of the other predictions of the method can also be verified experimentally. Although the predicted modifications of Arg, Lys, Cys, and Met were also searched for, only Met modifications were observed.
The combination of the results from both methods does not only help to gain deeper insights into the mechanism of plasma-induced modifications of the enzymes but also the SASA analysis method can help simplify target oriented evaluation of the MS data. In the future, further experimental studies should be carried out to test the general validity of the SASA model applying it to other systems.

\begin{acknowledgement}
C.J., J.E.B and T.J. gratefully acknowledge funding through the DFG (CRC 1316-2). The authors acknowledge support by the state of Baden-Württemberg through bwHPC and the DFG through grant no INST 40/575-1 FUGG (JUSTUS 2 cluster). J.E.B. is grateful to the State of North Rhine-Westphalia and the European Union - European Regional Development Fund "Investing in your Future", Research Infrastructure "Center for Systems Antibiotics Research - CESAR" for funding of the mass spectrometer.

\noindent The authors declare no conflict of interest.
\end{acknowledgement}

\begin{suppinfo}
\setcounter{figure}{0}  
\setcounter{table}{0}  

\subsection{Amino acid sequence of \textit{Cvi}UPO}
\label{cviUPO_sequence}
\begin{figure}[H]
    \centering
    \includegraphics[width=1\textwidth]{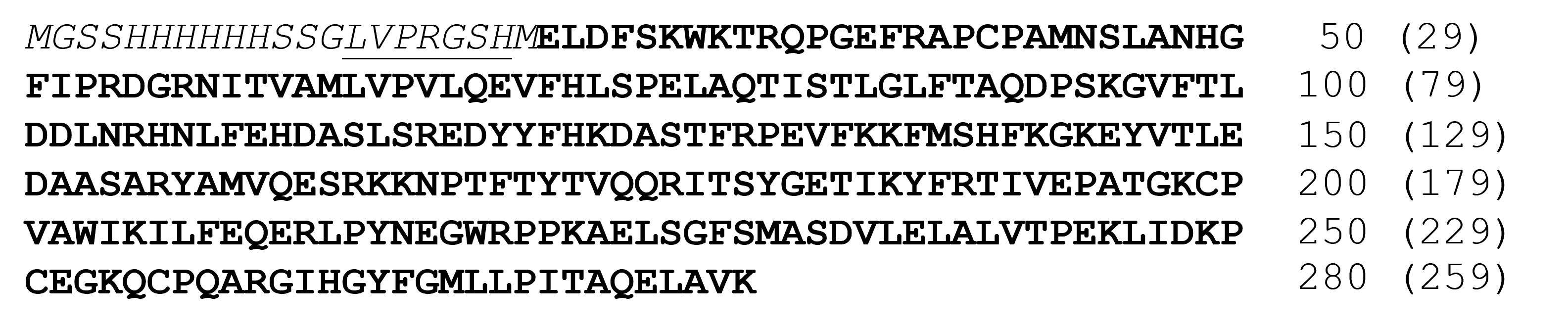}
    \caption{Amino acid sequence of the \textit{Collariella virescens} UPO used in the experimental study. The poly-His tail sequence is shown in italics and thrombin recognition sequence is underlined. The amino acid positions without the poly-His tail sequence (utilized in the text) are shown in brackets.}
    \label{sequence_cviUPO}
\end{figure}

\subsection{Explanation to the species nature in ReaxFF} \label{SI_Species}
It is important to note that, due to the limitations of the reactive molecular dynamics approach, modeling real plasma species remains a significant challenge.
Fundamentally, the primary limitation arises from the nature of the datasets used to parameterize force fields (FFs). High-quality FFs are typically derived from quantum mechanical (QM) calculations, which are inherently restricted to ground-state descriptions. Consequently, any molecular dynamics (MD)-scale representation of plasma species is necessarily limited in accuracy. While the explicit treatment of plasma species within the QM regime is a distinct and important topic, it falls outside the scope of the present study.
In a similar context, and most importantly, the ReaxFF employed in this study was not developed by us \cite{Monti2013Exploringconformationalreactive}. Therefore, we have no direct knowledge of whether specific species, such as OH radicals or OH ions, were included in the original training dataset. Even if such species were incorporated, the current ReaxFF framework does not explicitly treat electrons, making a rigorous distinction between radicals and ions fundamentally challenging. Efforts to address this issue, such as the development of eReaxFF, are ongoing, but they represent a separate and complex research direction.
By design, ReaxFF relies on the bond-order formalism to simulate bond dissociation and formation, setting it apart from conventional FFs. In addition to bond order, ReaxFF employs dynamic charge equilibration, allowing for the tracking of charge redistribution. For OH species, this means that charge variations in the surrounding environment or counter ions may provide some insights into partial charge changes \cite{Duin2001ReaxFFReactiveForce, Mortier1985Electronegativityequalizationapplication, Senftle2016ReaxFFreactiveforce}. 
For instance, an analysis of the bond order and partial charge variations among OH species could suggest possible differences in their chemical nature. However, at a fundamental level, the reliability of ReaxFF is entirely dependent on the quality of the training dataset and the methodology used in its parametrization. In particular, charge-related properties are known to be less accurate within this framework. As a result, distinguishing between OH ions and OH radicals in our study remains highly uncertain. \\
In summary, the ReaxFF employed here lacks the precision required to unambiguously differentiate OH radicals from OH ions. Consequently, any conclusions drawn from these simulations should be interpreted with caution. \\
While the limitations of the force field should be considered, they only moderately affect the interpretation of the results. This is because, in a real system, it can be assumed that species from the plasma effluent diffuse through several milliliters of solvent before interacting with the enzyme, likely losing their excitation and possibly their radical characteristics by the time they reach the enzyme.

\subsection{SASA interaction analysis} \label{SI_SASA}
\subsubsection{\textit{Cvi}UPO} \label{SI_SASA_Cvi}
\begin{figure}[H]
    \centering
    \includegraphics[width=1\textwidth]{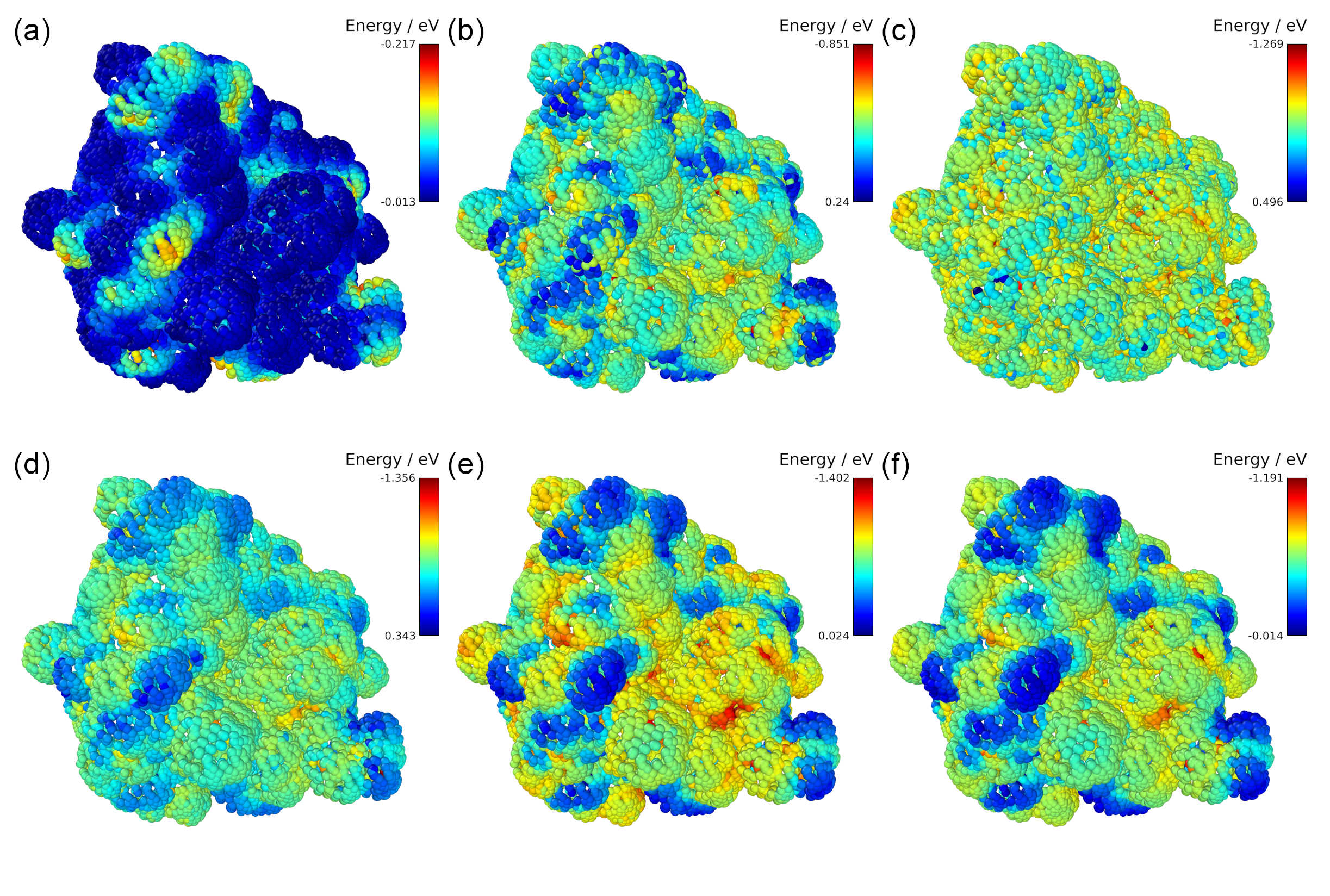}
    \caption{Interaction profiles from SASA for the enzyme \textit{Cvi}UPO at 300\,K with (a) \ch{H} and (b) \ch{OH}, (c) \ch{H2O2}, (d) \ch{NO}, (e) \ch{O2} and (f) \ch{O}.}
    \label{SI-SASA_Cvi}
\end{figure}

\begin{figure}[H]
    \renewcommand\thesubfigure{\roman{subfigure}}
    \centering
    \begin{subfigure}{0.49\textwidth}
    \includegraphics[width=1\textwidth]{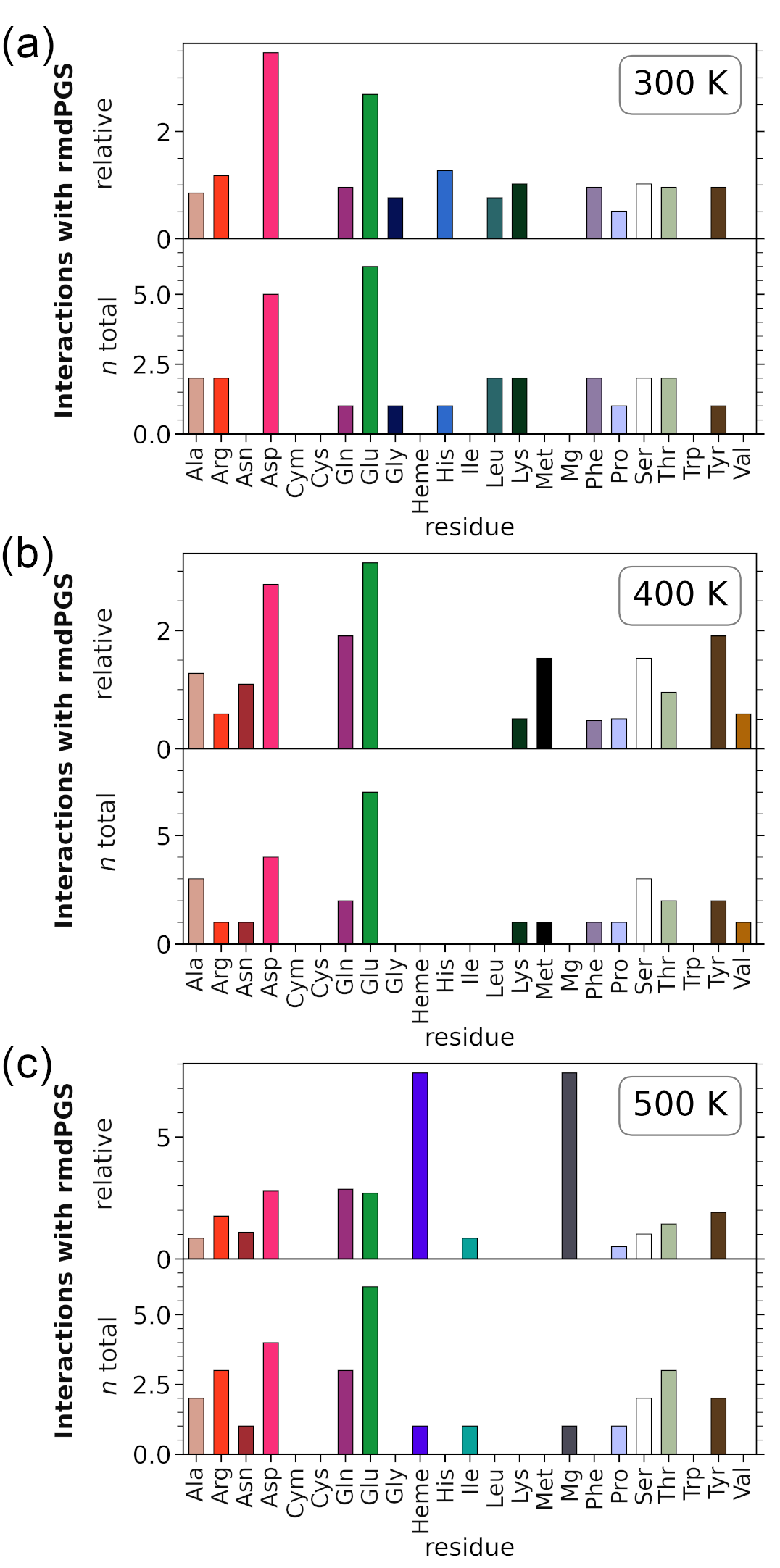}
    \caption{\ch{H}}
    \end{subfigure}
    \begin{subfigure}{0.49\textwidth}
    \includegraphics[width=1\textwidth]{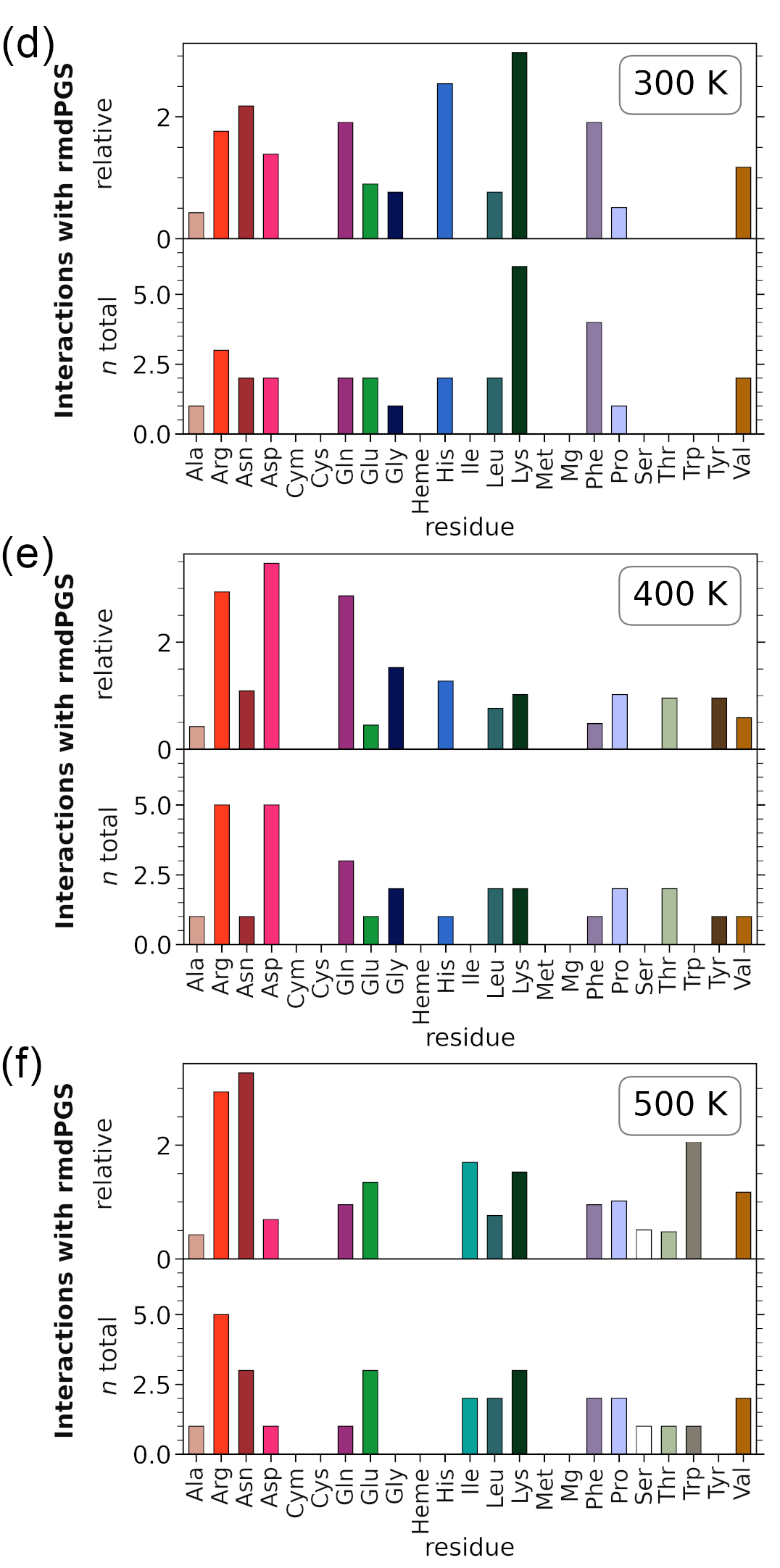}
    \caption{\ch{OH}}
    \end{subfigure}
    \caption{SASA interaction profile for \ch{H} and \ch{OH} with \textit{Cvi}UPO at different temperatures: (a) at 300\,K, (b) 400\,K, and (c) at 500\,K.}
    \label{SI-SASA-1}
\end{figure}

\begin{figure}[H]
    \renewcommand\thesubfigure{\roman{subfigure}}
    \centering
    \begin{subfigure}{0.49\textwidth}
    \includegraphics[width=1\textwidth]{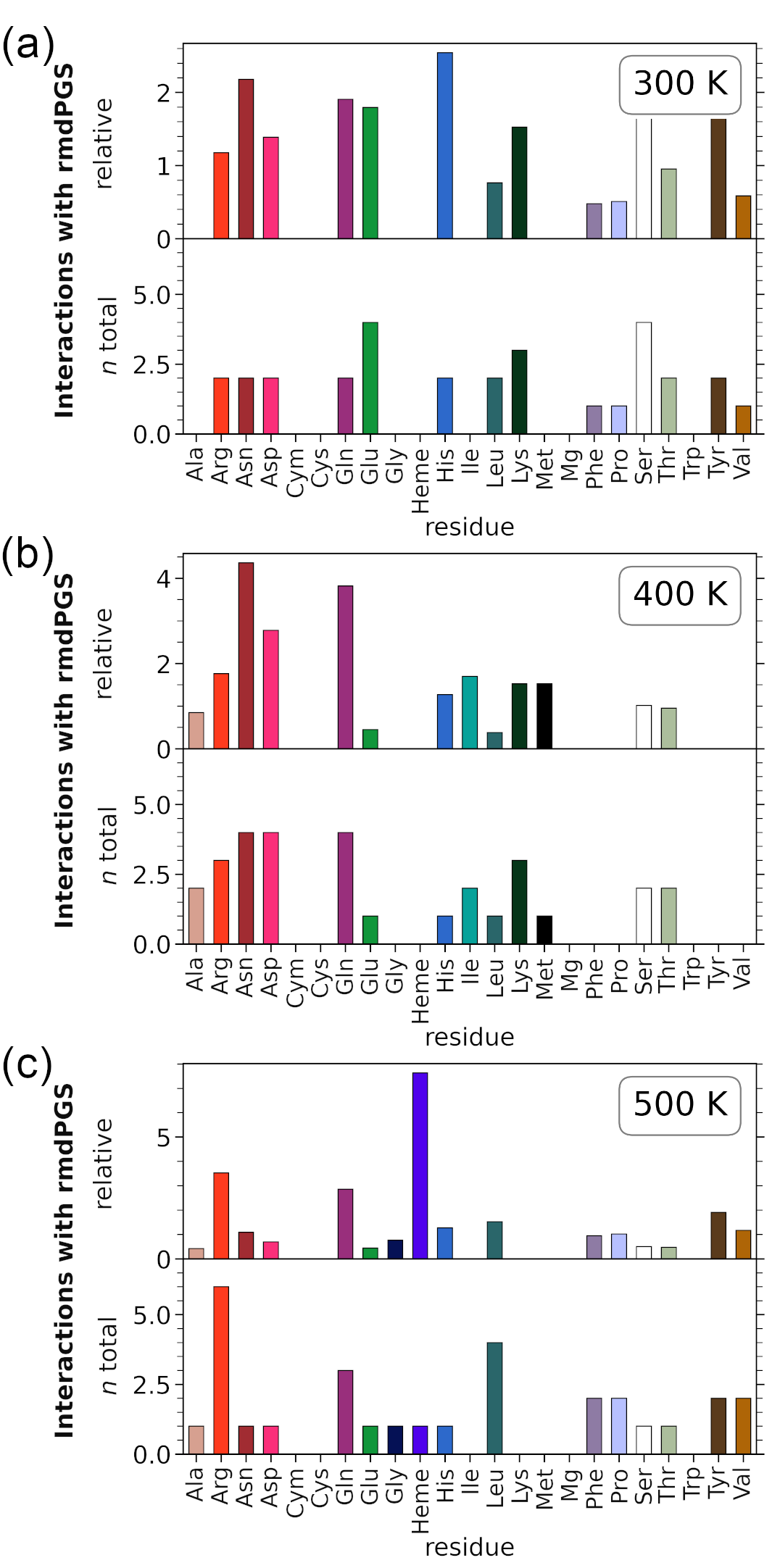}
    \caption{\ch{H2O2}}
    \end{subfigure}
    \begin{subfigure}{0.49\textwidth}
    \includegraphics[width=1\textwidth]{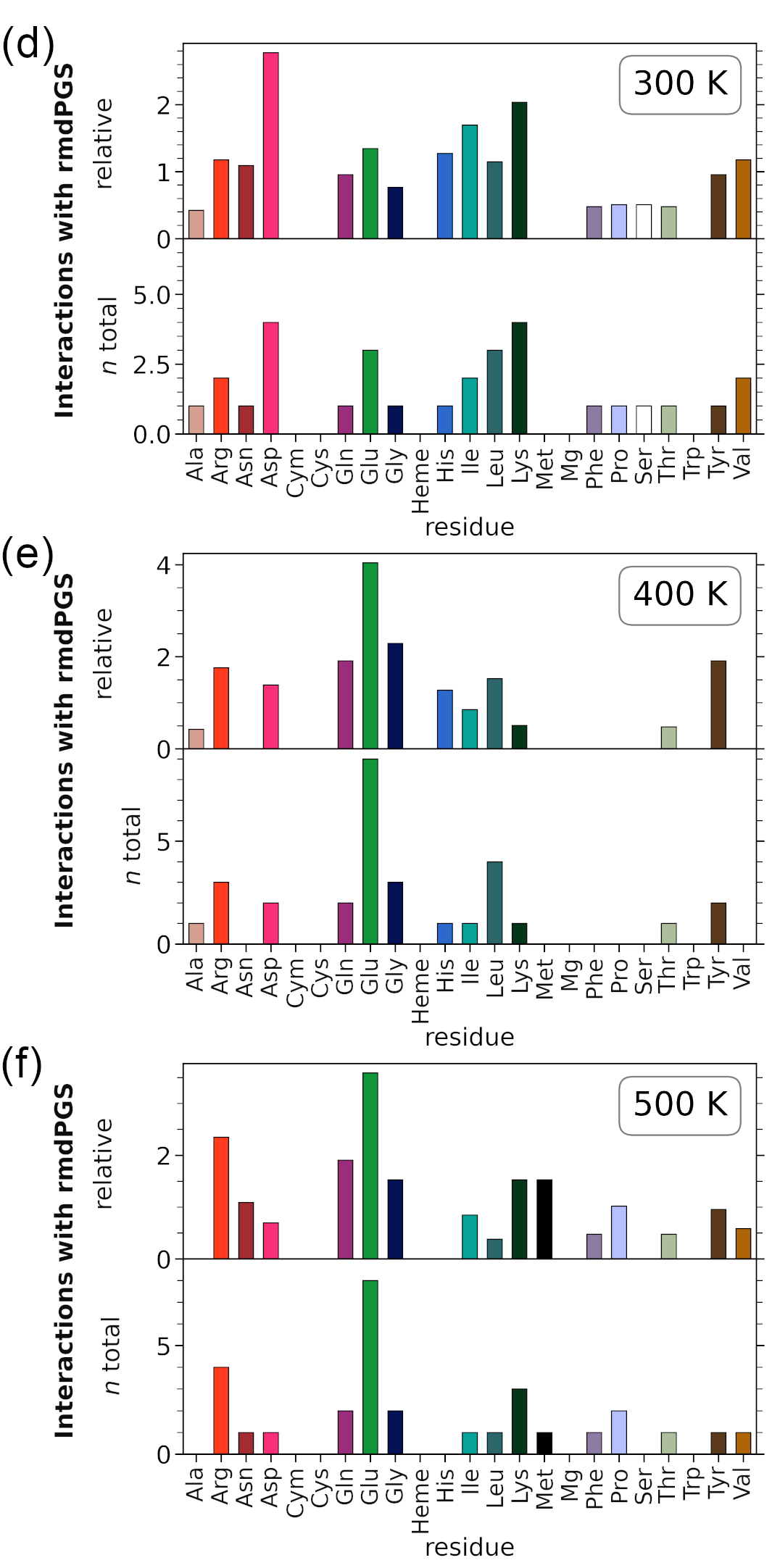}
    \caption{\ch{NO}}
    \end{subfigure}
    \caption{SASA interaction profile for \ch{H2O2} and \ch{NO} with \textit{Cvi}UPO at different temperatures: (a) at 300\,K, (b) 400\,K, and (c) at 500\,K.}
    \label{SI-SASA-2}
\end{figure}

\begin{figure}[H]
    \renewcommand\thesubfigure{\roman{subfigure}}
    \centering
    \begin{subfigure}{0.49\textwidth}
    \includegraphics[width=1\textwidth]{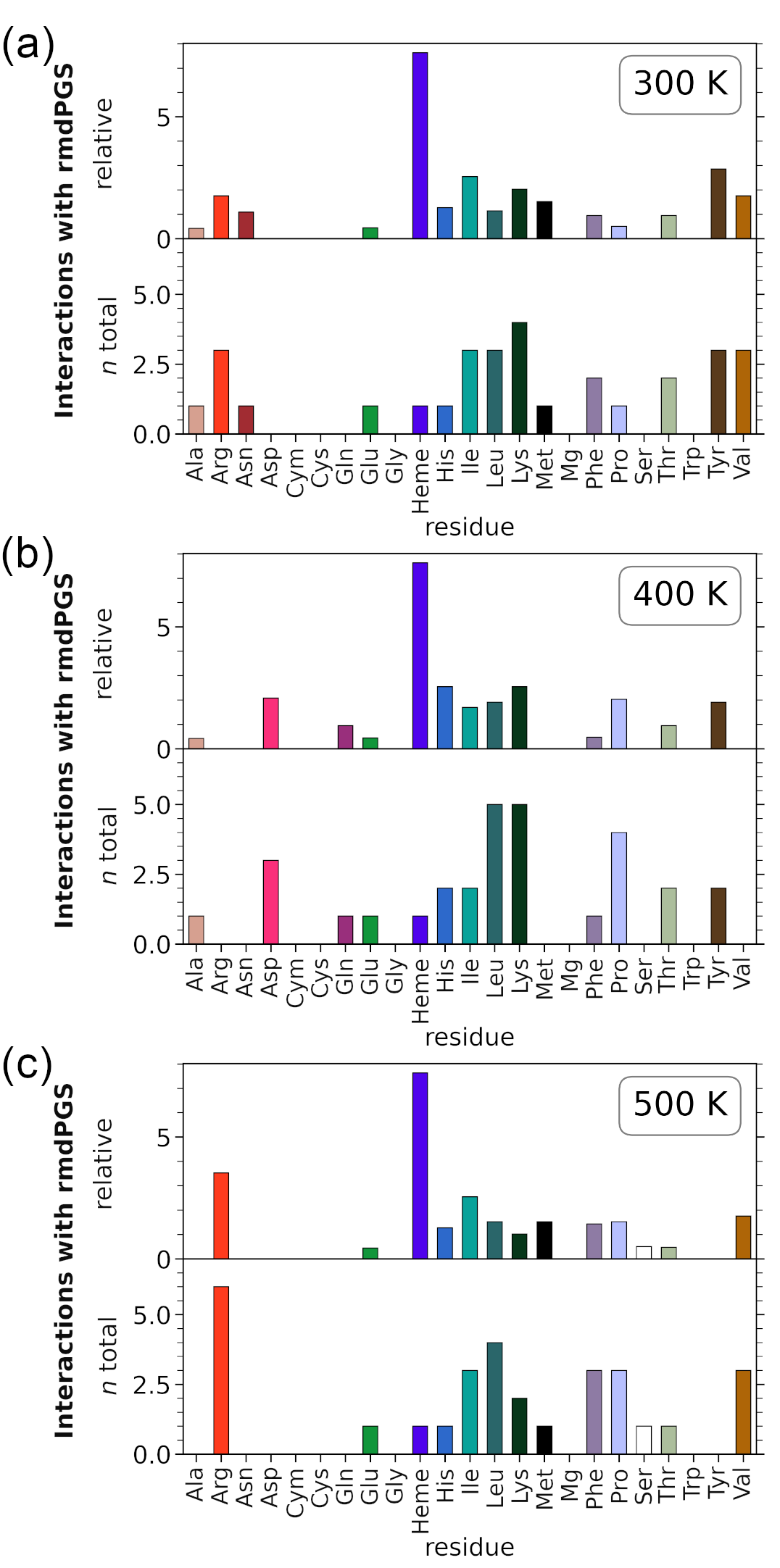}
    \caption{\ch{O2}}
    \end{subfigure}
    \begin{subfigure}{0.49\textwidth}
    \includegraphics[width=1\textwidth]{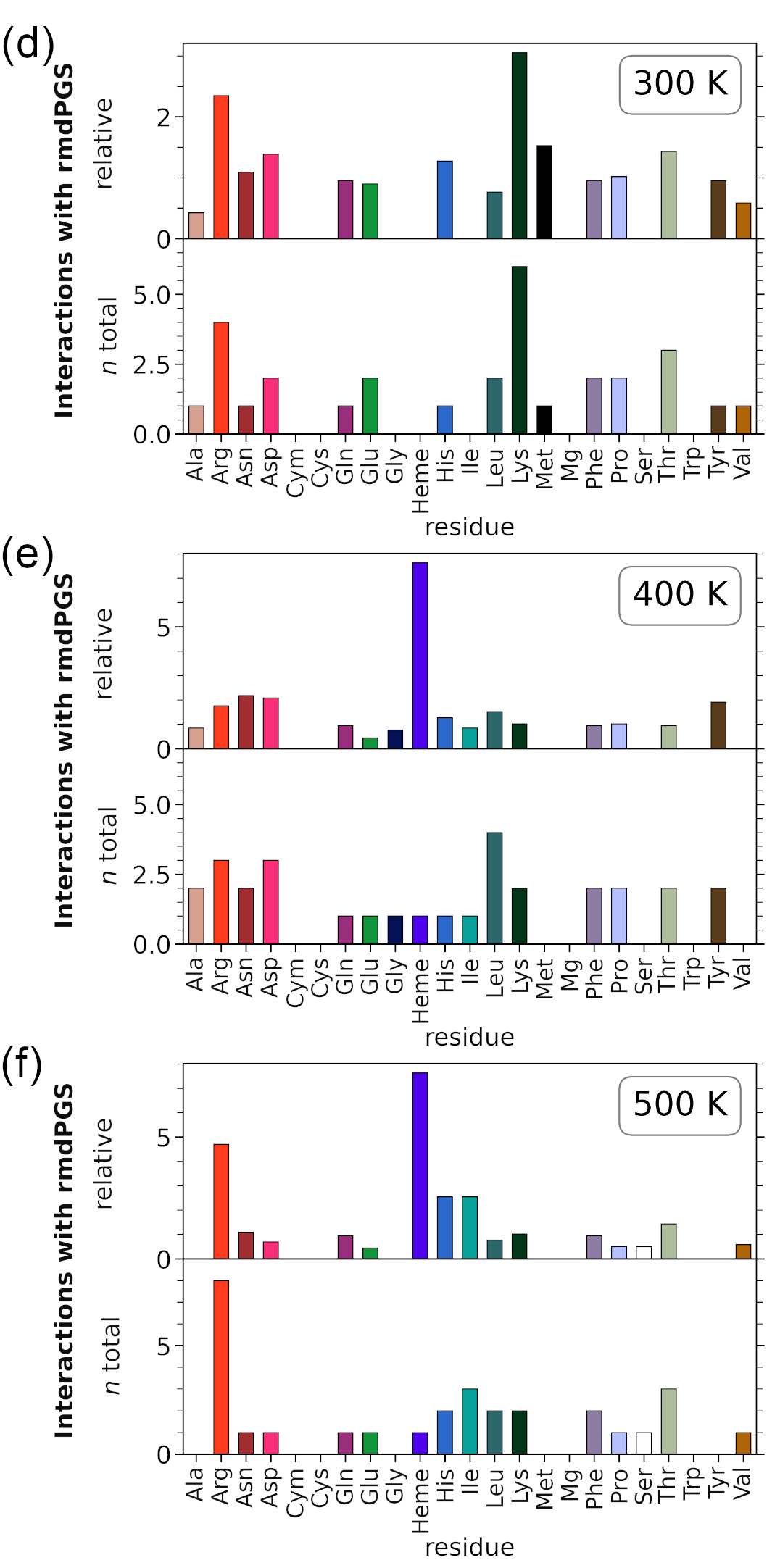}
    \caption{\ch{O}}
    \end{subfigure}
    \caption{SASA interaction profile for \ch{O2} and \ch{O} with \textit{Cvi}UPO at different temperatures: (a) at 300\,K, (b) 400\,K, and (c) at 500\,K.}
    \label{SI-SASA-3}
\end{figure}

\begin{table}[H]
\caption{Data for the short validation MD simulations from Section "SASA interaction analysis \textit{Cvi}UPO" \ref{SASA-Cvi} for all tested \textit{rmd}PGS at 300\,K in \textbf{vacuum}. Of the ten tested SASA positions (with the lowest interaction energies) only those SASA points are in the table listed where an immediate reaction occurred.  Therefore, the rows of \ch{H2O2} and \ch{O2} do not contain any data. The full data set is available and has been provided alongside the main paper.}
\label{tab:SI-Cvi_SASA_MD}
\resizebox{\textwidth}{!}{%
\begin{tabular}{lcccc}
\multicolumn{1}{c}{\textbf{H}}    & \textbf{xyz}                                                                              & \textbf{SASA closest residue} & \textbf{SASA E\_int / eV} & \textbf{MD interaction} \\ \hline
                                  & 0.025 2.826 -23.8                                                                         & Glu124                           & -0.217                    & OE1 (Glu124)                 \\
                                  & -15.765 -12.366 11.894                                                                    & Asp104                           & -0.212                    & OD1 (Asp104)                 \\
                                  & -0.381 19.075 1.27                                                                        & Thr221                           & -0.207                    & O (221)                 \\
                                  & -13.936 -12.142 12.812                                                                    & Lys103                           & -0.202                    & OD1 (Asp104)                 \\
                                  & 0.803 1.108 -22.84                                                                        & Tyr125                           & -0.201                    & OH (Tyr125)                 \\
                                  & -0.298 20.662 10.614                                                                      & Leu225                           & -0.191                    & O (Leu225)                 \\
                                  & 10.015 11.442 -7.978                                                                      & Glu216                           & -0.186                    & O (Glu216)                 \\
                                  & -11.175 0.619 5.562                                                                       & Arg96                           & -0.183                    & O (Arg96)                 \\
\multicolumn{1}{c}{\textbf{OH}}   & \textbf{xyz}                                                                              & \textbf{SASA closest residue} & \textbf{SASA E\_int / eV} & \textbf{MD interaction} \\ \hline
                                  & 1.107 21.172 3.397                                                                        & Leu225                           & -0.825                    & HN (Leu225)                 \\
                                  & -4.136 1.205 13.22                                                                        & Asp91                           & -0.793                    & HN (Asp91)                 \\
                                  & -2.516 -8.267 19.134                                                                      & Arg84                           & -0.728                    & HE (Arg84)                 \\
                                  & 2.024 12.392 16.171                                                                       & Val153                           & -0.728                    & HG23 (Val153)                 \\
\multicolumn{1}{c}{\textbf{H2O2}} & \textbf{xyz}                                                                              & \textbf{SASA closest residue} & \textbf{SASA E\_int / eV} & \textbf{MD interaction} \\ \hline
                                  & \multicolumn{1}{l}{} & \multicolumn{1}{c}{}          & \multicolumn{1}{l}{}      & \multicolumn{1}{c}{no reaction}    \\
\multicolumn{1}{c}{\textbf{NO}}   & \textbf{xyz}                                                                              & \textbf{SASA closest residue} & \textbf{SASA E\_int / eV} & \textbf{MD interaction} \\ \hline
                                  & -6.19 -22.057 -9.084                                                                      & Thr9                           & -1.01                     & HN (Thr9)                 \\
\multicolumn{1}{c}{\textbf{O}}    & \textbf{xyz}                                                                              & \textbf{SASA closest residue} & \textbf{SASA E\_int / eV} & \textbf{MD interaction} \\ \hline
                                  & 1.107 21.172 3.397                                                                        & Leu225                           & -1.175                    & HN (Leu225), HN (Ile18)                 \\
                                  & -18.717 -3.891 -8.046                                                                     & Lys114                           & -1.127                    & HN (Lys114)                 \\
                                  & -6.255 20.747 6.826                                                                       & Lys224                           & -1.125                    & HZ2 (Lys224)                 \\
                                  & -6.493 20.708 6.454                                                                       & Glu223                           & -1.12                     & HN (Lys224)                 \\
                                  & 3.565 -21.510 12.105                                                                      & Asn37                           & -1.081                    & HN (Asn37)                 \\
                                  & -14.064 -7.808 -13.948                                                                    & Lys184                           & -1.071                    & HZ2 (Lys184)                 \\
\multicolumn{1}{c}{\textbf{O2}}   & \textbf{xyz}                                                                              & \textbf{SASA closest residue} & \textbf{SASA E\_int / eV} & \textbf{MD interaction} \\ \hline
                                  & \multicolumn{1}{l}{} & \multicolumn{1}{l}{}          & \multicolumn{1}{l}{}      & \multicolumn{1}{c}{no reaction}   
\end{tabular}}
\end{table}

\begin{table}[H]
\caption{Data for the short validation MD simulations from Section "SASA interaction analysis \textit{Cvi}UPO" \ref{SASA-Cvi} for all tested \textit{rmd}PGS at 300\,K in \textbf{solvent}. Of the ten tested SASA positions (with the lowest interaction energies) only those SASA points are in the table listed where an immediate reaction occurred.  Therefore, the rows of \ch{H2O2} and \ch{O2} do not contain any data. The full data set is available and has been provided alongside the main paper.}
\label{tab:SI-Cvi_SASA_MD_solv}
\resizebox{\textwidth}{!}{%
\begin{tabular}{lcccc}
\multicolumn{1}{c}{\textbf{H}}    & \textbf{xyz}           & \textbf{SASA closest residue} & \textbf{SASA E\_int / eV} & \textbf{MD interaction} \\ \hline
                                  & -0.296 20.962 11.561   & Asp227                        & -0.19                     & OD2 (Asp227)            \\
\multicolumn{1}{c}{\textbf{OH}}   & \textbf{xyz}           & \textbf{SASA closest residue} & \textbf{SASA E\_int / eV} & \textbf{MD interaction} \\ \hline
                                  & \multicolumn{1}{l}{}   & \multicolumn{1}{l}{}          & \multicolumn{1}{l}{}      & no reaction             \\
\multicolumn{1}{c}{\textbf{H2O2}} & \textbf{xyz}           & \textbf{SASA closest residue} & \textbf{SASA E\_int / eV} & \textbf{MD interaction} \\ \hline
                                  & -4.654 3.113 10.773    & Gln155                        & -1.269                    & HE21 (Gln155)           \\
                                  & -4.425 0.721 13.495    & Asp91                         & -1.146                    & H (Asp91)               \\
                                  & -3.29 -0.177 13.716    & Hish85                        & -1.081                    & HE1 (Hish85)            \\
\multicolumn{1}{c}{\textbf{NO}}   & \textbf{xyz}           & \textbf{SASA closest residue} & \textbf{SASA E\_int / eV} & \textbf{MD interaction} \\ \hline
                                  & -0.437 -20.604 -15.493 & Lys8                          & -1.018                    & HG1, CG (Lys8)          \\
                                  & 2.763 -21.539 12.001   & Arg36                         & -1.001                    & HA (Arg36)              \\
\multicolumn{1}{c}{\textbf{O}}    & \textbf{xyz}           & \textbf{SASA closest residue} & \textbf{SASA E\_int / eV} & \textbf{MD interaction} \\ \hline
                                  & \multicolumn{1}{l}{}   & \multicolumn{1}{l}{}          & \multicolumn{1}{l}{}      & no reaction             \\
\multicolumn{1}{c}{\textbf{O2}}   & \textbf{xyz}           & \textbf{SASA closest residue} & \textbf{SASA E\_int / eV} & \textbf{MD interaction} \\ \hline
                                  & 2.79 -21.421 11.992    & Thr78                         & -1.368                    & HB (Thr78), HG1 (Arg36) \\
                                  & 2.763 -21.539 12.001   & Arg36                         & -1.357                    & HB (Thr78), HG1 (Arg36)
\end{tabular}}
\end{table}

The results for higher temperatures were excluded from this section to minimize the length of the Supplementary Information. However, the corresponding data are available and have been provided alongside the main paper.

\subsubsection{\textit{Aae}UPO} \label{SI_SASA_Aae}

\begin{figure}[H]
    \centering
    \includegraphics[width=1\textwidth]{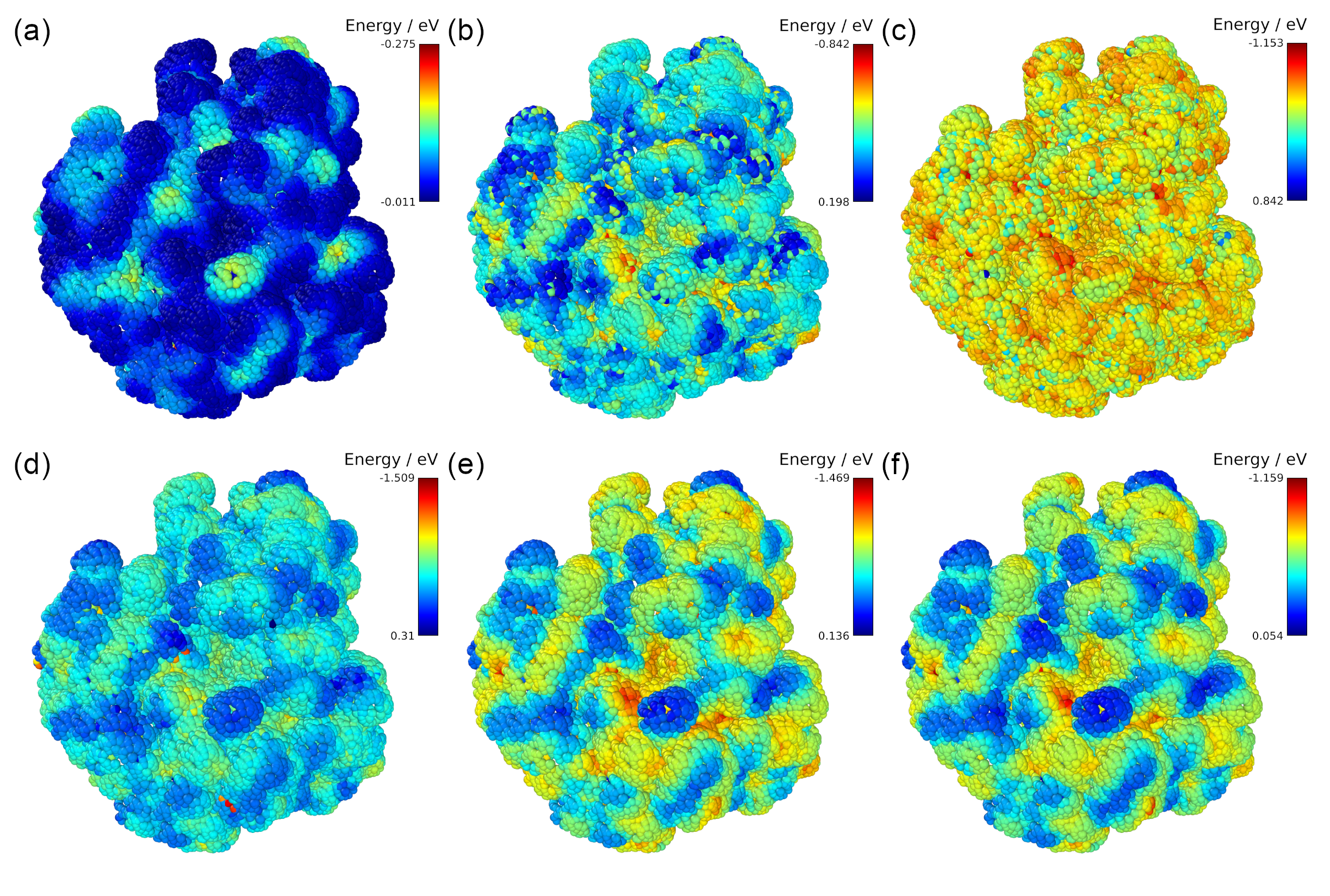}
    \caption{Interaction profiles from SASA for the enzyme \textit{Aae}UPO at 300\,K. (a) shows the interaction map with \ch{H} and (b), (c), (d), (e) and (f) show the interaction maps for \ch{OH}, \ch{H2O2}, \ch{NO}, \ch{O2} and \ch{O}, respectively.}
    \label{SI-SASA_Aae}
\end{figure}

\begin{figure}[H]
    \renewcommand\thesubfigure{\roman{subfigure}}
    \centering
    \begin{subfigure}{0.49\textwidth}
    \includegraphics[width=1\textwidth]{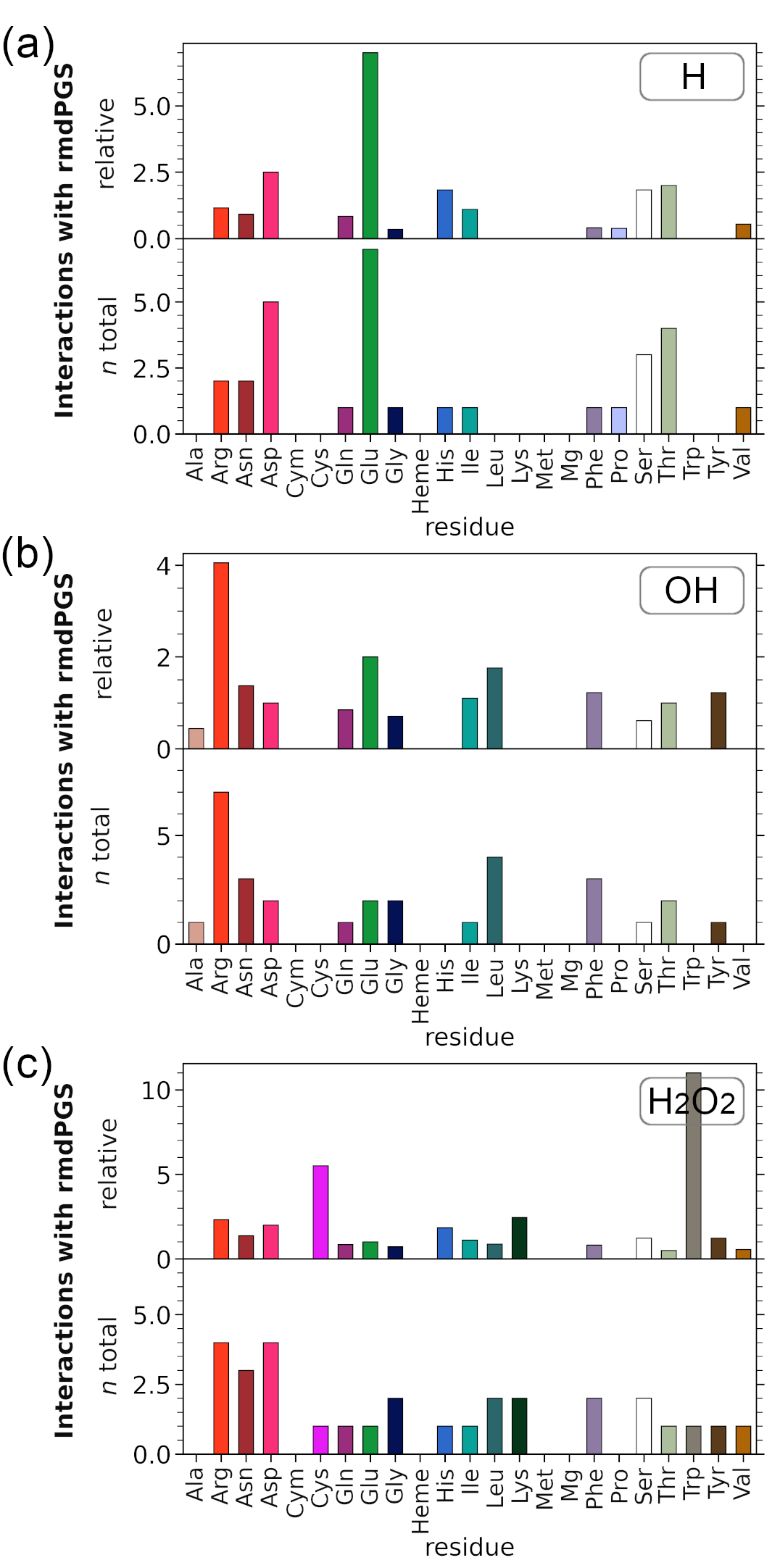}
    \end{subfigure}
    \begin{subfigure}{0.49\textwidth}
    \includegraphics[width=1\textwidth]{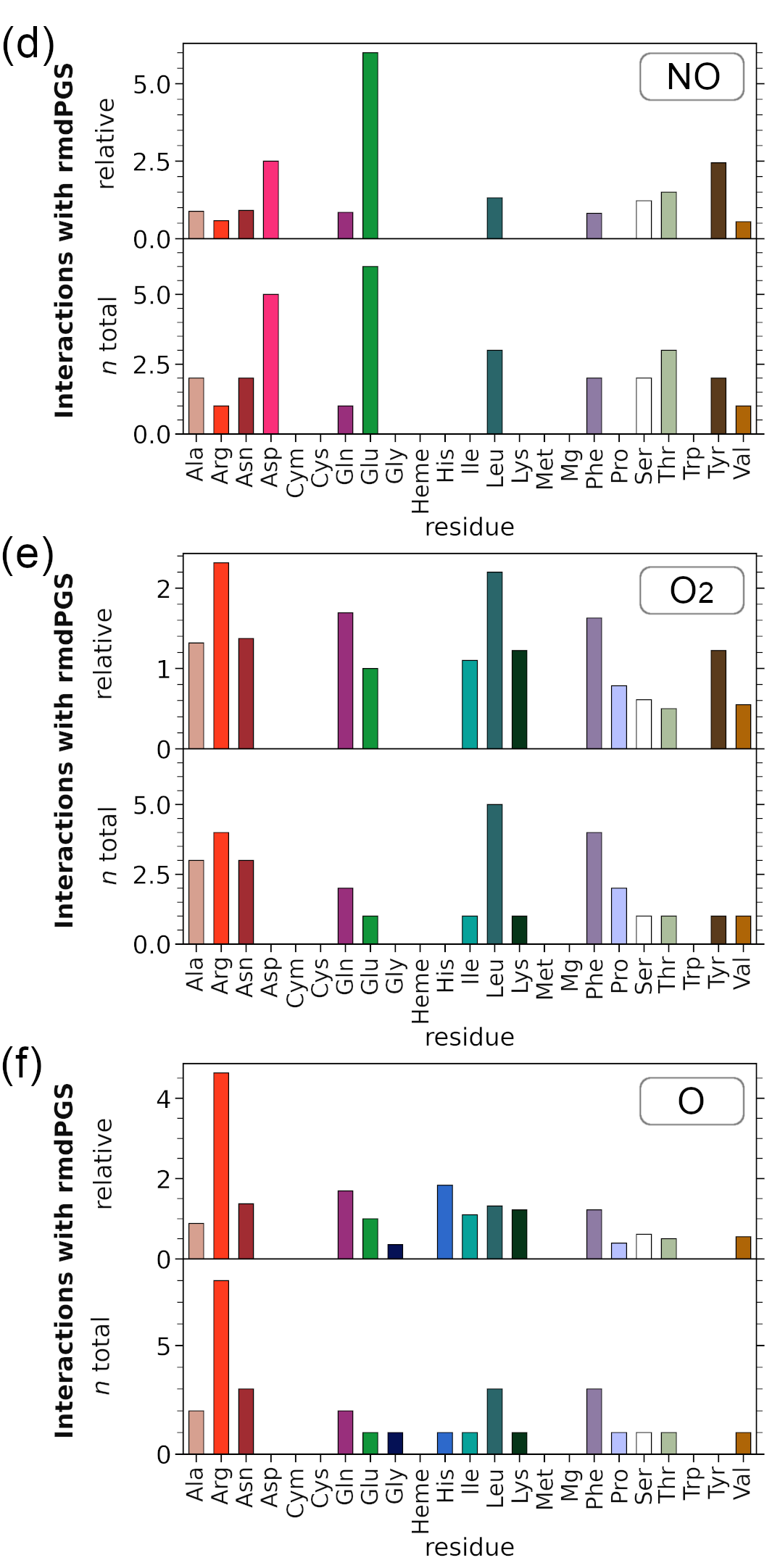}
    \end{subfigure}
    \caption{SASA total interactions per residue at 300\,K for \textit{Aae}UPO. (a) shows the interactions with \ch{H} and (b), (c), (d), (e) and (f) show the interactions for \ch{OH}, \ch{H2O2}, \ch{NO}, \ch{O2} and \ch{O}, respectively.}
    \label{SI-SASA-Aae-2}
\end{figure}

\begin{table}[H]
\caption{Data for the short validation MD simulations from Section "Comparison to \textit{Aae}UPO and GapA"\ref{SASA-Comp} for all tested \textit{rmd}PGS at 300\,K in \textbf{vacuum}. Of the ten tested SASA positions (with the lowest interaction energies) only those SASA points are in the table listed where an immediate reaction occurred.  Therefore, the rows of \ch{H2O2}, \ch{NO} and \ch{O2} do not contain any data. The full data set is available and has been provided alongside the main paper.}
\label{tab:SI-Aae_SASA_MD}
\resizebox{\textwidth}{!}{%
\begin{tabular}{lcccc}
\multicolumn{1}{c}{\textbf{H}}    & \textbf{xyz}          & \textbf{SASA closest residue} & \textbf{SASA E\_int / eV} & \textbf{MD interaction}           \\ \hline
                                  & 1.941 -7.982 -26.657  & Glu146                        & -0.275                    & OE2 (Glu146)                      \\
                                  & 1.026 -7.549 -26.845  & Glu142                        & -0.247                    & OE2 (Glu146)                      \\
                                  & -2.554 -5.693 19.374  & Asp91                         & -0.223                    & OD1 (Asp91)                       \\
                                  & 4.14 14.58 12.59      & Gln313                        & -0.192                    & O (Phe321)                        \\
                                  & -2.328 -8.81 18.342   & Thr90                         & -0.189                    & O (Thr90)                         \\
                                  & 3.225 15.249 12.77    & Val315                        & -0.183                    & O (Phe321)                        \\
                                  & -11.902 21.905 -3.468 & Ser271                        & -0.182                    & O  (Ser271)                       \\
                                  & 14.725 -5.608 8.976   & Asn61                         & -0.180                    & OD1 (Asn61)                       \\
\multicolumn{1}{c}{\textbf{OH}}   & \textbf{xyz}          & \textbf{SASA closest residue} & \textbf{SASA E\_int / eV} & \textbf{MD interaction}           \\ \hline
                                  & -7.076 6.975 -8.468   & Arg189                        & -0.842                    & HH21 (Arg189)                     \\
                                  & 15.438 6.689 6.478    & Leu9                          & -0.832                    & HN (Leu9)                         \\
                                  & 12.739 12.509 -13.144 & Leu162                        & -0.804                    & HN (Leu162)                       \\
                                  & -21.881 4.895 -20.556 & Ile262                        & -0.769                    & HN (Ile262)                       \\
\multicolumn{1}{c}{\textbf{H2O2}} & \textbf{xyz}          & \textbf{SASA closest residue} & \textbf{SASA E\_int / eV} & \textbf{MD interaction}           \\ \hline
                                  & \multicolumn{1}{l}{}  & \multicolumn{1}{l}{}          & \multicolumn{1}{l}{}      & no reactions                      \\
\multicolumn{1}{c}{\textbf{NO}}   & \textbf{xyz}          & \textbf{SASA closest residue} & \textbf{SASA E\_int / eV} & \textbf{MD interaction}           \\ \hline
                                  & \multicolumn{1}{l}{}  & \multicolumn{1}{l}{}          & \multicolumn{1}{l}{}      & no reactions                      \\
\multicolumn{1}{c}{\textbf{O2}}   & \textbf{xyz}          & \textbf{SASA closest residue} & \textbf{SASA E\_int / eV} & \textbf{MD interaction}           \\ \hline
                                  & \multicolumn{1}{l}{}  & \multicolumn{1}{l}{}          & \multicolumn{1}{l}{}      & no reactions                      \\
\multicolumn{1}{c}{\textbf{O}}    & \textbf{xyz}          & \textbf{SASA closest residue} & \textbf{SASA E\_int / eV} & \textbf{MD interaction}           \\ \hline
                                  & -5.719 20.127 -7.11   & Phe274                        & -1.159                    & HD1	Phe274                        \\
                                  & -21.881 4.895 -20.556 & Ile262                        & -1.137                    & HN	Ile262                         \\
                                  & 20.19 2.179 9.391     & Glu10                         & -1.118                    & HN (Glu10), HN (Asn11)            \\
                                  & -12.742 6.762 24.991  & Asn328                        & -1.112                    & HD2	(Arg301), HD21 (Asn328)       \\
                                  & 11.861 13.57 -1.451   & Ser240                        & -1.09                     & HN (Ser240)                       \\
                                  & -5.344 20.631 -6.851  & Phe274                        & -1.057                    & HN        (Phe274), HH21 (Arg257)
\end{tabular}}
\end{table}

No interactions between the \textit{rmd}PGS and \textit{Aae}UPO were detected in the short SASA MDs with solvent. 

The results for higher temperatures were excluded from this section to minimize the length of the Supplementary Information. However, the corresponding data are available and have been provided alongside the main paper.

\subsubsection{GapA} \label{SI_SASA_GapA}

\begin{figure}[H]
    \centering
    \includegraphics[width=1\textwidth]{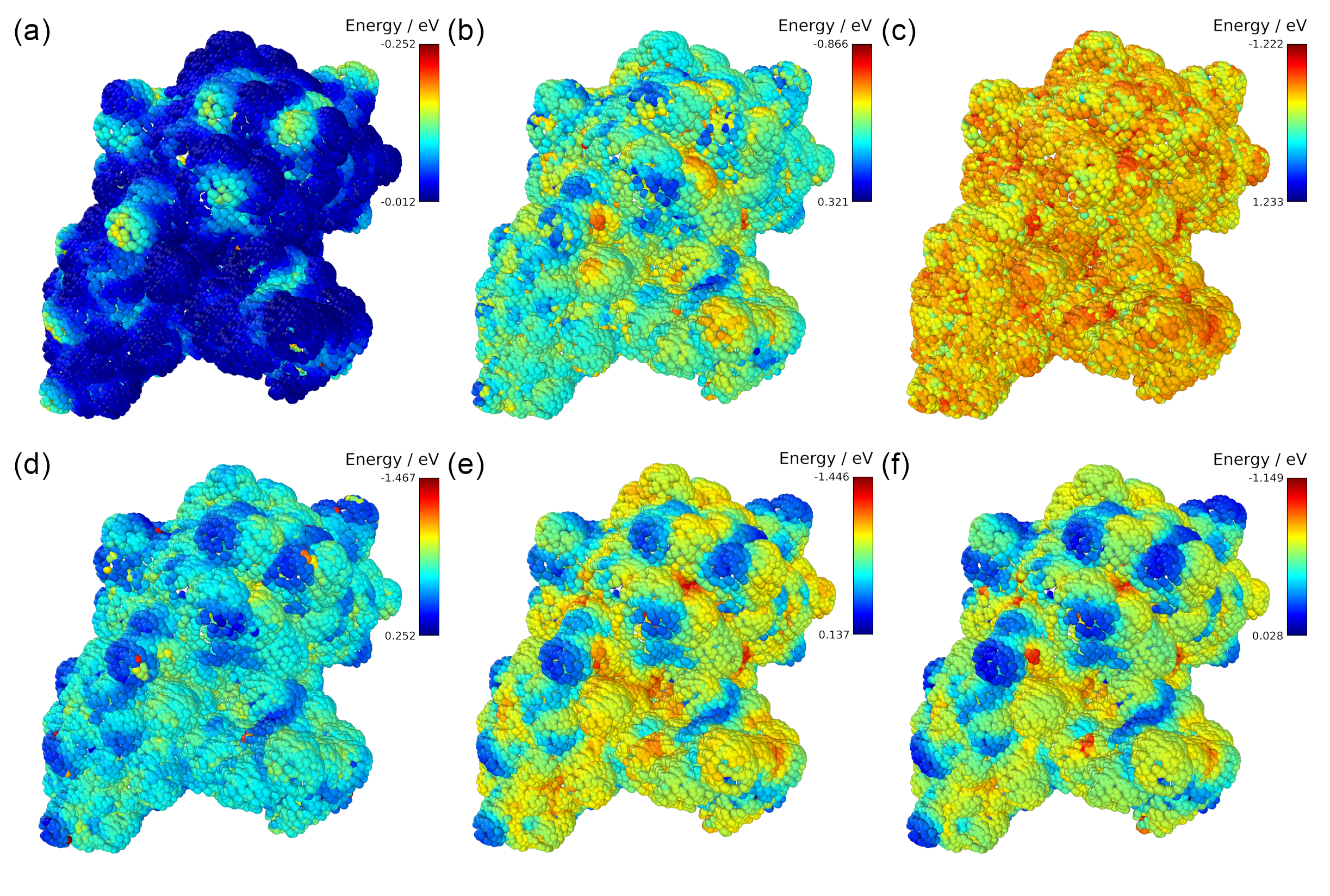}
    \caption{Interaction profiles from SASA for the enzyme GapA at 300\,K. (a) shows the interaction map with \ch{H} and (b), (c), (d), (e) and (f) show the interaction maps for \ch{OH}, \ch{H2O2}, \ch{NO}, \ch{O2} and \ch{O}, respectively.}
    \label{SI-SASA_GapA}
\end{figure}

\begin{figure}[H]
    \renewcommand\thesubfigure{\roman{subfigure}}
    \centering
    \begin{subfigure}{0.49\textwidth}
    \includegraphics[width=1\textwidth]{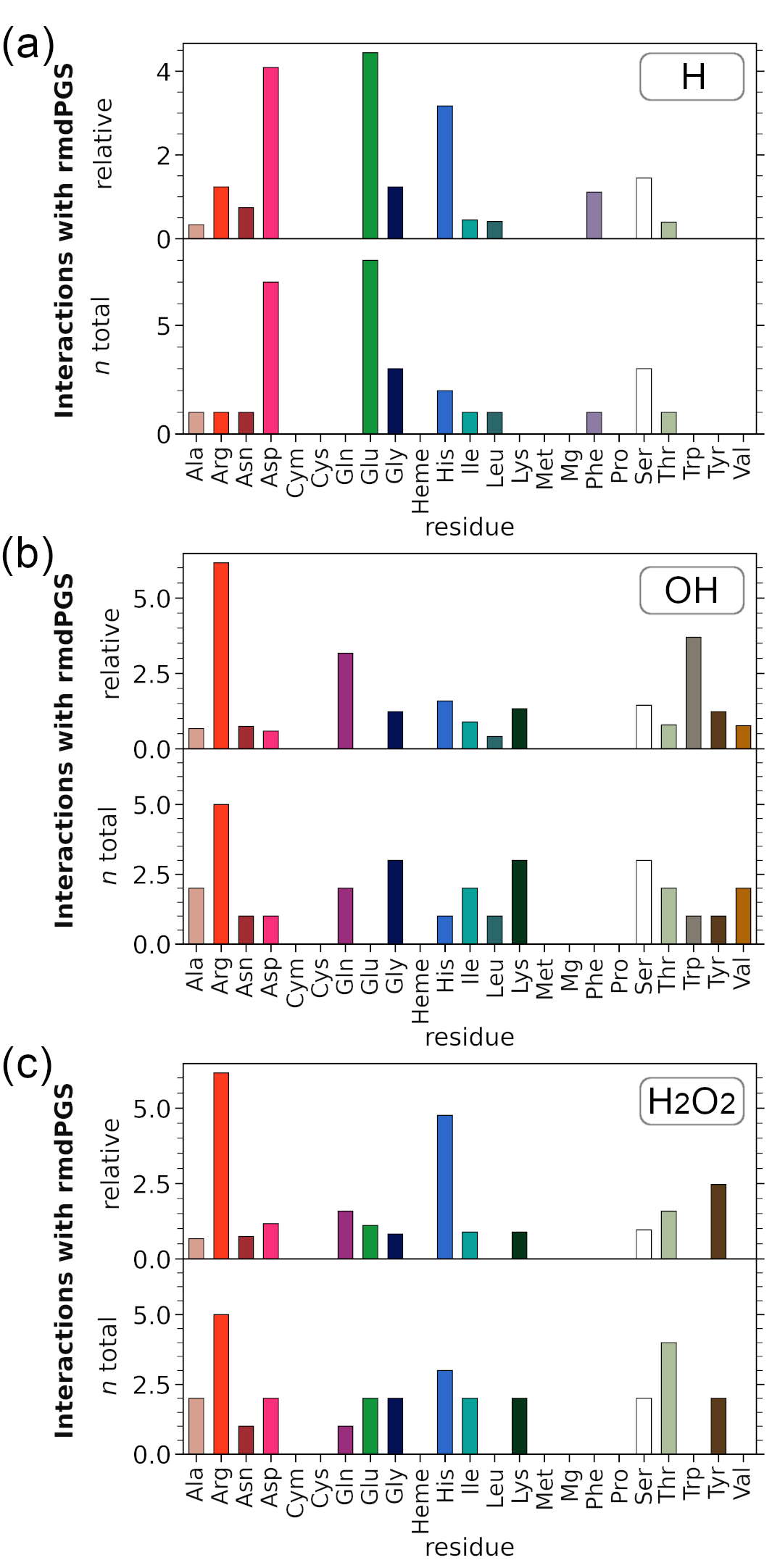}
    \end{subfigure}
    \begin{subfigure}{0.49\textwidth}
    \includegraphics[width=1\textwidth]{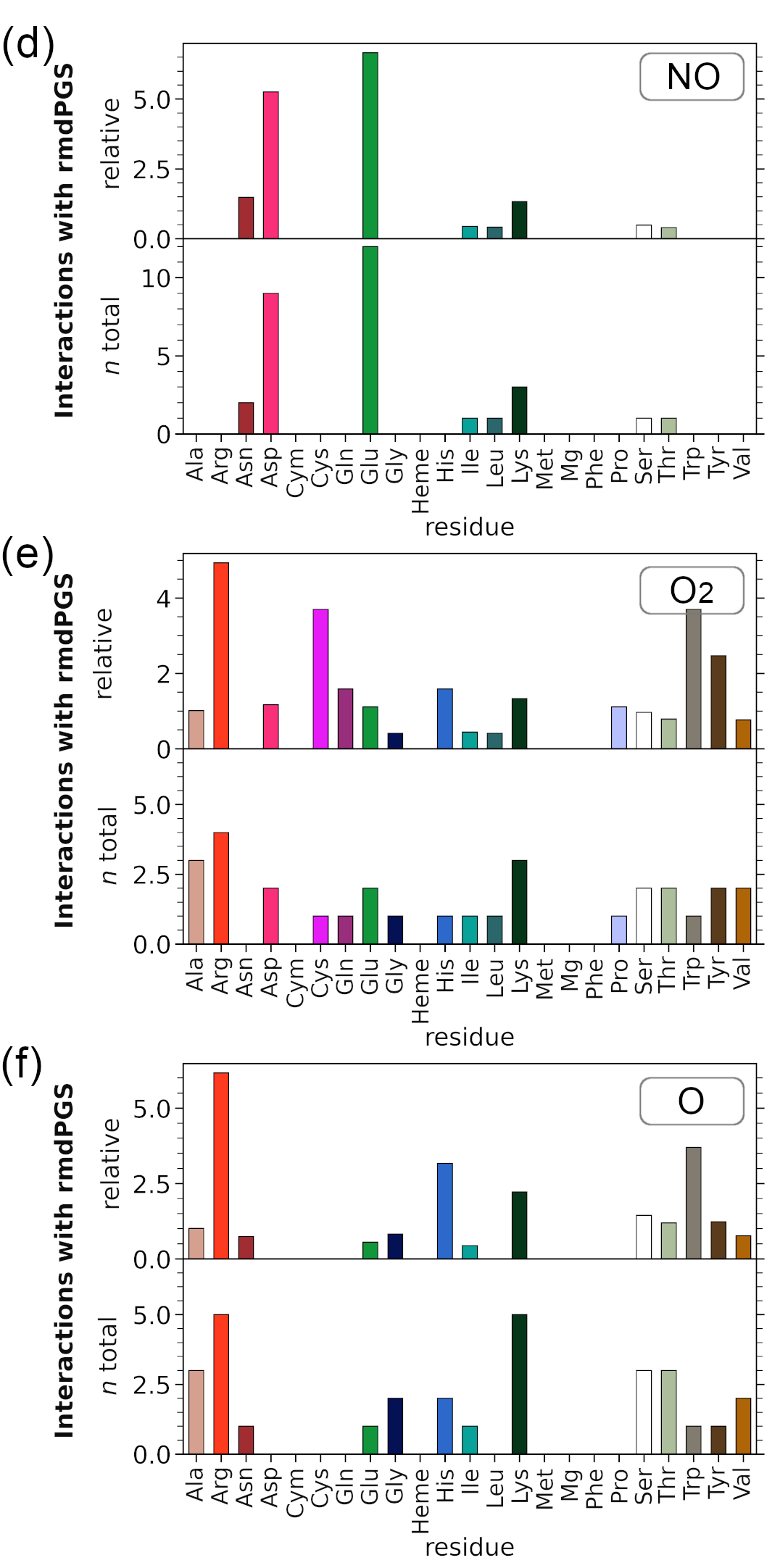}
    \end{subfigure}
    \caption{SASA total interactions per residue at 300\,K for GapA. (a) shows the interactions with \ch{H} and (b), (c), (d), (e) and (f) show the interactions for \ch{OH}, \ch{H2O2}, \ch{NO}, \ch{O2} and \ch{O}, respectively.}
    \label{SI-SASA-GapA-2}
\end{figure}

\begin{table}[H]
\caption{Data for the short validation MD simulations from Section "Comparison to \textit{Aae}UPO and GapA"\ref{SASA-Comp} for all tested \textit{rmd}PGS at 300\,K in \textbf{vacuum}. Of the ten tested SASA positions (with the lowest interaction energies) only those SASA points are in the table listed where an immediate reaction occurred.  Therefore, the rows of \ch{H2O2}, \ch{NO} and \ch{O2} do not contain any data. The full data set is available and has been provided alongside the main paper.}
\label{tab:SI-GapA_SASA_MD}
\resizebox{\textwidth}{!}{%
\begin{tabular}{lcccc}
\multicolumn{1}{c}{\textbf{H}}    & \textbf{xyz}           & \textbf{SASA closest residue} & \textbf{SASA E\_int / eV} & \textbf{MD interaction}           \\ \hline
                                  & -14.174 13.575 -22.124 & Glu202                        & -0.234                    & OD1 (Asp187)                      \\
                                  & 16.83 2.245 21.372     & Leu333                        & -0.213                    & OT1 (Leu333)                      \\
                                  & 16.632 2.676 21.313    & Gly140                        & -0.211                    & OT1 (Leu333)                      \\
                                  & -11.523 -7.879 2.094   & Ser285                        & -0.207                    & O (Asp281)                        \\
                                  & -3.328 -20.437 -13.4   & Thr296                        & -0.197                    & O (Thr296)                        \\
                                  & -14.856 -16.476 -5.369 & Glu276                        & -0.195                    & O (Glu276)                        \\
                                  & -12.534 -7.989 2.371   & Asp281                        & -0.191                    & O (Asp281)                        \\
                                  & -18.562 18.793 22.722  & Asp59                         & -0.191                    & OD1 (Asp59)                       \\
                                  & -12.437 -7.571 2.860   & Ile83                         & -0.190                    & O (Ile283)                        \\
                                  & 20.023 -18.895 -15.649 & Glu169                        & -0.188                    & OE1 (Glu169)                      \\
\multicolumn{1}{c}{\textbf{OH}}   & \textbf{xyz}           & \textbf{SASA closest residue} & \textbf{SASA E\_int / eV} & \textbf{MD interaction}           \\ \hline
                                  & -2.05 -5.305 12.117    & Arg19                         & -0.807                    & HH11 (Arg19)                      \\
                                  & -12.752 -7.05 -10.723  & Ser279                        & -0.799                    & HN (Ser279)                       \\
                                  & -10.964 13.683 2.709   & Arg14                         & -0.73                     & HH22 (Arg14)                      \\
                                  & -12.16 -7.333 -10.803  & Trp312                        & -0.711                    & HN (Ser279)                       \\
                                  & 8.812 22.837 8.457     & Lys81                         & -0.704                    & HN (Lys81)                        \\
                                  & -14.238 4.07 8.004     & Arg18                         & -0.702                    & HH21 (Arg18)                      \\
\multicolumn{1}{c}{\textbf{H2O2}} & \textbf{xyz}           & \textbf{SASA closest residue} & \textbf{SASA E\_int / eV} & \textbf{MD interaction}           \\ \hline
                                  & \multicolumn{1}{l}{}   & \multicolumn{1}{l}{}          & \multicolumn{1}{l}{}      & no reactions                      \\
\multicolumn{1}{c}{\textbf{NO}}   & \textbf{xyz}           & \textbf{SASA closest residue} & \textbf{SASA E\_int / eV} & \textbf{MD interaction}           \\ \hline
                                  & \multicolumn{1}{l}{}   & \multicolumn{1}{l}{}          & \multicolumn{1}{l}{}      & no reactions                      \\
\multicolumn{1}{c}{\textbf{O2}}   & \textbf{xyz}           & \textbf{SASA closest residue} & \textbf{SASA E\_int / eV} & \textbf{MD interaction}           \\ \hline
                                  & \multicolumn{1}{l}{}   & \multicolumn{1}{l}{}          & \multicolumn{1}{l}{}      & no reactions                      \\
\multicolumn{1}{c}{\textbf{O}}    & \textbf{xyz}           & \textbf{SASA closest residue} & \textbf{SASA E\_int / eV} & \textbf{MD interaction}           \\ \hline
                                  & -10.109 13.349 2.4     & Arg14                         & -1.097                    & HH12 (Arg14)                      \\
                                  & -12.752 -7.05 -10.723  & Ser279                        & -1.085                    & HN (Ser279), HG1  Ser279)         \\
                                  & -5.333 24.677 -17.099  & Arg190                        & -1.07                     & HD2 (Arg190)                      \\
                                  & 8.637 11.106 -16.115   & Arg195                        & -1.058                    & \multicolumn{1}{c}{HH11 (Arg195)} \\
                                  & 8.812 22.837 8.457     & Lys81                         & -1.047                    & HN (Lys81)                        \\
                                  & -12.160 -7.333 -10.803 & Trp312                        & -1.032                    & HN (Ser279)                      
\end{tabular}}
\end{table}
No interactions between the \textit{rmd}PGS and GapA were detected in the short SASA MDs with solvent.

The results for higher temperatures were excluded from this section to minimize the length of the Supplementary Information. However, the corresponding data are available and have been provided alongside the main paper.

\subsection{Interactions with high PGS concentrations} \label{SI_Concentration}
\subsubsection{\textit{Cvi}UPO} \label{SI_Conc-Cvi}

\begin{figure}[H]
    \renewcommand\thesubfigure{\roman{subfigure}}
    \centering
    \begin{subfigure}{0.49\textwidth}
    \includegraphics[width=1\textwidth]{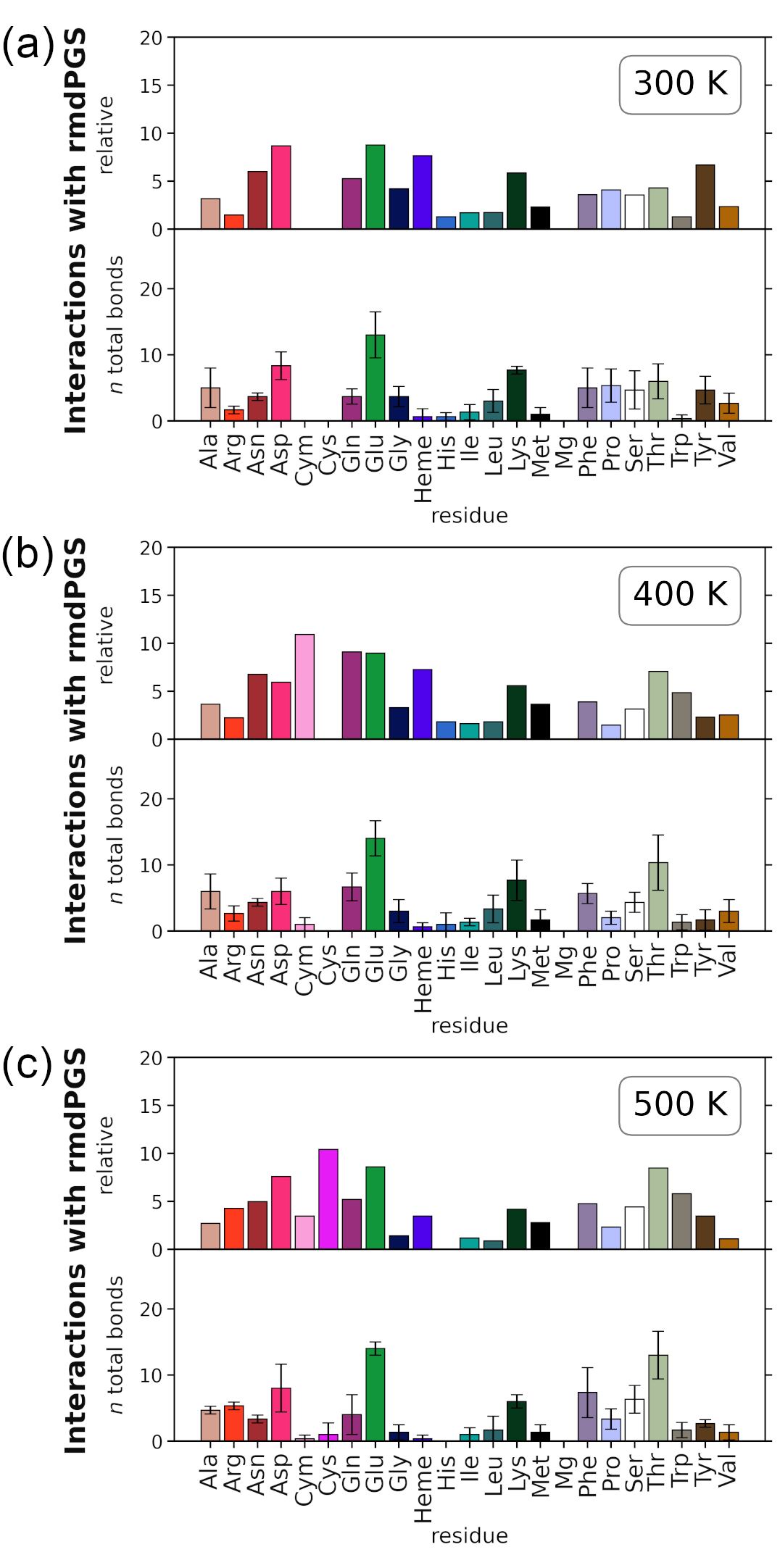}
    \caption{Without solvent}
    \end{subfigure}
    \begin{subfigure}{0.49\textwidth}
    \includegraphics[width=1\textwidth]{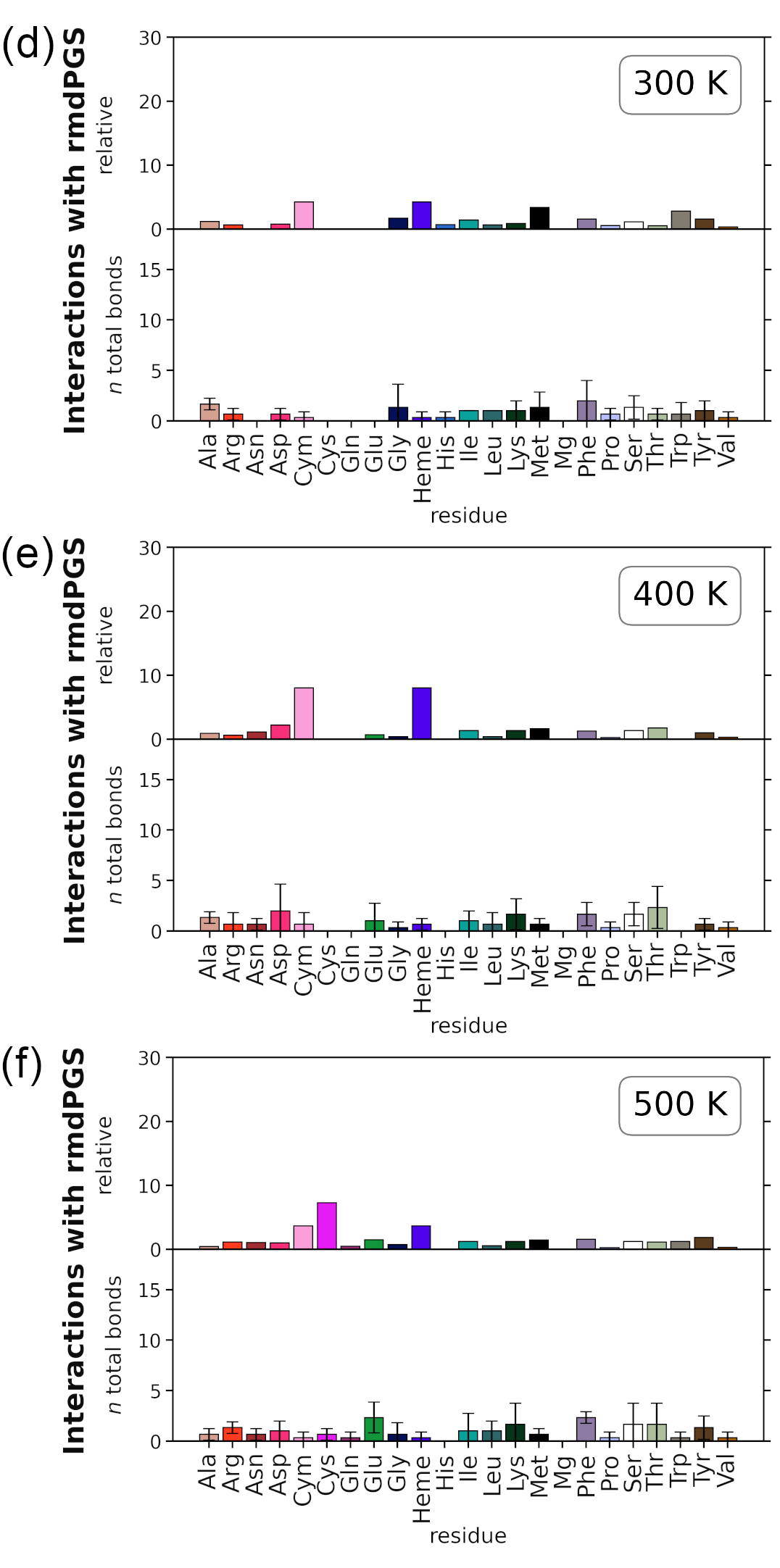}
    \caption{With solvent}
    \end{subfigure}
    \caption{Bond analysis for high concentrations of \ch{H} with \textit{Cvi}UPO. All panels show the total number of bonds to an additional hydrogen atom per residue and the relative interactions calculated following equation \ref{eq:2.2}. The panels (a) and (d) show the interactions at 300\,K, without and with solvent, respectively. The panels (b) and (e) show the interactions at at 400\,K and the panels (c) and (f) show the interactions at at 500\,K}
    \label{SI-Conc_MD_Cvi-h}
\end{figure}

\begin{figure}[H]    \renewcommand\thesubfigure{\roman{subfigure}}
    \centering
    \begin{subfigure}{0.49\textwidth}
    \includegraphics[width=1\textwidth]{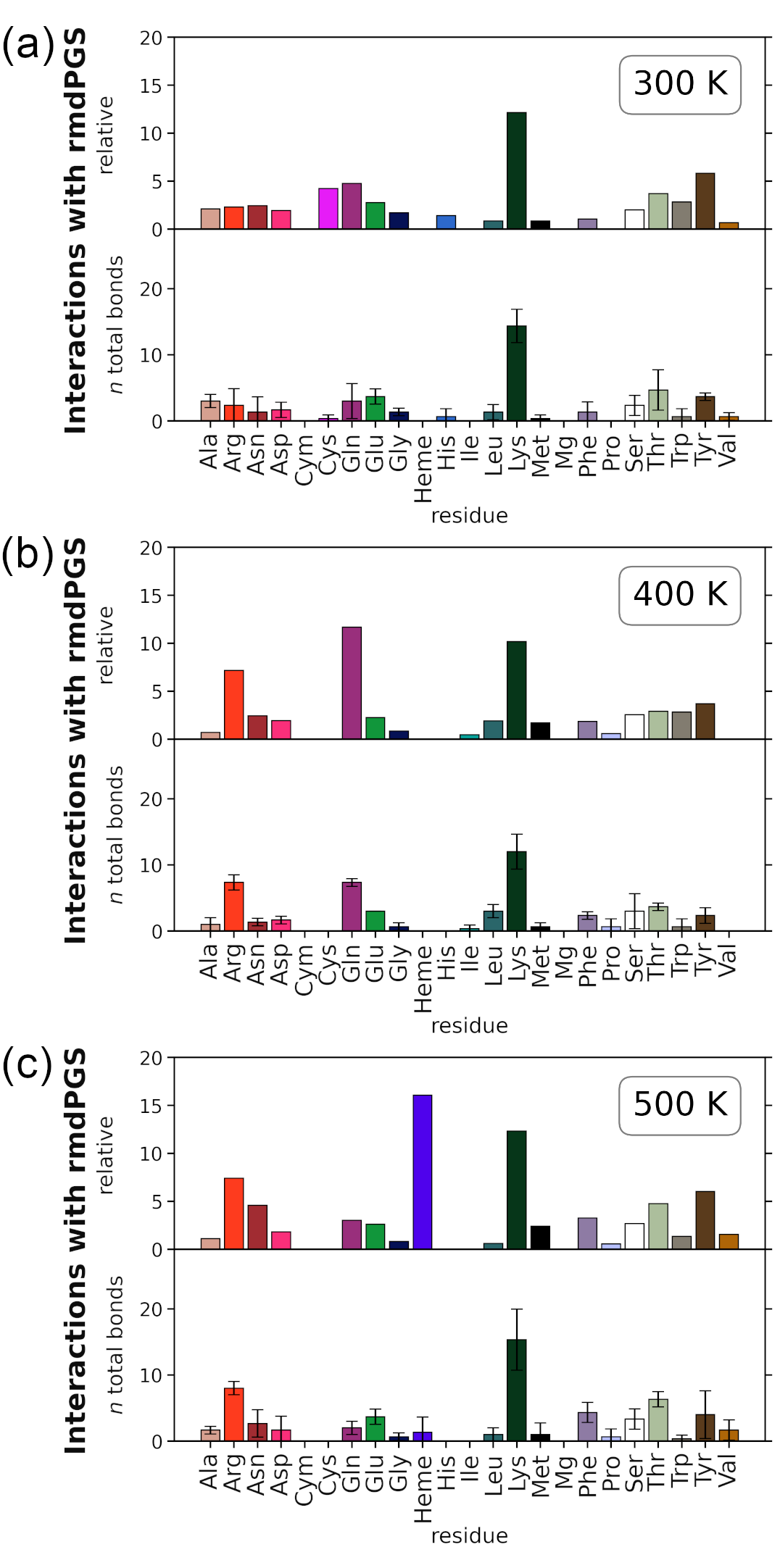}
    \caption{Without solvent}
    \end{subfigure}
    \begin{subfigure}{0.49\textwidth}
    \includegraphics[width=1\textwidth]{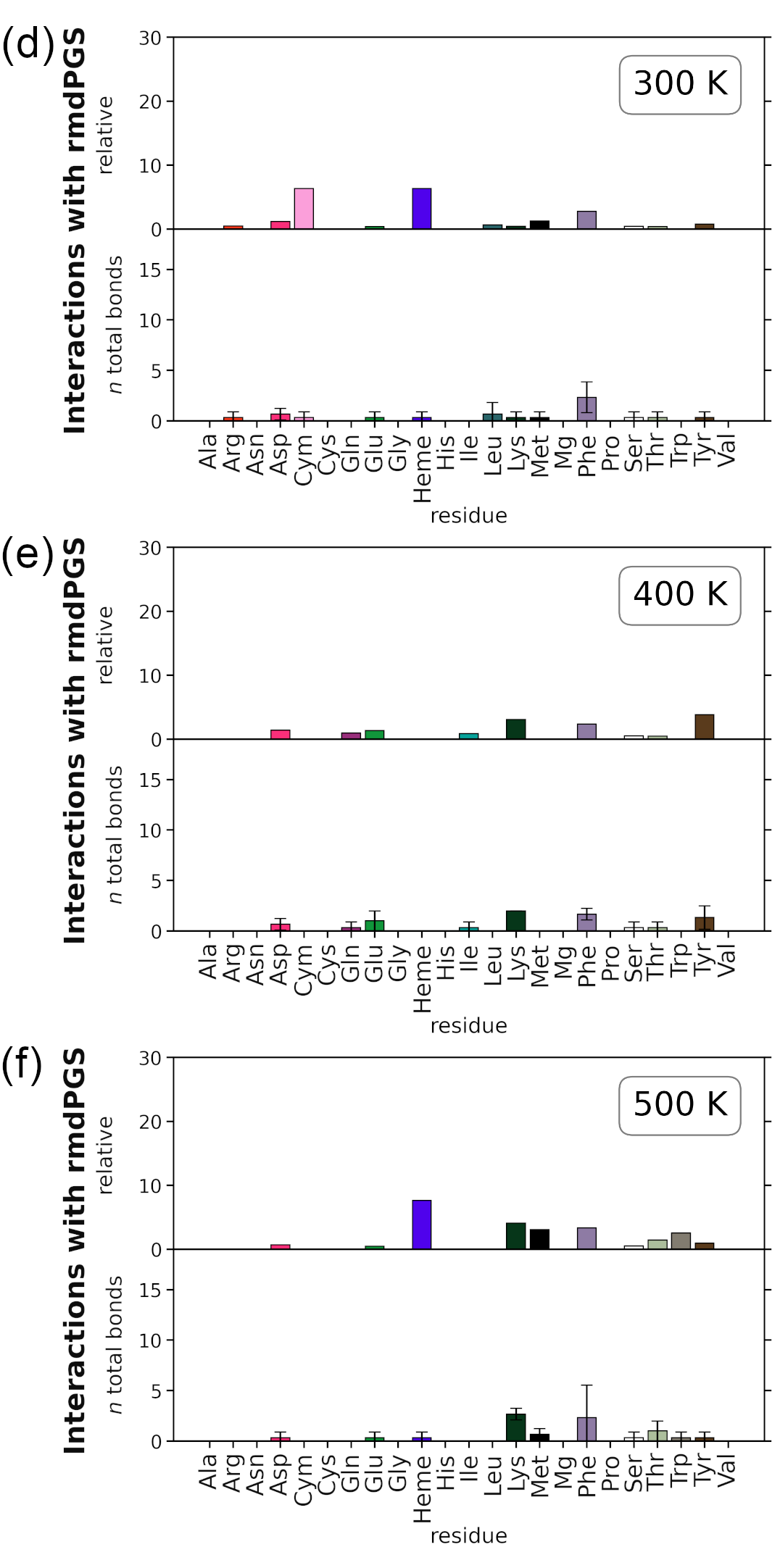}
    \caption{With solvent}
    \end{subfigure}
    \caption{Bond analysis for high concentrations of \ch{H2O2} with \textit{Cvi}UPO. All panels show the total number of bonds to an additional hydrogen atom per residue and the relative interactions calculated following equation \ref{eq:2.2}. The panels (a) and (d) show the interactions at 300\,K, without and with solvent, respectively. The panels (b) and (e) show the interactions at at 400\,K and the panels (c) and (f) show  the interactions at at 500\,K}
    \label{SI-Conc_MD_Cvi-h2o2}
\end{figure}

\begin{figure}[H]
    \renewcommand\thesubfigure{\roman{subfigure}}
    \centering
    \begin{subfigure}{0.49\textwidth}
    \includegraphics[width=1\textwidth]{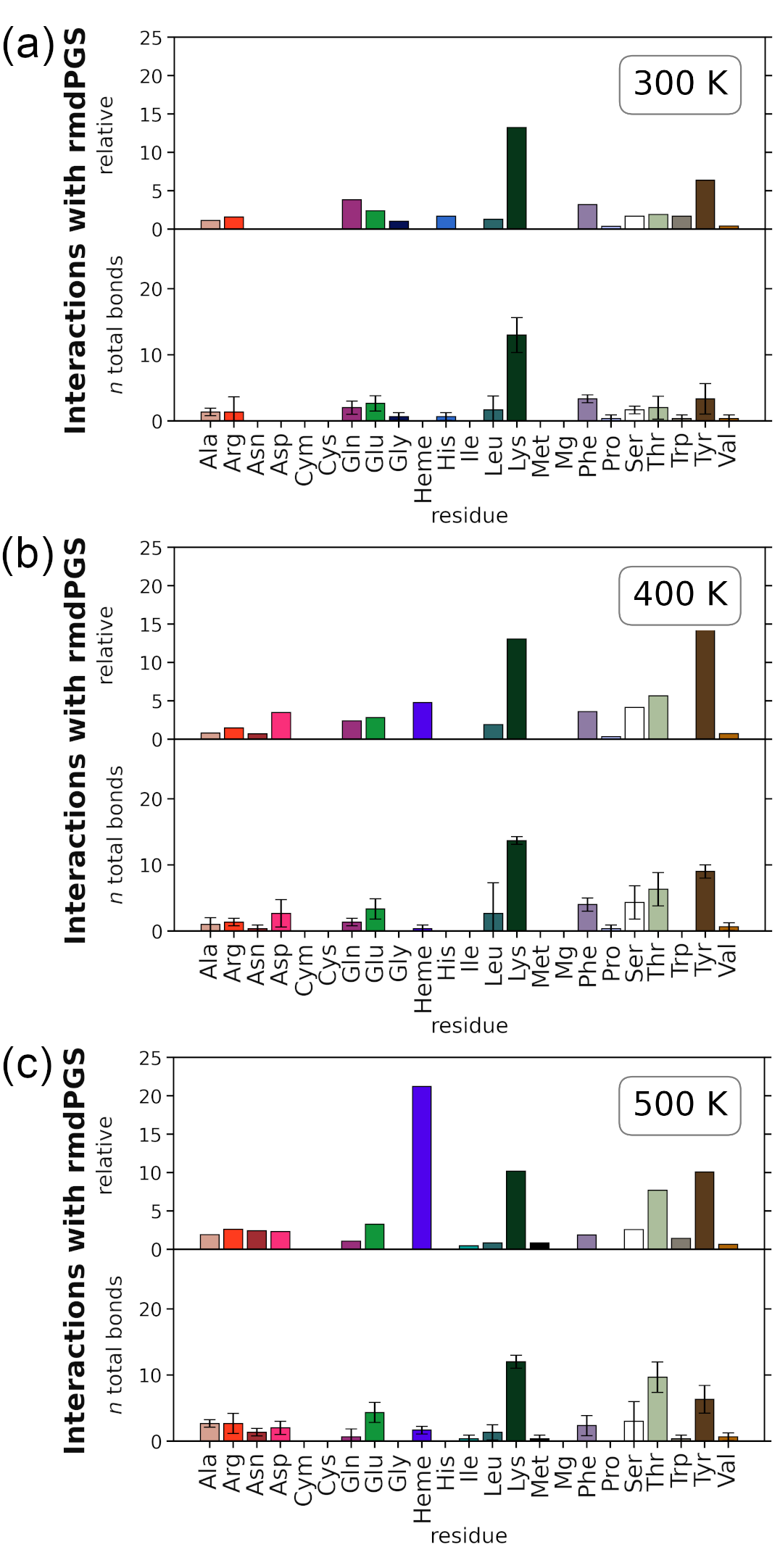}
    \caption{Without solvent}
    \end{subfigure}
    \begin{subfigure}{0.49\textwidth}
    \includegraphics[width=1\textwidth]{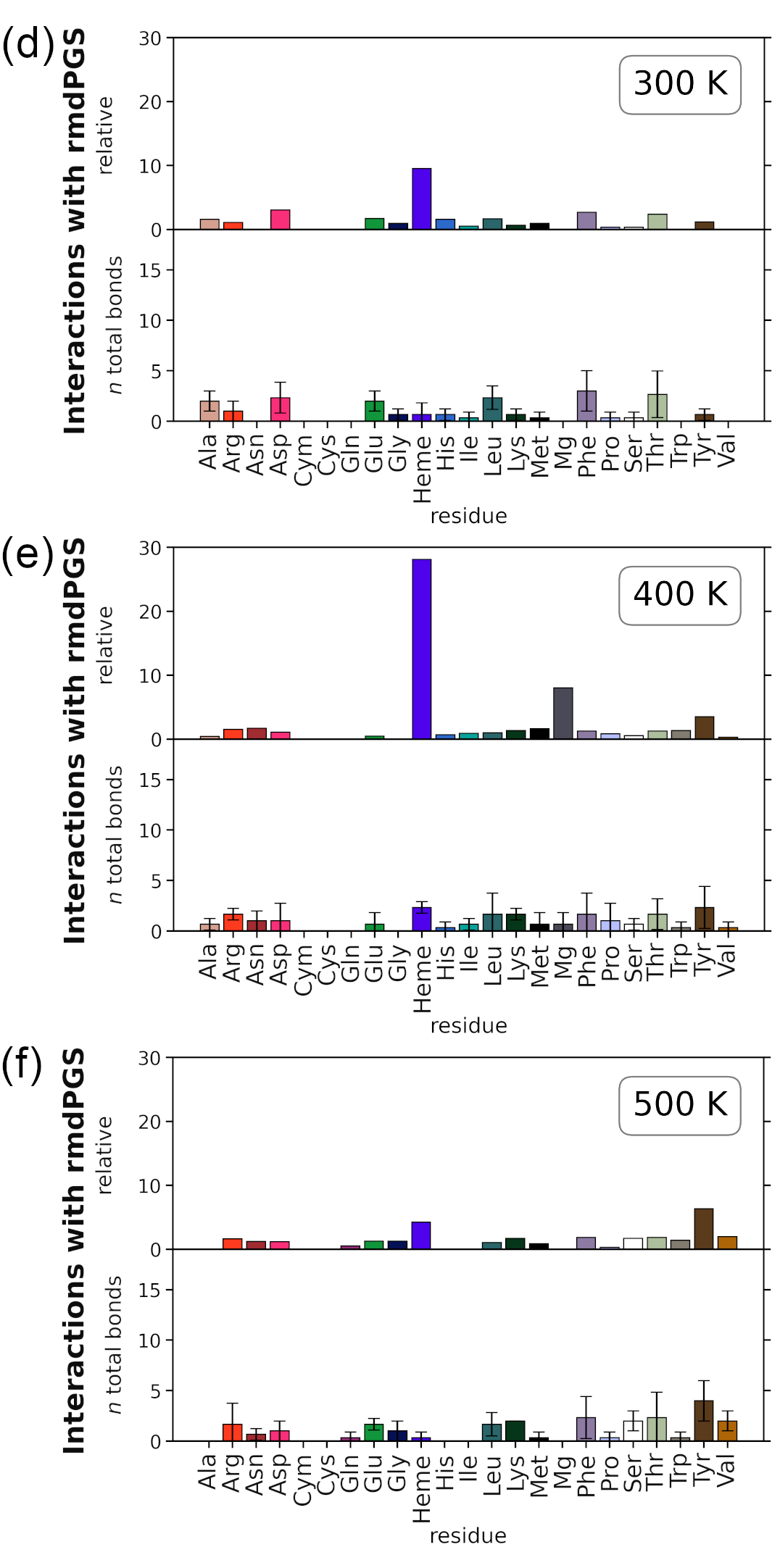}
    \caption{With solvent}
    \end{subfigure}
    \caption{Bond analysis for high concentrations of \ch{NO} with \textit{Cvi}UPO. All panels show the total number of bonds to an additional hydrogen atom per residue and the relative interactions calculated following equation \ref{eq:2.2}. The panels (a) and (d) show the interactions at 300\,K, without and with solvent, respectively. The panels (b) and (e) show the interactions at at 400\,K and the panels (c) and (f) show the interactions at at 500\,K}
    \label{SI-Conc_MD_Cvi-no}
\end{figure}

\begin{figure}[H]
    \renewcommand\thesubfigure{\roman{subfigure}}
    \centering
    \begin{subfigure}{0.49\textwidth}
    \includegraphics[width=1\textwidth]{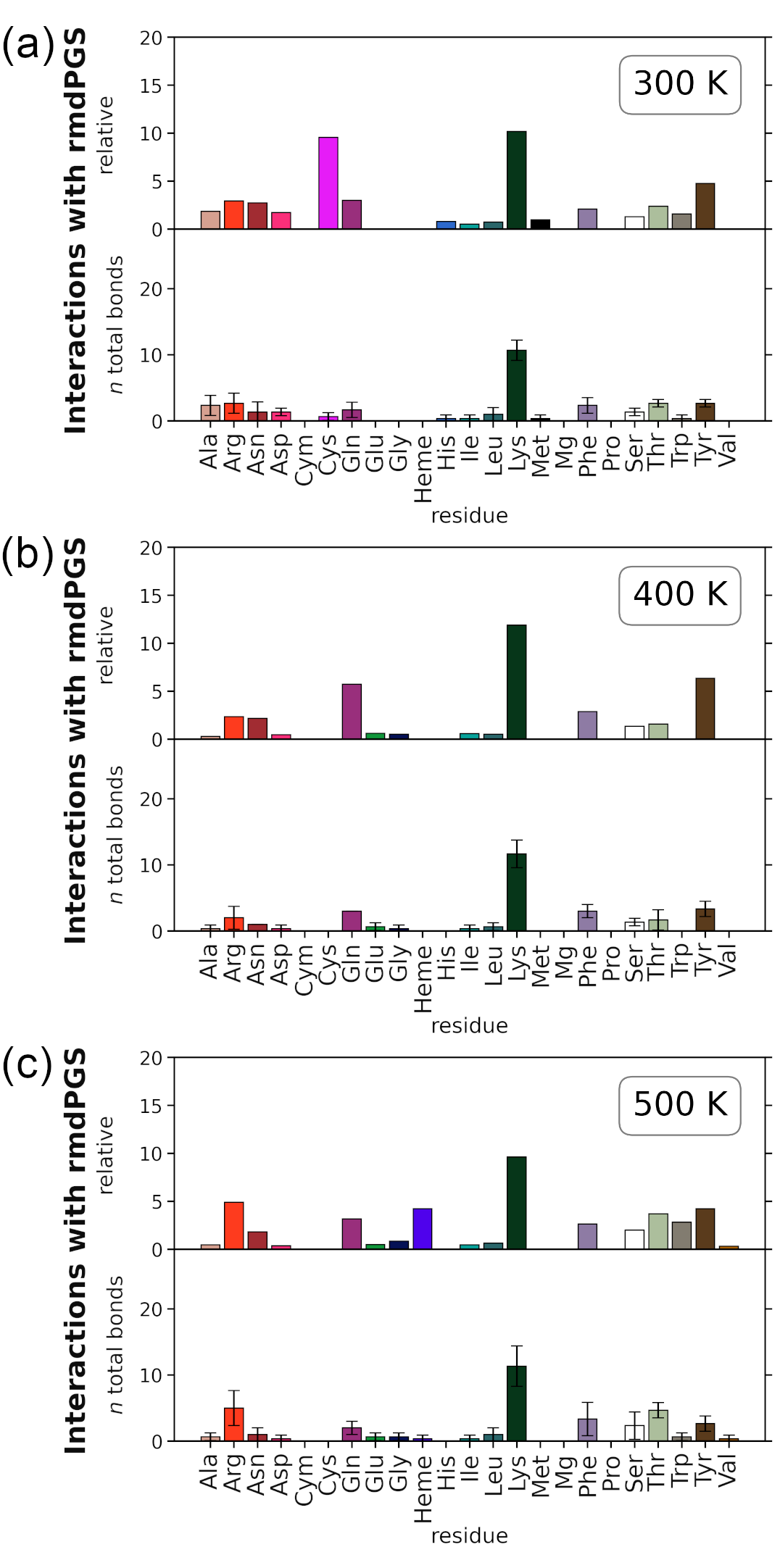}
    \caption{Without solvent}
    \end{subfigure}
    \begin{subfigure}{0.49\textwidth}
    \includegraphics[width=1\textwidth]{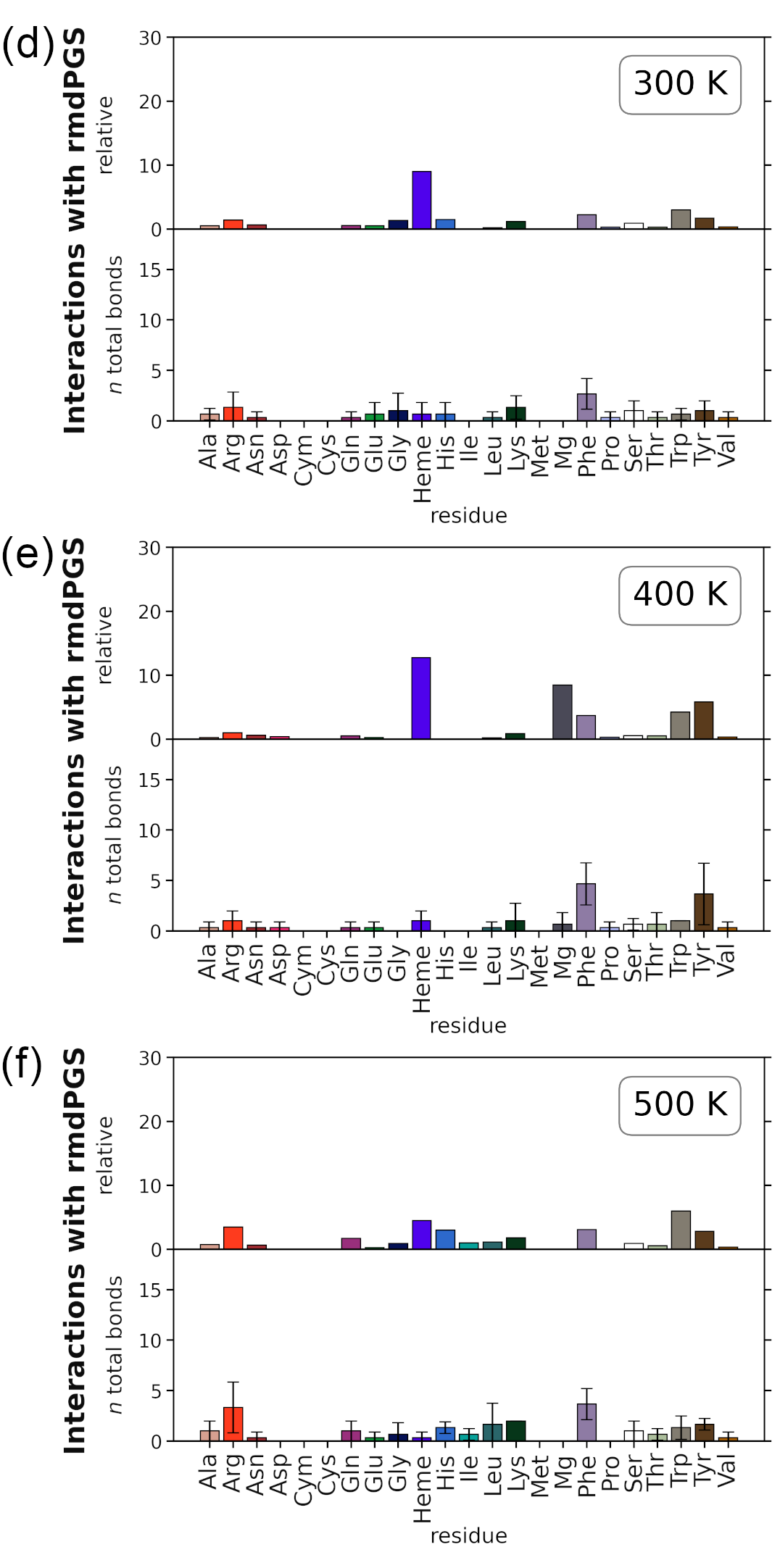}
    \caption{With solvent}
    \end{subfigure}
    \caption{Bond analysis for high concentrations of \ch{O2} with \textit{Cvi}UPO. All panels show the total number of bonds to an additional hydrogen atom per residue and the relative interactions calculated following equation \ref{eq:2.2}. The panels (a) and (d) show the interactions at 300\,K, without and with solvent, respectively. The panels (b) and (e) show the interactions at at 400\,K and the panels (c) and (f) show the interactions at at 500\,K}
    \label{SI-Conc_MD_Cvi-o2}
\end{figure}

\begin{figure}[H]
    \renewcommand\thesubfigure{\roman{subfigure}}
    \centering
    \begin{subfigure}{0.49\textwidth}
    \includegraphics[width=1\textwidth]{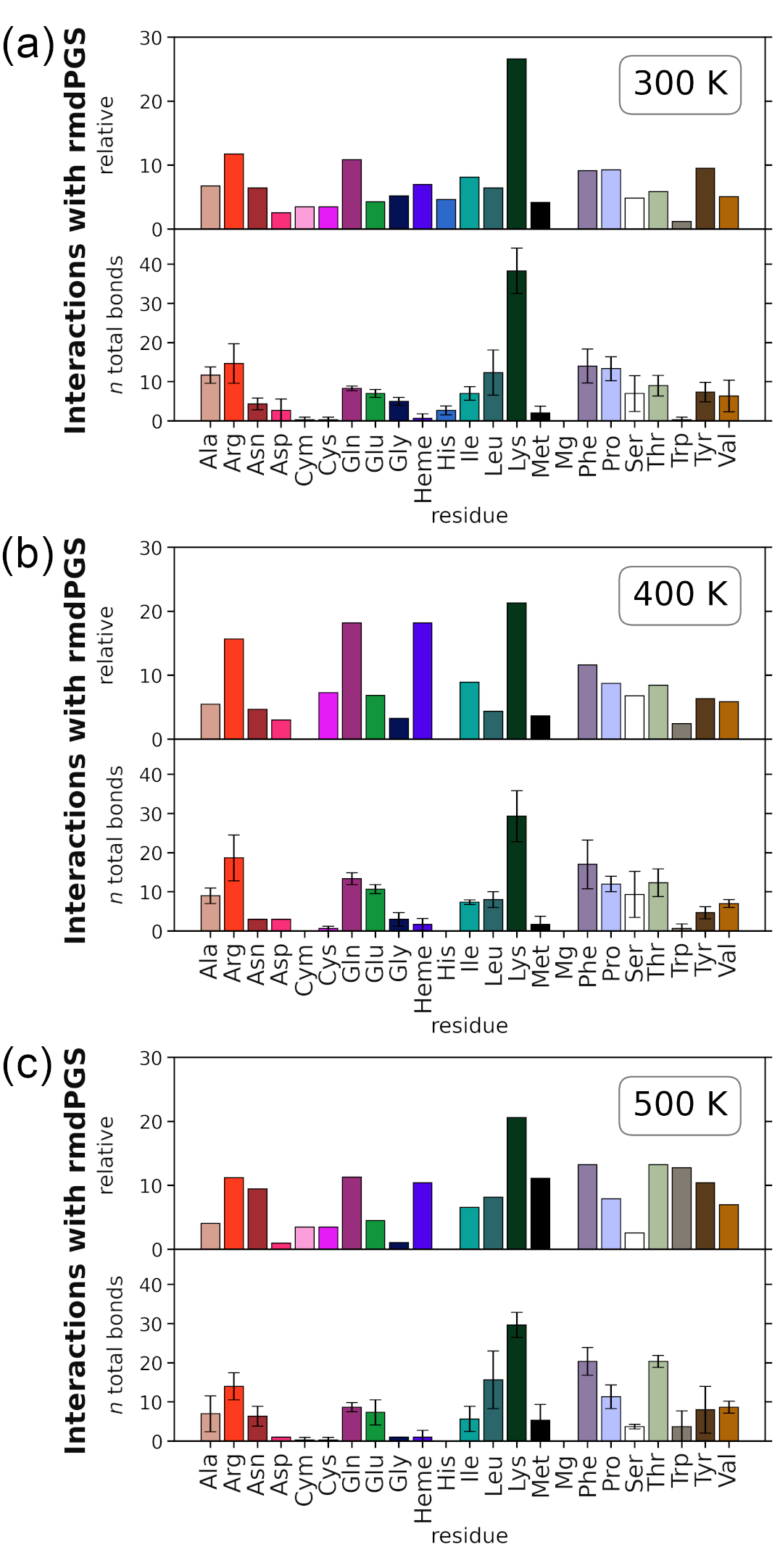}
    \caption{Without solvent}
    \end{subfigure}
    \begin{subfigure}{0.49\textwidth}
    \includegraphics[width=1\textwidth]{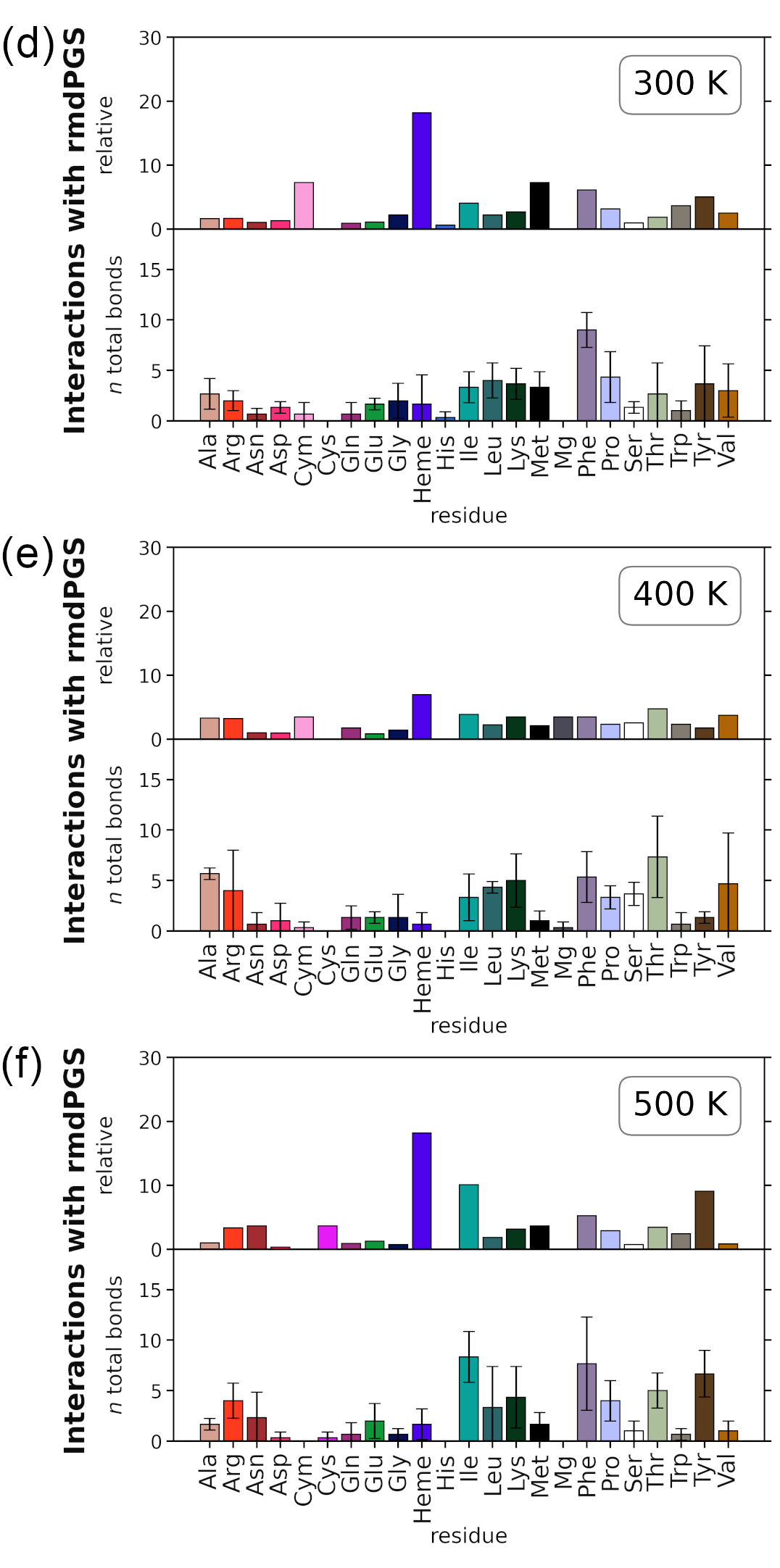}
    \caption{With solvent}
    \end{subfigure}
    \caption{Bond analysis for high concentrations of oxygen with \textit{Cvi}UPO. All panels show the total number of bonds to an additional hydrogen atom per residue and the relative interactions calculated following equation \ref{eq:2.2}. The panels (a) and (d) show the interactions at 300\,K, without and with solvent, respectively. The panels (b) and (e) show the interactions at at 400\,K and the panels (c) and (f) show the interactions at at 500\,K}
    \label{SI-Conc_MD_Cvi-o}
\end{figure}

\subsubsection{\textit{Aae}UPO} \label{SI_Conc-Aae}
The interaction behavior of the PGS with \textit{Aae}UPO shows more pronounced differences to the SASA predictions. While the SASA analysis showed various interactions between PGS and Arg, these are much reduced in the MD simulations. It is particularly striking that all PGS show a strong tendency towards proline (Pro) in the MD simulations, which might be due to the fact that Pro is the second most frequent amino acid in the protein. Similar to the other two enzymes temperature increase does not seem to change the behavior significantly. Interestingly the interaction profiles of the solvent simulations of \textit{Aae}UPO are closer to the SASA predictions than the vacuum results. One explanation for this could be that the PGS in the solvent have the opportunity to diffuse to their preferred interaction partners. Nevertheless, the results should also be treated with caution here, as the error bars are still very high.

\begin{figure}[H]
    \renewcommand\thesubfigure{\roman{subfigure}}
    \centering
    \begin{subfigure}{0.49\textwidth}
    \includegraphics[width=1\textwidth]{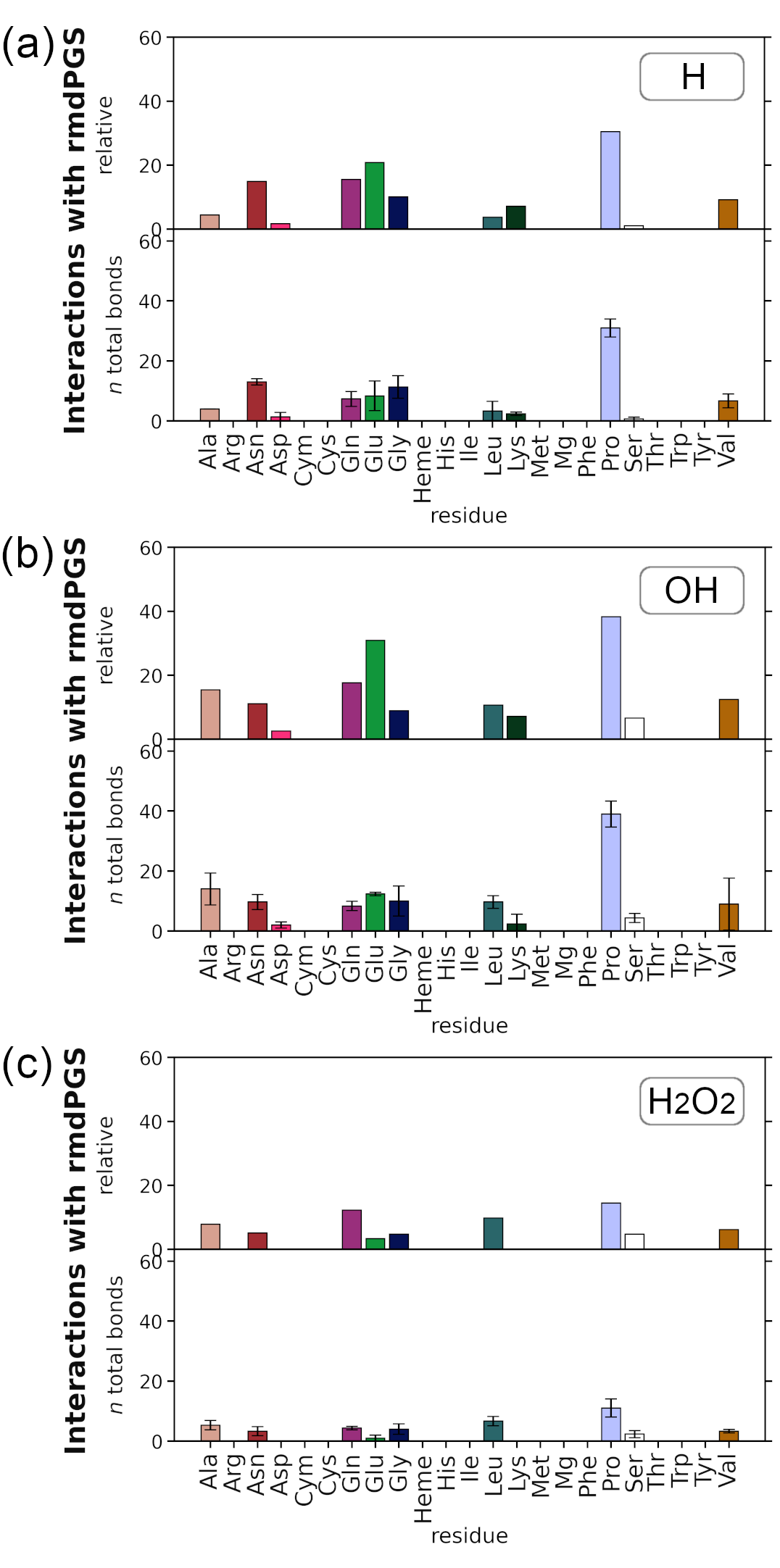}
    \caption{Without solvent}
    \end{subfigure}
    \begin{subfigure}{0.49\textwidth}
    \includegraphics[width=1\textwidth]{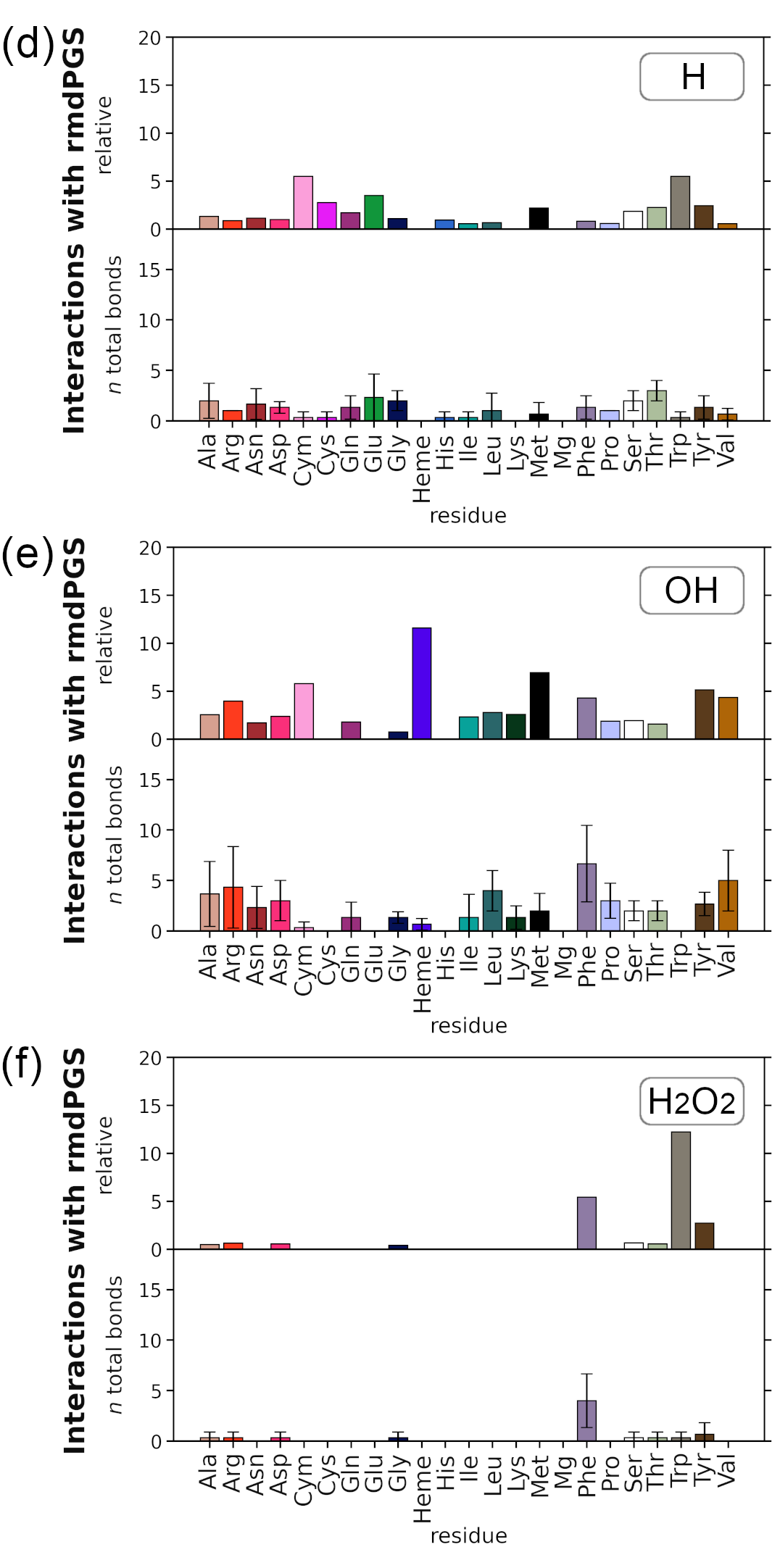}
    \caption{With solvent}
    \end{subfigure}
    \caption{Bond analysis for high concentrations of PGS with \textit{Aae}UPO at 300\,K. All panels show the total number of bonds to an additional hydrogen atom per residue and the relative interactions calculated following equation \ref{eq:2.2}. (a) shows the bonds to \ch{H} without solvent while (b) and (c) show the bonds to \ch{OH}, \ch{H2O2}. The panel (d) shows the bonds to \ch{H} with solvent. (b) and (c) show the bonds with solvent for \ch{OH}, \ch{H2O2}, respectively.}
    \label{SI-Conc_MD_Aae-1}
\end{figure}

\begin{figure}[H]
    \renewcommand\thesubfigure{\roman{subfigure}}
    \centering
    \begin{subfigure}{0.49\textwidth}
    \includegraphics[width=1\textwidth]{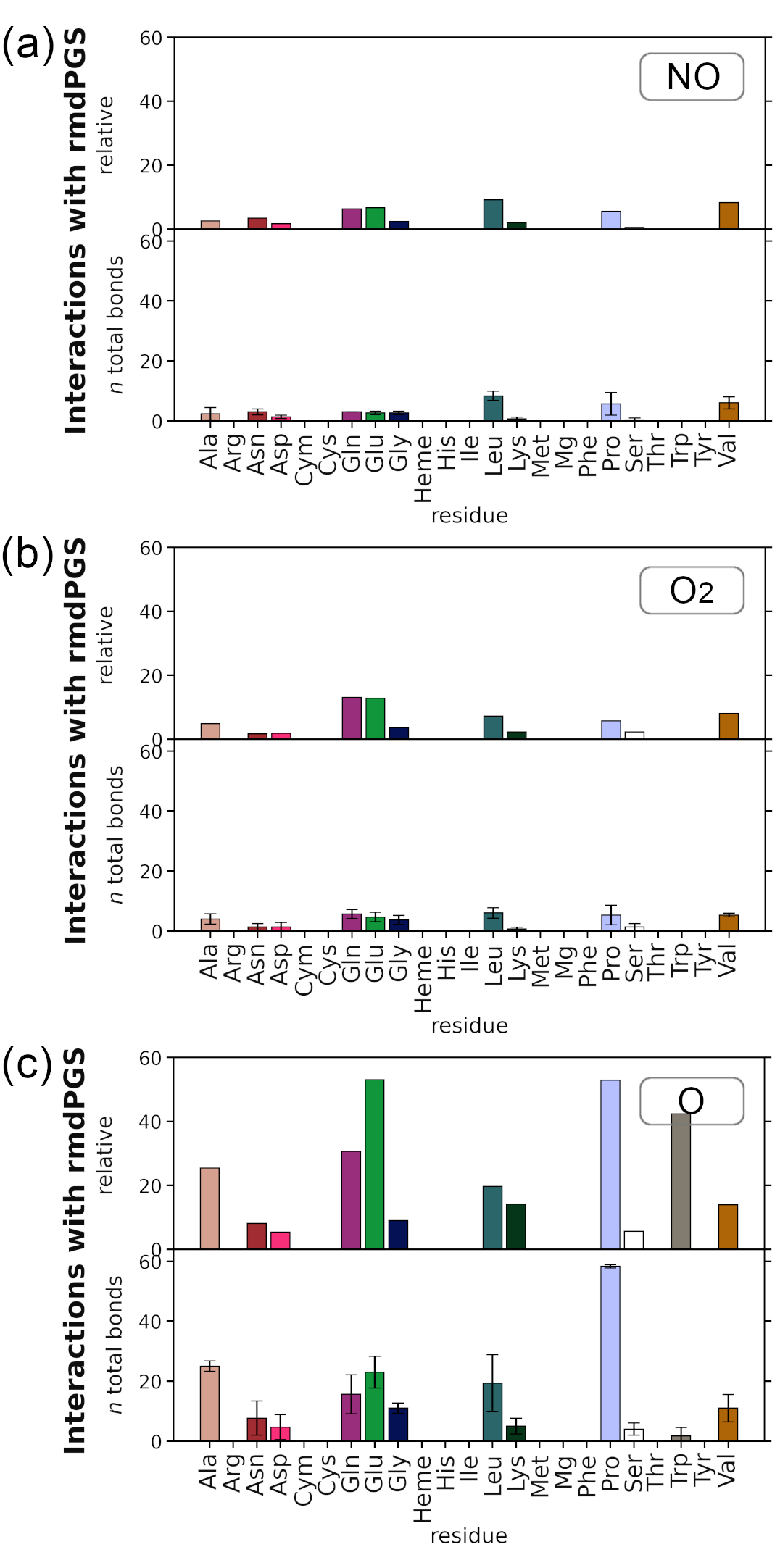}
    \caption{Without solvent}
    \end{subfigure}
    \begin{subfigure}{0.49\textwidth}
    \includegraphics[width=1\textwidth]{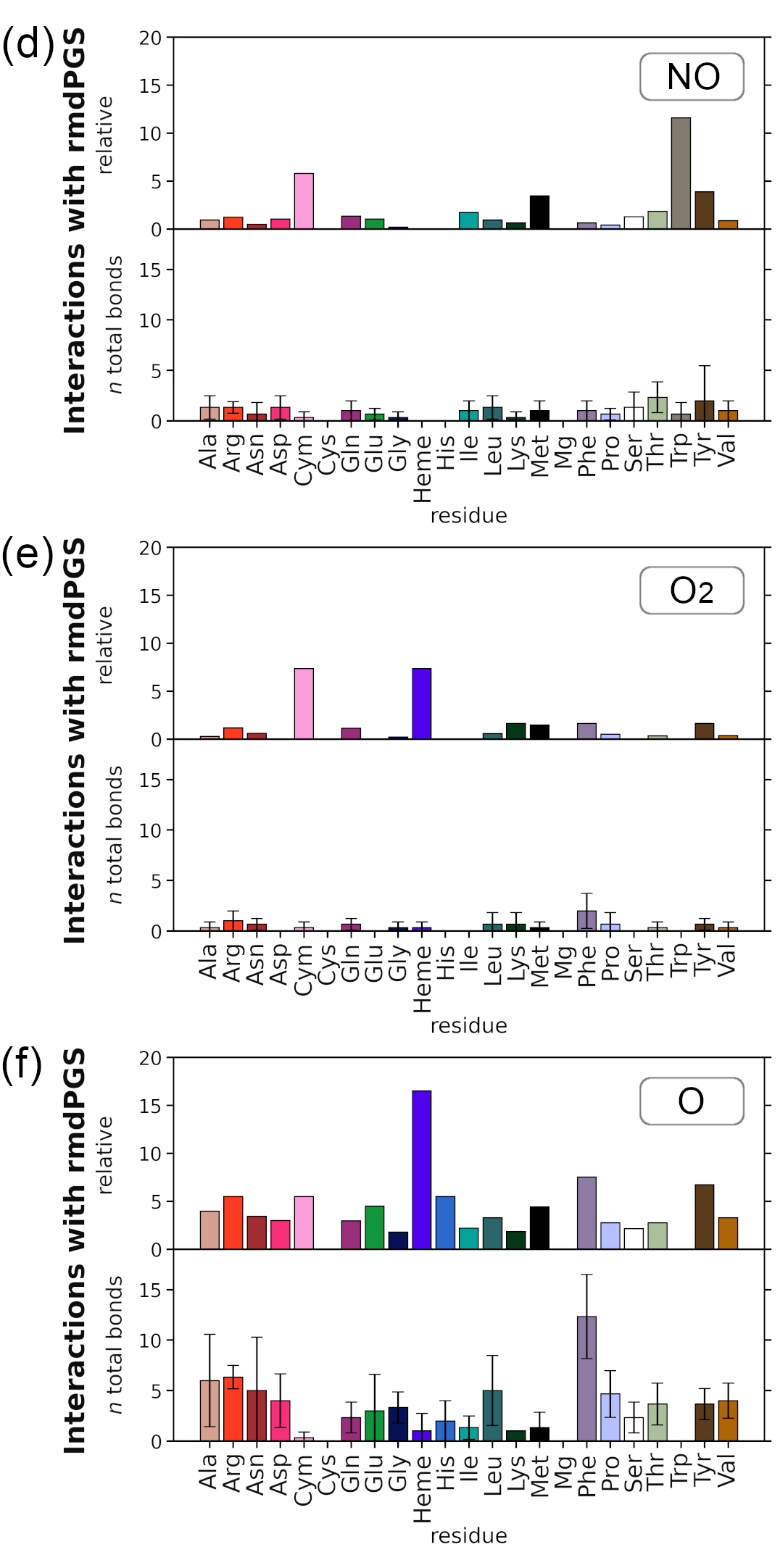}
    \caption{With solvent}
    \end{subfigure}
    \caption{Bond analysis for high concentrations of PGS with \textit{Aae}UPO at 300\,K. All panels show the total number of bonds to an additional hydrogen atom per residue and the relative interactions calculated following equation \ref{eq:2.2}. (a) shows the bonds to \ch{NO} without solvent while (b) and (c) show the bonds to \ch{O2}, \ch{O}. The panel (d) shows the bonds to \ch{NO} with solvent.( b) and (c) show the bonds with solvent for \ch{O2}, \ch{O}, respectively.}
    \label{SI-Conc_MD_Aae-2}
\end{figure}

The results for higher temperatures were excluded from this section to minimize the length of the Supplementary Information. However, the corresponding data are available and have been provided alongside the main paper

\subsubsection{ GapA} \label{SI_Conc-GapA}
For GapA, the MD simulations also follow the trend of the SASA predictions. For most PGS the results of the MD simulations look very similar with slight changes in the ratios. Overall, Lys is the most attacked amino acid for all tested PGS, followed by Thr and Arg. Interestingly, 
NO interacts with more residues in the MD simulations than predicted by the SASA analysis, although the variety of different residues is smaller compared to the other PGS. As with \textit{Cvi}UPO, the solvent results are even closer to the SASA predictions.

\begin{figure}[H]
    \renewcommand\thesubfigure{\roman{subfigure}}
    \centering
    \begin{subfigure}{0.49\textwidth}
    \includegraphics[width=1\textwidth]{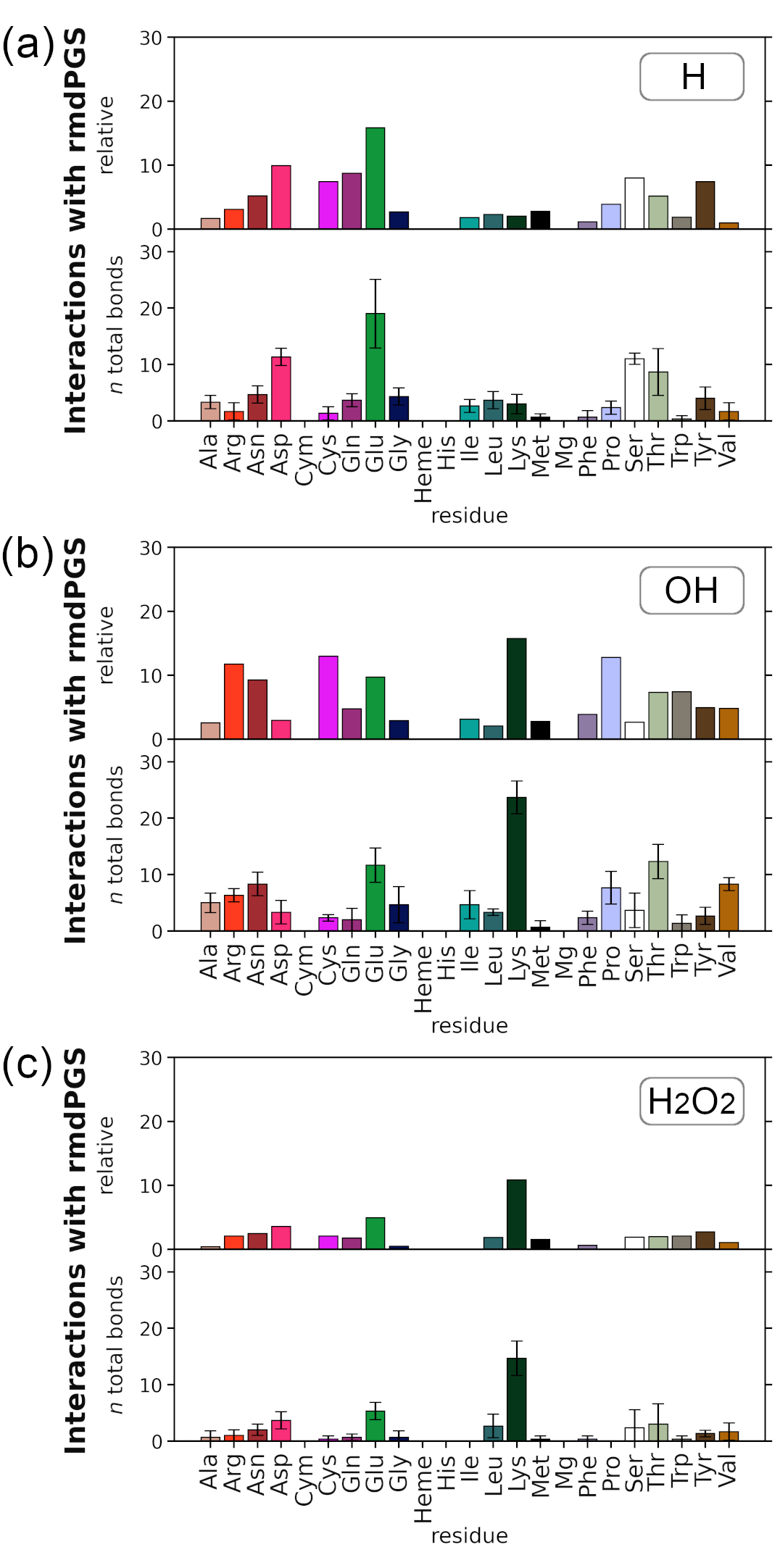}
    \caption{Without solvent}
    \end{subfigure}
    \begin{subfigure}{0.49\textwidth}
    \includegraphics[width=1\textwidth]{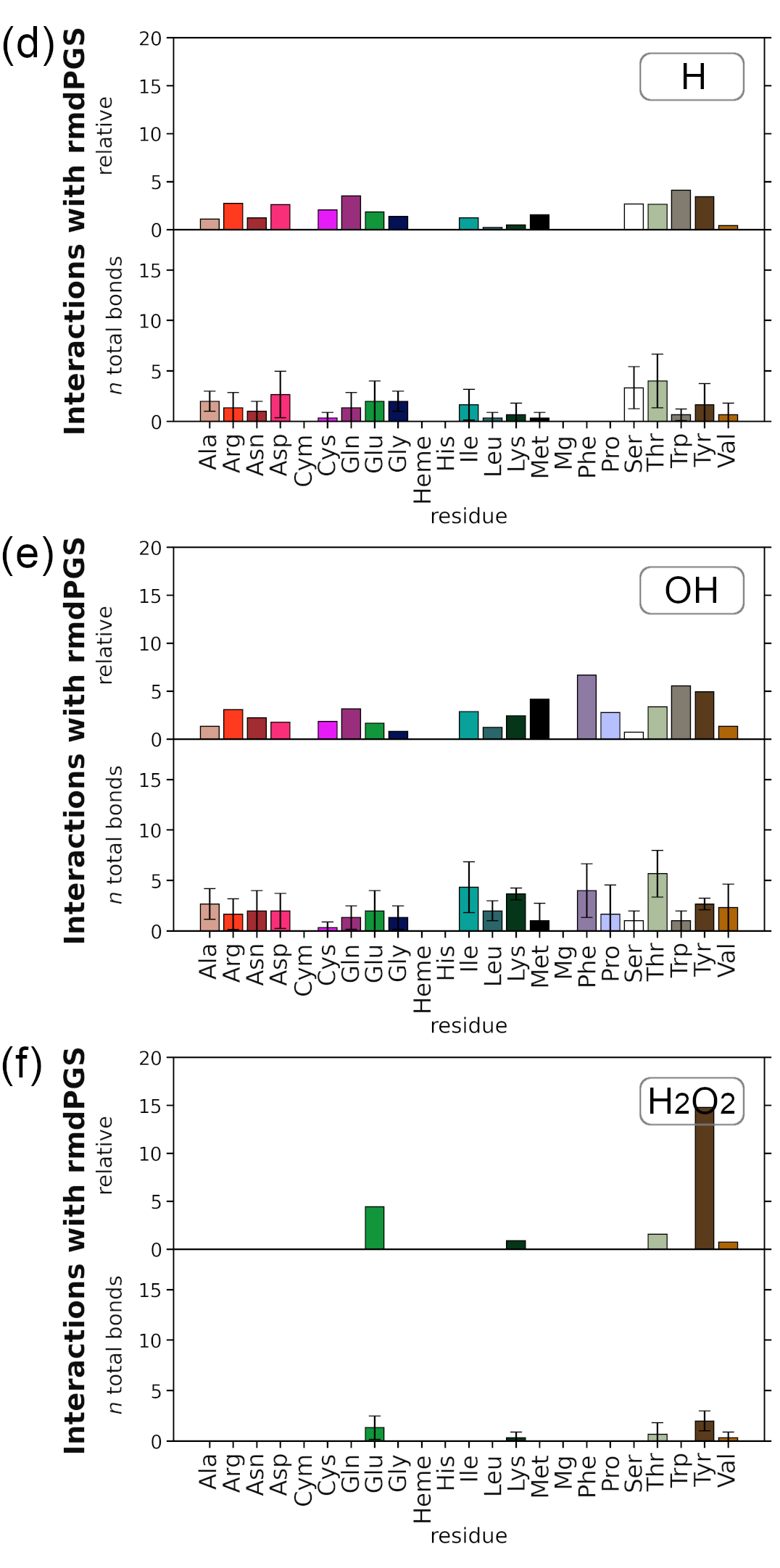}
    \caption{With solvent}
    \end{subfigure}
    \caption{Bond analysis for high concentrations of PGS with GapA at 300\,K. All panels show the total number of bonds to an additional hydrogen atom per residue and the relative interactions calculated following equation \ref{eq:2.2}. (a) shows the bonds to \ch{H} without solvent while (b) and (c) show the bonds to \ch{OH}, \ch{H2O2}. The panel (d) shows the bonds to \ch{H} with solvent. (b) and (c) show the bonds with solvent for \ch{OH}, \ch{H2O2}, respectively.}
    \label{SI-Conc_MD_GapA-1}
\end{figure}

\begin{figure}[H]
    \renewcommand\thesubfigure{\roman{subfigure}}
    \centering
    \begin{subfigure}{0.49\textwidth}
    \includegraphics[width=1\textwidth]{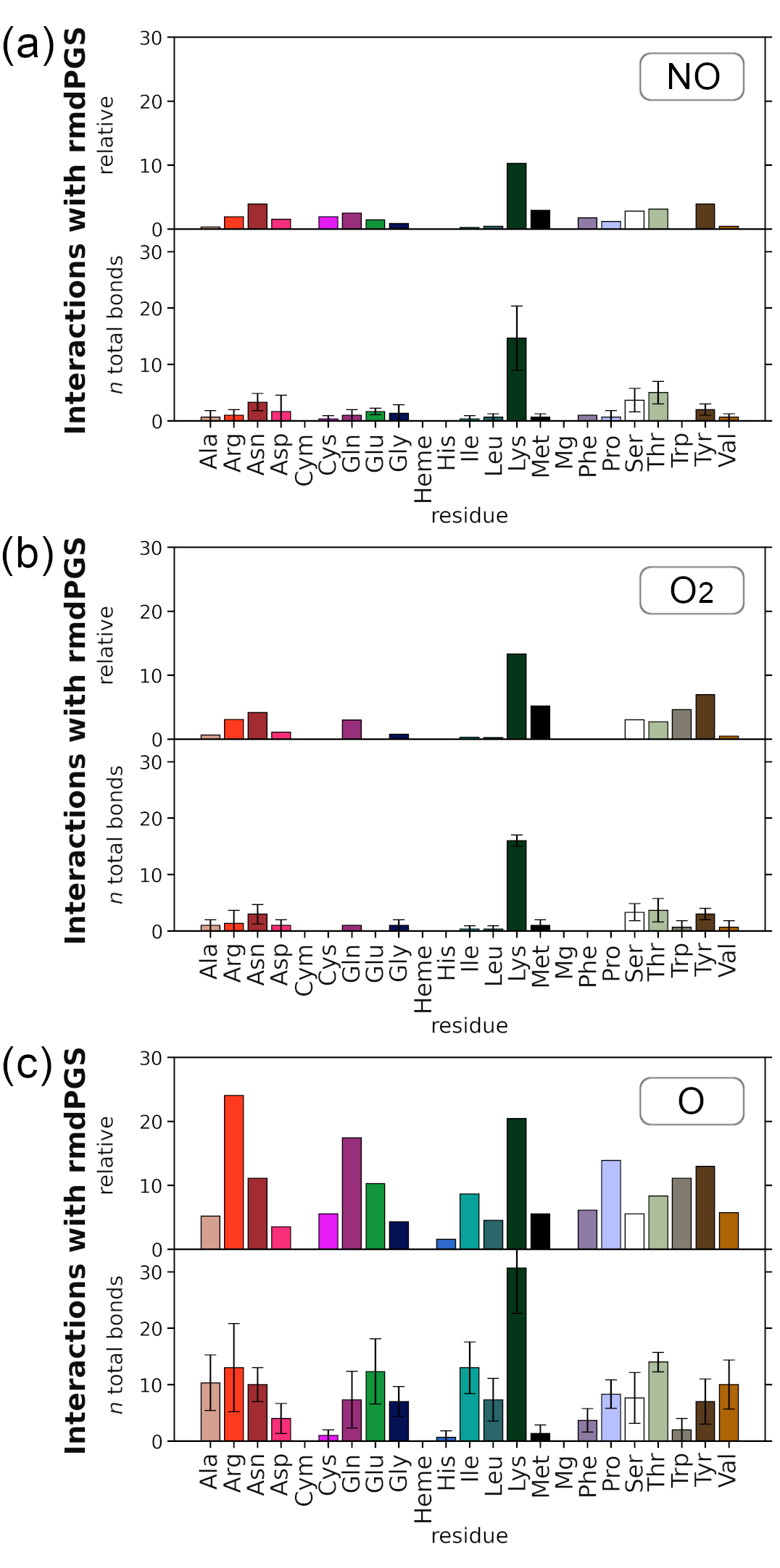}
    \caption{Without solvent}
    \end{subfigure}
    \begin{subfigure}{0.49\textwidth}
    \includegraphics[width=1\textwidth]{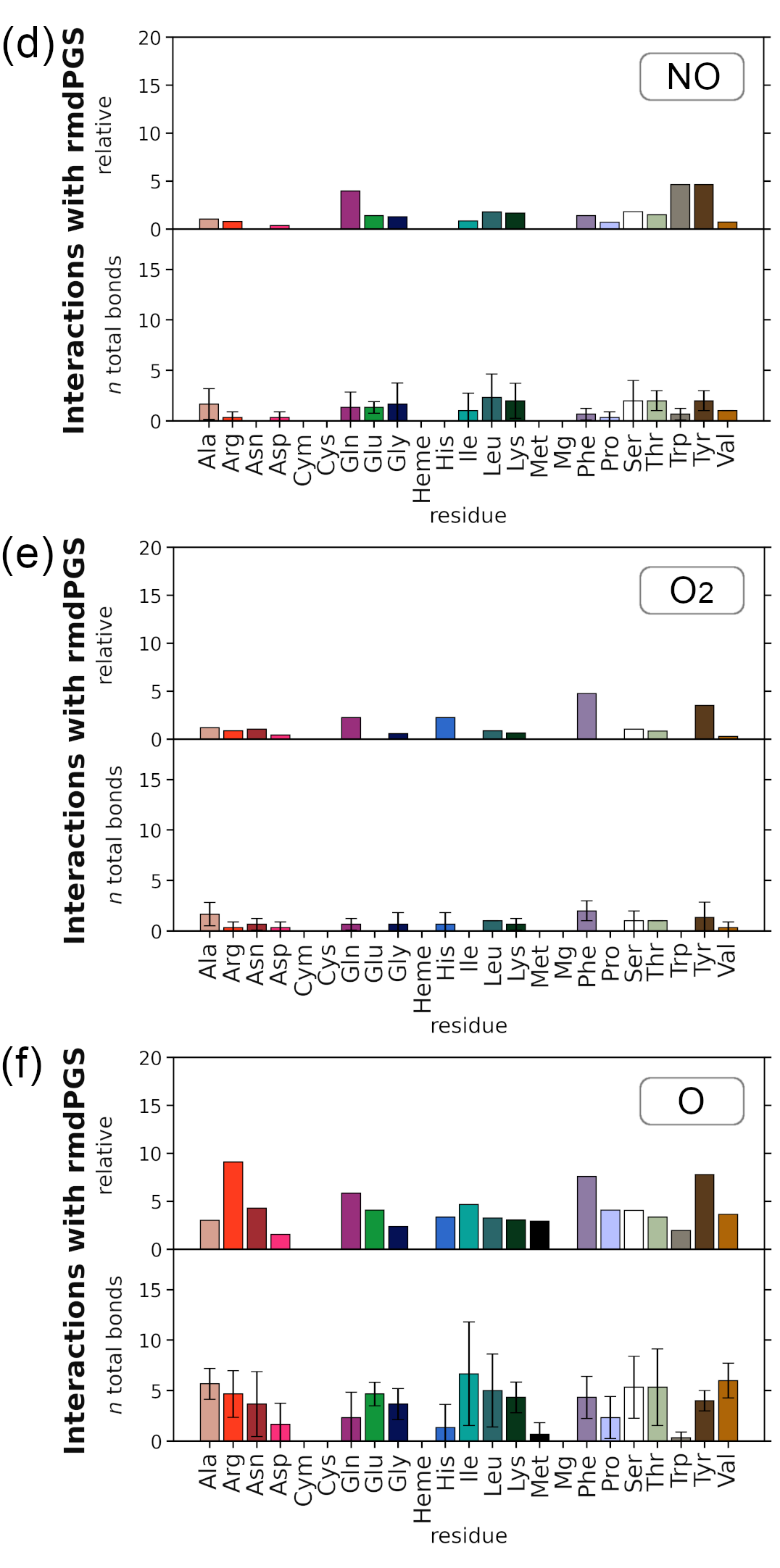}
    \caption{With solvent}
    \end{subfigure}
    \caption{Bond analysis for high concentrations of PGS with GapA at 300\,K. All panels show the total number of bonds to an additional hydrogen atom per residue and the relative interactions calculated following equation \ref{eq:2.2}. (a) shows the bonds to \ch{NO} without solvent while (b) and (c) show the bonds to \ch{O2}, \ch{O}. The panel (d) shows the bonds to \ch{NO} with solvent. (b) and (c) show the bonds with solvent for \ch{O2}, \ch{O}, respectively.}
    \label{SI-Conc_MD_GapA-2}
\end{figure}

The results for higher temperatures were excluded from this section to minimize the length of the Supplementary Information. However, the corresponding data are available and have been provided alongside the main paper
%%%%%%%%%%%%%%%%%%%%%%%%%%%%%%%%%%%%%%%%%%%%%%%%%%%%%%%%%%%%%%%%%%%%%%%%%%

\end{suppinfo}

\bibliography{Lit_Bio-Catalysis}

\providecommand{\latin}[1]{#1}
\makeatletter
\providecommand{\doi}
  {\begingroup\let\do\@makeother\dospecials
  \catcode`\{=1 \catcode`\}=2 \doi@aux}
\providecommand{\doi@aux}[1]{\endgroup\texttt{#1}}
\makeatother
\providecommand*\mcitethebibliography{\thebibliography}
\csname @ifundefined\endcsname{endmcitethebibliography}  {\let\endmcitethebibliography\endthebibliography}{}
\begin{mcitethebibliography}{49}
\providecommand*\natexlab[1]{#1}
\providecommand*\mciteSetBstSublistMode[1]{}
\providecommand*\mciteSetBstMaxWidthForm[2]{}
\providecommand*\mciteBstWouldAddEndPuncttrue
  {\def\EndOfBibitem{\unskip.}}
\providecommand*\mciteBstWouldAddEndPunctfalse
  {\let\EndOfBibitem\relax}
\providecommand*\mciteSetBstMidEndSepPunct[3]{}
\providecommand*\mciteSetBstSublistLabelBeginEnd[3]{}
\providecommand*\EndOfBibitem{}
\mciteSetBstSublistMode{f}
\mciteSetBstMaxWidthForm{subitem}{(\alph{mcitesubitemcount})}
\mciteSetBstSublistLabelBeginEnd
  {\mcitemaxwidthsubitemform\space}
  {\relax}
  {\relax}

\bibitem[Rothenberg(2017)]{Rothenberg2017CatalysisConceptsGreen}
Rothenberg,~G. \emph{Catalysis: Concepts and Green Applications}, 2nd ed.; Wiley-VCH Verlag, 2017\relax
\mciteBstWouldAddEndPuncttrue
\mciteSetBstMidEndSepPunct{\mcitedefaultmidpunct}
{\mcitedefaultendpunct}{\mcitedefaultseppunct}\relax
\EndOfBibitem
\bibitem[Berg \latin{et~al.}(2018)Berg, Tymoczko, Gatto, and Stryer]{Berg2018Biochemie}
Berg,~J.~M.; Tymoczko,~J.~L.; Gatto,~G.~J.; Stryer,~L. \emph{Biochemie}; Springer Berlin Heidelberg, 2018\relax
\mciteBstWouldAddEndPuncttrue
\mciteSetBstMidEndSepPunct{\mcitedefaultmidpunct}
{\mcitedefaultendpunct}{\mcitedefaultseppunct}\relax
\EndOfBibitem
\bibitem[Bell \latin{et~al.}(2021)Bell, Finnigan, France, Green, Hayes, Hepworth, Lovelock, Niikura, Osuna, Romero, Ryan, Turner, and Flitsch]{Bell2021Biocatalysis}
Bell,~E.~L.; Finnigan,~W.; France,~S.~P.; Green,~A.~P.; Hayes,~M.~A.; Hepworth,~L.~J.; Lovelock,~S.~L.; Niikura,~H.; Osuna,~S.; Romero,~E.; Ryan,~K.~S.; Turner,~N.~J.; Flitsch,~S.~L. Biocatalysis. \emph{Nature Reviews Methods Primers} \textbf{2021}, \emph{1}, 1--21\relax
\mciteBstWouldAddEndPuncttrue
\mciteSetBstMidEndSepPunct{\mcitedefaultmidpunct}
{\mcitedefaultendpunct}{\mcitedefaultseppunct}\relax
\EndOfBibitem
\bibitem[Wu \latin{et~al.}(2021)Wu, Snajdrova, Moore, Baldenius, and Bornscheuer]{Wu2021BiocatalysisEnzymaticSynthesis}
Wu,~S.; Snajdrova,~R.; Moore,~J.~C.; Baldenius,~K.; Bornscheuer,~U.~T. Biocatalysis: Enzymatic Synthesis for Industrial Applications. \emph{Angewandte Chemie International Edition} \textbf{2021}, \emph{60}, 88--119\relax
\mciteBstWouldAddEndPuncttrue
\mciteSetBstMidEndSepPunct{\mcitedefaultmidpunct}
{\mcitedefaultendpunct}{\mcitedefaultseppunct}\relax
\EndOfBibitem
\bibitem[Vasudevan and Weiland(1990)Vasudevan, and Weiland]{Vasudevan1990Deactivationcatalasehydrogen}
Vasudevan,~P.~T.; Weiland,~R.~H. Deactivation of catalase by hydrogen peroxide. \emph{Biotechnology and Bioengineering} \textbf{1990}, \emph{36}, 783--789\relax
\mciteBstWouldAddEndPuncttrue
\mciteSetBstMidEndSepPunct{\mcitedefaultmidpunct}
{\mcitedefaultendpunct}{\mcitedefaultseppunct}\relax
\EndOfBibitem
\bibitem[Valderrama \latin{et~al.}(2002)Valderrama, Ayala, and Vazquez-Duhalt]{Valderrama2002SuicideInactivationPeroxidases}
Valderrama,~B.; Ayala,~M.; Vazquez-Duhalt,~R. Suicide {Inactivation} of {Peroxidases} and the {Challenge} of {Engineering} {More} {Robust} {Enzymes}. \emph{Chemistry \& Biology} \textbf{2002}, \emph{9}, 555--565\relax
\mciteBstWouldAddEndPuncttrue
\mciteSetBstMidEndSepPunct{\mcitedefaultmidpunct}
{\mcitedefaultendpunct}{\mcitedefaultseppunct}\relax
\EndOfBibitem
\bibitem[Karich \latin{et~al.}(2016)Karich, Scheibner, Ullrich, and Hofrichter]{Karich2016Exploringcatalaseactivity}
Karich,~A.; Scheibner,~K.; Ullrich,~R.; Hofrichter,~M. Exploring the catalase activity of unspecific peroxygenases and the mechanism of peroxide-dependent heme destruction. \emph{Journal of Molecular Catalysis B: Enzymatic} \textbf{2016}, \emph{134}, 238--246\relax
\mciteBstWouldAddEndPuncttrue
\mciteSetBstMidEndSepPunct{\mcitedefaultmidpunct}
{\mcitedefaultendpunct}{\mcitedefaultseppunct}\relax
\EndOfBibitem
\bibitem[Wapshott-Stehli and Grunden(2021)Wapshott-Stehli, and Grunden]{WapshottStehli2021situH2O2generation}
Wapshott-Stehli,~H.~L.; Grunden,~A.~M. In situ {H2O2} generation methods in the context of enzyme biocatalysis. \emph{Enzyme and Microbial Technology} \textbf{2021}, \emph{145}, 109744\relax
\mciteBstWouldAddEndPuncttrue
\mciteSetBstMidEndSepPunct{\mcitedefaultmidpunct}
{\mcitedefaultendpunct}{\mcitedefaultseppunct}\relax
\EndOfBibitem
\bibitem[Yayci \latin{et~al.}(2020)Yayci, Álvaro Gómez~Baraibar, Krewing, Fueyo, Hollmann, Alcalde, Kourist, and Bandow]{Yayci2020PlasmaDrivenSitu}
Yayci,~A.; Álvaro Gómez~Baraibar; Krewing,~M.; Fueyo,~E.~F.; Hollmann,~F.; Alcalde,~M.; Kourist,~R.; Bandow,~J.~E. Plasma-Driven in Situ Production of Hydrogen Peroxide for Biocatalysis. \emph{ChemSusChem} \textbf{2020}, \emph{13}, 2072--2079\relax
\mciteBstWouldAddEndPuncttrue
\mciteSetBstMidEndSepPunct{\mcitedefaultmidpunct}
{\mcitedefaultendpunct}{\mcitedefaultseppunct}\relax
\EndOfBibitem
\bibitem[Yayci \latin{et~al.}(2020)Yayci, Dirks, Kogelheide, Alcalde, Hollmann, Awakowicz, and Bandow]{Yayci2020MicroscaleAtmosphericPressure}
Yayci,~A.; Dirks,~T.; Kogelheide,~F.; Alcalde,~M.; Hollmann,~F.; Awakowicz,~P.; Bandow,~J.~E. Microscale Atmospheric Pressure Plasma Jet as a Source for Plasma-Driven Biocatalysis. \emph{ChemCatChem} \textbf{2020}, \emph{12}, 5893--5897\relax
\mciteBstWouldAddEndPuncttrue
\mciteSetBstMidEndSepPunct{\mcitedefaultmidpunct}
{\mcitedefaultendpunct}{\mcitedefaultseppunct}\relax
\EndOfBibitem
\bibitem[{Schr\"odinger, LLC}(2015)]{PyMOL}
{Schr\"odinger, LLC} \relax
\mciteBstWouldAddEndPunctfalse
\mciteSetBstMidEndSepPunct{\mcitedefaultmidpunct}
{}{\mcitedefaultseppunct}\relax
\EndOfBibitem
\bibitem[Dickenson \latin{et~al.}(2018)Dickenson, Britun, Nikiforov, Leys, Hasan, and Walsh]{Dickenson2018generationtransportreactive}
Dickenson,~A.; Britun,~N.; Nikiforov,~A.; Leys,~C.; Hasan,~M.~I.; Walsh,~J.~L. The generation and transport of reactive nitrogen species from a low temperature atmospheric pressure air plasma source. \emph{Physical Chemistry Chemical Physics} \textbf{2018}, \emph{20}, 28499--28510\relax
\mciteBstWouldAddEndPuncttrue
\mciteSetBstMidEndSepPunct{\mcitedefaultmidpunct}
{\mcitedefaultendpunct}{\mcitedefaultseppunct}\relax
\EndOfBibitem
\bibitem[202(2022)]{2022GROMACS2022.2Source}
{GROMACS} 2022.2 {Source} code. 2022; \url{https://zenodo.org/records/6637571}\relax
\mciteBstWouldAddEndPuncttrue
\mciteSetBstMidEndSepPunct{\mcitedefaultmidpunct}
{\mcitedefaultendpunct}{\mcitedefaultseppunct}\relax
\EndOfBibitem
\bibitem[Abraham \latin{et~al.}(2015)Abraham, Murtola, Schulz, P{\'{a}}ll, Smith, Hess, and Lindahl]{Abraham2015GROMACSHighperformance}
Abraham,~M.~J.; Murtola,~T.; Schulz,~R.; P{\'{a}}ll,~S.; Smith,~J.~C.; Hess,~B.; Lindahl,~E. {GROMACS}: {High} performance molecular simulations through multi-level parallelism from laptops to supercomputers. \emph{SoftwareX} \textbf{2015}, \emph{1-2}, 19--25\relax
\mciteBstWouldAddEndPuncttrue
\mciteSetBstMidEndSepPunct{\mcitedefaultmidpunct}
{\mcitedefaultendpunct}{\mcitedefaultseppunct}\relax
\EndOfBibitem
\bibitem[Berendsen \latin{et~al.}(1995)Berendsen, van~der Spoel, and van Drunen]{Berendsen1995GROMACSmessagepassing}
Berendsen,~H. J.~C.; van~der Spoel,~D.; van Drunen,~R. {GROMACS}: {A} message-passing parallel molecular dynamics implementation. \emph{Computer Physics Communications} \textbf{1995}, \emph{91}, 43--56\relax
\mciteBstWouldAddEndPuncttrue
\mciteSetBstMidEndSepPunct{\mcitedefaultmidpunct}
{\mcitedefaultendpunct}{\mcitedefaultseppunct}\relax
\EndOfBibitem
\bibitem[Hess \latin{et~al.}(2008)Hess, Kutzner, van~der Spoel, and Lindahl]{Hess2008GROMACS4Algorithms}
Hess,~B.; Kutzner,~C.; van~der Spoel,~D.; Lindahl,~E. {GROMACS} 4:\hspace{0.167em} {Algorithms} for {Highly} {Efficient}, {Load}-{Balanced}, and {Scalable} {Molecular} {Simulation}. \emph{Journal of Chemical Theory and Computation} \textbf{2008}, \emph{4}, 435--447\relax
\mciteBstWouldAddEndPuncttrue
\mciteSetBstMidEndSepPunct{\mcitedefaultmidpunct}
{\mcitedefaultendpunct}{\mcitedefaultseppunct}\relax
\EndOfBibitem
\bibitem[Lindahl \latin{et~al.}(2001)Lindahl, Hess, and van~der Spoel]{Lindahl2001GROMACS3.0package}
Lindahl,~E.; Hess,~B.; van~der Spoel,~D. {GROMACS} 3.0: a package for molecular simulation and trajectory analysis. \emph{Molecular modeling annual} \textbf{2001}, \emph{7}, 306--317\relax
\mciteBstWouldAddEndPuncttrue
\mciteSetBstMidEndSepPunct{\mcitedefaultmidpunct}
{\mcitedefaultendpunct}{\mcitedefaultseppunct}\relax
\EndOfBibitem
\bibitem[P{\'{a}}ll \latin{et~al.}(2015)P{\'{a}}ll, Abraham, Kutzner, Hess, and Lindahl]{Pall2015TacklingExascaleSoftware}
P{\'{a}}ll,~S.; Abraham,~M.~J.; Kutzner,~C.; Hess,~B.; Lindahl,~E. Tackling {Exascale} {Software} {Challenges} in {Molecular} {Dynamics} {Simulations} with {GROMACS}. Solving {Software} {Challenges} for {Exascale}. Cham, 2015; pp 3--27\relax
\mciteBstWouldAddEndPuncttrue
\mciteSetBstMidEndSepPunct{\mcitedefaultmidpunct}
{\mcitedefaultendpunct}{\mcitedefaultseppunct}\relax
\EndOfBibitem
\bibitem[P{\'{a}}ll \latin{et~al.}(2020)P{\'{a}}ll, Zhmurov, Bauer, Abraham, Lundborg, Gray, Hess, and Lindahl]{Pall2020Heterogeneousparallelizationacceleration}
P{\'{a}}ll,~S.; Zhmurov,~A.; Bauer,~P.; Abraham,~M.; Lundborg,~M.; Gray,~A.; Hess,~B.; Lindahl,~E. Heterogeneous parallelization and acceleration of molecular dynamics simulations in {GROMACS}. \emph{The Journal of Chemical Physics} \textbf{2020}, \emph{153}, 134110\relax
\mciteBstWouldAddEndPuncttrue
\mciteSetBstMidEndSepPunct{\mcitedefaultmidpunct}
{\mcitedefaultendpunct}{\mcitedefaultseppunct}\relax
\EndOfBibitem
\bibitem[Pronk \latin{et~al.}(2013)Pronk, P{\'{a}}ll, Schulz, Larsson, Bjelkmar, Apostolov, Shirts, Smith, Kasson, van~der Spoel, Hess, and Lindahl]{Pronk2013GROMACS4.5high}
Pronk,~S.; P{\'{a}}ll,~S.; Schulz,~R.; Larsson,~P.; Bjelkmar,~P.; Apostolov,~R.; Shirts,~M.~R.; Smith,~J.~C.; Kasson,~P.~M.; van~der Spoel,~D.; Hess,~B.; Lindahl,~E. {GROMACS} 4.5: a high-throughput and highly parallel open source molecular simulation toolkit. \emph{Bioinformatics} \textbf{2013}, \emph{29}, 845--854\relax
\mciteBstWouldAddEndPuncttrue
\mciteSetBstMidEndSepPunct{\mcitedefaultmidpunct}
{\mcitedefaultendpunct}{\mcitedefaultseppunct}\relax
\EndOfBibitem
\bibitem[Van Der~Spoel \latin{et~al.}(2005)Van Der~Spoel, Lindahl, Hess, Groenhof, Mark, and Berendsen]{VanDerSpoel2005GROMACSFastflexible}
Van Der~Spoel,~D.; Lindahl,~E.; Hess,~B.; Groenhof,~G.; Mark,~A.~E.; Berendsen,~H. J.~C. {GROMACS}: {Fast}, flexible, and free. \emph{Journal of Computational Chemistry} \textbf{2005}, \emph{26}, 1701--1718\relax
\mciteBstWouldAddEndPuncttrue
\mciteSetBstMidEndSepPunct{\mcitedefaultmidpunct}
{\mcitedefaultendpunct}{\mcitedefaultseppunct}\relax
\EndOfBibitem
\bibitem[Soteras~Guti{\'{e}}rrez \latin{et~al.}(2016)Soteras~Guti{\'{e}}rrez, Lin, Vanommeslaeghe, Lemkul, Armacost, Brooks, and MacKerell]{SoterasGutierrez2016Parametrizationhalogenbonds}
Soteras~Guti{\'{e}}rrez,~I.; Lin,~F.-Y.; Vanommeslaeghe,~K.; Lemkul,~J.~A.; Armacost,~K.~A.; Brooks,~C.~L.; MacKerell,~A.~D. Parametrization of halogen bonds in the {CHARMM} general force field: {Improved} treatment of ligand--protein interactions. \emph{Bioorganic \& Medicinal Chemistry} \textbf{2016}, \emph{24}, 4812--4825\relax
\mciteBstWouldAddEndPuncttrue
\mciteSetBstMidEndSepPunct{\mcitedefaultmidpunct}
{\mcitedefaultendpunct}{\mcitedefaultseppunct}\relax
\EndOfBibitem
\bibitem[Vanommeslaeghe \latin{et~al.}(2010)Vanommeslaeghe, Hatcher, Acharya, Kundu, Zhong, Shim, Darian, Guvench, Lopes, Vorobyov, and Mackerell~Jr.]{Vanommeslaeghe2010CHARMMgeneralforce}
Vanommeslaeghe,~K.; Hatcher,~E.; Acharya,~C.; Kundu,~S.; Zhong,~S.; Shim,~J.; Darian,~E.; Guvench,~O.; Lopes,~P.; Vorobyov,~I.; Mackerell~Jr.,~A.~D. {CHARMM} general force field: {A} force field for drug-like molecules compatible with the {CHARMM} all-atom additive biological force fields. \emph{Journal of Computational Chemistry} \textbf{2010}, \emph{31}, 671--690\relax
\mciteBstWouldAddEndPuncttrue
\mciteSetBstMidEndSepPunct{\mcitedefaultmidpunct}
{\mcitedefaultendpunct}{\mcitedefaultseppunct}\relax
\EndOfBibitem
\bibitem[Vanommeslaeghe and MacKerell(2012)Vanommeslaeghe, and MacKerell]{Vanommeslaeghe2012AutomationCHARMMGeneral}
Vanommeslaeghe,~K.; MacKerell,~A. D.~J. Automation of the {CHARMM} {General} {Force} {Field} ({CGenFF}) {I}: {Bond} {Perception} and {Atom} {Typing}. \emph{Journal of Chemical Information and Modeling} \textbf{2012}, \emph{52}, 3144--3154\relax
\mciteBstWouldAddEndPuncttrue
\mciteSetBstMidEndSepPunct{\mcitedefaultmidpunct}
{\mcitedefaultendpunct}{\mcitedefaultseppunct}\relax
\EndOfBibitem
\bibitem[Vanommeslaeghe \latin{et~al.}(2012)Vanommeslaeghe, Raman, and MacKerell]{Vanommeslaeghe2012AutomationCHARMMGenerala}
Vanommeslaeghe,~K.; Raman,~E.~P.; MacKerell,~A. D.~J. Automation of the {CHARMM} {General} {Force} {Field} ({CGenFF}) {II}: {Assignment} of {Bonded} {Parameters} and {Partial} {Atomic} {Charges}. \emph{Journal of Chemical Information and Modeling} \textbf{2012}, \emph{52}, 3155--3168\relax
\mciteBstWouldAddEndPuncttrue
\mciteSetBstMidEndSepPunct{\mcitedefaultmidpunct}
{\mcitedefaultendpunct}{\mcitedefaultseppunct}\relax
\EndOfBibitem
\bibitem[Yu \latin{et~al.}(2012)Yu, He, Vanommeslaeghe, and MacKerell~Jr.]{Yu2012ExtensionCHARMMgeneral}
Yu,~W.; He,~X.; Vanommeslaeghe,~K.; MacKerell~Jr.,~A.~D. Extension of the {CHARMM} general force field to sulfonyl-containing compounds and its utility in biomolecular simulations. \emph{Journal of Computational Chemistry} \textbf{2012}, \emph{33}, 2451--2468\relax
\mciteBstWouldAddEndPuncttrue
\mciteSetBstMidEndSepPunct{\mcitedefaultmidpunct}
{\mcitedefaultendpunct}{\mcitedefaultseppunct}\relax
\EndOfBibitem
\bibitem[Lemkul(2019)]{Lemkul2019ProteinsPerturbedHamiltonians}
Lemkul,~J. From Proteins to Perturbed Hamiltonians: A Suite of Tutorials for the {GROMACS}-2018 Molecular Simulation Package. \emph{Living Journal of Computational Molecular Science} \textbf{2019}, \emph{1}\relax
\mciteBstWouldAddEndPuncttrue
\mciteSetBstMidEndSepPunct{\mcitedefaultmidpunct}
{\mcitedefaultendpunct}{\mcitedefaultseppunct}\relax
\EndOfBibitem
\bibitem[202(2023)]{2023LAMMPSLargescale}
{LAMMPS}: {Large}-scale {Atomic}/{Molecular} {Massively} {Parallel} {Simulator}. 2023; \url{https://zenodo.org/records/10806852}\relax
\mciteBstWouldAddEndPuncttrue
\mciteSetBstMidEndSepPunct{\mcitedefaultmidpunct}
{\mcitedefaultendpunct}{\mcitedefaultseppunct}\relax
\EndOfBibitem
\bibitem[Aktulga \latin{et~al.}(2012)Aktulga, Fogarty, Pandit, and Grama]{Aktulga2012Parallelreactivemoleculara}
Aktulga,~H.~M.; Fogarty,~J.~C.; Pandit,~S.~A.; Grama,~A.~Y. Parallel reactive molecular dynamics: {Numerical} methods and algorithmic techniques. \emph{Parallel Computing} \textbf{2012}, \emph{38}, 245--259\relax
\mciteBstWouldAddEndPuncttrue
\mciteSetBstMidEndSepPunct{\mcitedefaultmidpunct}
{\mcitedefaultendpunct}{\mcitedefaultseppunct}\relax
\EndOfBibitem
\bibitem[Plimpton(1995)]{Plimpton1995FastParallelAlgorithms}
Plimpton,~S. Fast {Parallel} {Algorithms} for {Short}-{Range} {Molecular} {Dynamics}. \emph{Journal of Computational Physics} \textbf{1995}, \emph{117}, 1--19\relax
\mciteBstWouldAddEndPuncttrue
\mciteSetBstMidEndSepPunct{\mcitedefaultmidpunct}
{\mcitedefaultendpunct}{\mcitedefaultseppunct}\relax
\EndOfBibitem
\bibitem[Thompson \latin{et~al.}(2022)Thompson, Aktulga, Berger, Bolintineanu, Brown, Crozier, in~'t Veld, Kohlmeyer, Moore, Nguyen, Shan, Stevens, Tranchida, Trott, and Plimpton]{Thompson2022LAMMPSflexiblesimulationa}
Thompson,~A.~P.; Aktulga,~H.~M.; Berger,~R.; Bolintineanu,~D.~S.; Brown,~W.~M.; Crozier,~P.~S.; in~'t Veld,~P.~J.; Kohlmeyer,~A.; Moore,~S.~G.; Nguyen,~T.~D.; Shan,~R.; Stevens,~M.~J.; Tranchida,~J.; Trott,~C.; Plimpton,~S.~J. {LAMMPS} - a flexible simulation tool for particle-based materials modeling at the atomic, meso, and continuum scales. \emph{Computer Physics Communications} \textbf{2022}, \emph{271}, 108171\relax
\mciteBstWouldAddEndPuncttrue
\mciteSetBstMidEndSepPunct{\mcitedefaultmidpunct}
{\mcitedefaultendpunct}{\mcitedefaultseppunct}\relax
\EndOfBibitem
\bibitem[Duin \latin{et~al.}(2001)Duin, Dasgupta, Lorant, and Goddard]{Duin2001ReaxFFReactiveForce}
Duin,~A. C.~V.; Dasgupta,~S.; Lorant,~F.; Goddard,~W.~A. ReaxFF: A Reactive Force Field for Hydrocarbons. \emph{Journal of Physical Chemistry A} \textbf{2001}, \emph{105}, 9396--9409\relax
\mciteBstWouldAddEndPuncttrue
\mciteSetBstMidEndSepPunct{\mcitedefaultmidpunct}
{\mcitedefaultendpunct}{\mcitedefaultseppunct}\relax
\EndOfBibitem
\bibitem[Monti \latin{et~al.}(2013)Monti, Corozzi, Fristrup, Joshi, Shin, Oelschlaeger, Duin, and Barone]{Monti2013Exploringconformationalreactive}
Monti,~S.; Corozzi,~A.; Fristrup,~P.; Joshi,~K.~L.; Shin,~Y.~K.; Oelschlaeger,~P.; Duin,~A. C.~V.; Barone,~V. Exploring the conformational and reactive dynamics of biomolecules in solution using an extended version of the glycine reactive force field. \emph{Physical Chemistry Chemical Physics} \textbf{2013}, \emph{15}, 15062--15077\relax
\mciteBstWouldAddEndPuncttrue
\mciteSetBstMidEndSepPunct{\mcitedefaultmidpunct}
{\mcitedefaultendpunct}{\mcitedefaultseppunct}\relax
\EndOfBibitem
\bibitem[Humphrey \latin{et~al.}(1996)Humphrey, Dalke, and Schulten]{HUMP96}
Humphrey,~W.; Dalke,~A.; Schulten,~K. {VMD} -- {V}isual {M}olecular {D}ynamics. \emph{Journal of Molecular Graphics} \textbf{1996}, \emph{14}, 33--38\relax
\mciteBstWouldAddEndPuncttrue
\mciteSetBstMidEndSepPunct{\mcitedefaultmidpunct}
{\mcitedefaultendpunct}{\mcitedefaultseppunct}\relax
\EndOfBibitem
\bibitem[Varshney \latin{et~al.}(1994)Varshney, Brooks, and Wright]{VARSH1994}
Varshney,~A.; Brooks,~F.~P.; Wright,~W.~V. Linearly Scalable Computation of Smooth Molecular Surfaces. \emph{IEEE Computer Graphics and Applications} \textbf{1994}, \emph{14}, 19--25\relax
\mciteBstWouldAddEndPuncttrue
\mciteSetBstMidEndSepPunct{\mcitedefaultmidpunct}
{\mcitedefaultendpunct}{\mcitedefaultseppunct}\relax
\EndOfBibitem
\bibitem[Stukowski(2009)]{Stukowski2009Visualizationanalysisatomistic}
Stukowski,~A. Visualization and analysis of atomistic simulation data with {OVITO}–the {Open} {Visualization} {Tool}. \emph{Modelling and Simulation in Materials Science and Engineering} \textbf{2009}, \emph{18}, 015012\relax
\mciteBstWouldAddEndPuncttrue
\mciteSetBstMidEndSepPunct{\mcitedefaultmidpunct}
{\mcitedefaultendpunct}{\mcitedefaultseppunct}\relax
\EndOfBibitem
\bibitem[Studier(2005)]{Studier.2005}
Studier,~F.~W. Protein production by auto-induction in high-density shaking cultures. \emph{Protein Expression and Purification} \textbf{2005}, \emph{41}, 207--234\relax
\mciteBstWouldAddEndPuncttrue
\mciteSetBstMidEndSepPunct{\mcitedefaultmidpunct}
{\mcitedefaultendpunct}{\mcitedefaultseppunct}\relax
\EndOfBibitem
\bibitem[Solar(1985)]{Solar1985ReactionOHphenylalanine}
Solar,~S. Reaction of {OH} with phenylalanine in neutral aqueous solution. \emph{Radiation Physics and Chemistry (1977)} \textbf{1985}, \emph{26}, 103--108\relax
\mciteBstWouldAddEndPuncttrue
\mciteSetBstMidEndSepPunct{\mcitedefaultmidpunct}
{\mcitedefaultendpunct}{\mcitedefaultseppunct}\relax
\EndOfBibitem
\bibitem[Takai \latin{et~al.}(2014)Takai, Kitamura, Kuwabara, Ikawa, Yoshizawa, Shiraki, Kawasaki, Arakawa, and Kitano]{Takai2014Chemicalmodificationamino}
Takai,~E.; Kitamura,~T.; Kuwabara,~J.; Ikawa,~S.; Yoshizawa,~S.; Shiraki,~K.; Kawasaki,~H.; Arakawa,~R.; Kitano,~K. Chemical modification of amino acids by atmospheric-pressure cold plasma in aqueous solution. \emph{Journal of Physics D: Applied Physics} \textbf{2014}, \emph{47}, 285403\relax
\mciteBstWouldAddEndPuncttrue
\mciteSetBstMidEndSepPunct{\mcitedefaultmidpunct}
{\mcitedefaultendpunct}{\mcitedefaultseppunct}\relax
\EndOfBibitem
\bibitem[Guo \latin{et~al.}(2023)Guo, Tian, and Zhang]{Guo2023Reactivemoleculardynamics}
Guo,~J.-S.; Tian,~S.-Q.; Zhang,~Y.-T. Reactive molecular dynamics simulations on interaction mechanisms of cold atmospheric plasmas and peptides. \emph{Physics of Plasmas} \textbf{2023}, \emph{30}, 043512\relax
\mciteBstWouldAddEndPuncttrue
\mciteSetBstMidEndSepPunct{\mcitedefaultmidpunct}
{\mcitedefaultendpunct}{\mcitedefaultseppunct}\relax
\EndOfBibitem
\bibitem[Lackmann \latin{et~al.}(2018)Lackmann, Wende, Verlackt, Golda, Volzke, Kogelheide, Held, Bekeschus, Bogaerts, Schulz-von~der Gathen, and Stapelmann]{Lackmann2018Chemicalfingerprintscold}
Lackmann,~J.-W.; Wende,~K.; Verlackt,~C.; Golda,~J.; Volzke,~J.; Kogelheide,~F.; Held,~J.; Bekeschus,~S.; Bogaerts,~A.; Schulz-von~der Gathen,~V.; Stapelmann,~K. Chemical fingerprints of cold physical plasmas – an experimental and computational study using cysteine as tracer compound. \emph{Scientific Reports} \textbf{2018}, \emph{8}, 7736\relax
\mciteBstWouldAddEndPuncttrue
\mciteSetBstMidEndSepPunct{\mcitedefaultmidpunct}
{\mcitedefaultendpunct}{\mcitedefaultseppunct}\relax
\EndOfBibitem
\bibitem[Gonz{\'a}lez-Benjumea \latin{et~al.}(2020)Gonz{\'a}lez-Benjumea, Carro, Renau-M{\'i}nguez, Linde, Fern{\'a}ndez-Fueyo, Guti{\'e}rrez, and Mart{\'i}nez]{GonzalezBenjumea.2020}
Gonz{\'a}lez-Benjumea,~A.; Carro,~J.; Renau-M{\'i}nguez,~C.; Linde,~D.; Fern{\'a}ndez-Fueyo,~E.; Guti{\'e}rrez,~A.; Mart{\'i}nez,~A.~T. Fatty acid epoxidation by Collariella virescens peroxygenase and heme-channel variants. \emph{Catalysis Science {\&} Technology} \textbf{2020}, \emph{10}, 717--725\relax
\mciteBstWouldAddEndPuncttrue
\mciteSetBstMidEndSepPunct{\mcitedefaultmidpunct}
{\mcitedefaultendpunct}{\mcitedefaultseppunct}\relax
\EndOfBibitem
\bibitem[Takai \latin{et~al.}(2014)Takai, Kitamura, Kuwabara, Ikawa, Yoshizawa, Shiraki, Kawasaki, Arakawa, and Kitano]{Takai.2014}
Takai,~E.; Kitamura,~T.; Kuwabara,~J.; Ikawa,~S.; Yoshizawa,~S.; Shiraki,~K.; Kawasaki,~H.; Arakawa,~R.; Kitano,~K. Chemical modification of amino acids by atmospheric-pressure cold plasma in aqueous solution. \emph{Plasma Sources Science and Technology} \textbf{2014}, \emph{47}, 285403\relax
\mciteBstWouldAddEndPuncttrue
\mciteSetBstMidEndSepPunct{\mcitedefaultmidpunct}
{\mcitedefaultendpunct}{\mcitedefaultseppunct}\relax
\EndOfBibitem
\bibitem[Liu \latin{et~al.}(2021)Liu, Zhou, Wang, Tan, Cheng, Bekhit, Aadil, and Liu]{Liu.2021}
Liu,~Z.-W.; Zhou,~Y.-X.; Wang,~F.; Tan,~Y.-C.; Cheng,~J.-H.; Bekhit,~A. E.-D.; Aadil,~R.~M.; Liu,~X.-B. Oxidation induced by dielectric barrier discharge (DBD) plasma treatment reduces IgG/IgE binding capacity and improves the functionality of glycinin. \emph{Food chemistry} \textbf{2021}, \emph{363}, 130300\relax
\mciteBstWouldAddEndPuncttrue
\mciteSetBstMidEndSepPunct{\mcitedefaultmidpunct}
{\mcitedefaultendpunct}{\mcitedefaultseppunct}\relax
\EndOfBibitem
\bibitem[Lackmann \latin{et~al.}(2015)Lackmann, Baldus, Steinborn, Edengeiser, Kogelheide, Langklotz, Schneider, Leichert, Benedikt, Awakowicz, and Bandow]{Lackmann.2015}
Lackmann,~J.-W.; Baldus,~S.; Steinborn,~E.; Edengeiser,~E.; Kogelheide,~F.; Langklotz,~S.; Schneider,~S.; Leichert,~L. I.~O.; Benedikt,~J.; Awakowicz,~P.; Bandow,~J.~E. A dielectric barrier discharge terminally inactivates RNase A by oxidizing sulfur-containing amino acids and breaking structural disulfide bonds. \emph{Plasma Sources Science and Technology} \textbf{2015}, \emph{48}, 494003\relax
\mciteBstWouldAddEndPuncttrue
\mciteSetBstMidEndSepPunct{\mcitedefaultmidpunct}
{\mcitedefaultendpunct}{\mcitedefaultseppunct}\relax
\EndOfBibitem
\bibitem[Yayci \latin{et~al.}(2020)Yayci, Dirks, Kogelheide, Alcalde, Hollmann, Awakowicz, and Bandow]{Yayci2020Protectionstrategiesbiocatalytic}
Yayci,~A.; Dirks,~T.; Kogelheide,~F.; Alcalde,~M.; Hollmann,~F.; Awakowicz,~P.; Bandow,~J.~E. Protection strategies for biocatalytic proteins under plasma treatment. \emph{Journal of Physics D: Applied Physics} \textbf{2020}, \emph{54}, 035204\relax
\mciteBstWouldAddEndPuncttrue
\mciteSetBstMidEndSepPunct{\mcitedefaultmidpunct}
{\mcitedefaultendpunct}{\mcitedefaultseppunct}\relax
\EndOfBibitem
\bibitem[Mortier \latin{et~al.}(1985)Mortier, Van~Genechten, and Gasteiger]{Mortier1985Electronegativityequalizationapplication}
Mortier,~W.~J.; Van~Genechten,~K.; Gasteiger,~J. Electronegativity equalization: application and parametrization. \emph{Journal of the American Chemical Society} \textbf{1985}, \emph{107}, 829--835\relax
\mciteBstWouldAddEndPuncttrue
\mciteSetBstMidEndSepPunct{\mcitedefaultmidpunct}
{\mcitedefaultendpunct}{\mcitedefaultseppunct}\relax
\EndOfBibitem
\bibitem[Senftle \latin{et~al.}(2016)Senftle, Hong, Islam, Kylasa, Zheng, Shin, Junkermeier, Engel-Herbert, Janik, Aktulga, Verstraelen, Grama, and Duin]{Senftle2016ReaxFFreactiveforce}
Senftle,~T.~P.; Hong,~S.; Islam,~M.~M.; Kylasa,~S.~B.; Zheng,~Y.; Shin,~Y.~K.; Junkermeier,~C.; Engel-Herbert,~R.; Janik,~M.~J.; Aktulga,~H.~M.; Verstraelen,~T.; Grama,~A.; Duin,~A. C.~V. The ReaxFF reactive force-field: development, applications and future directions. \emph{npj Computational Materials} \textbf{2016}, \emph{2}, 1--14\relax
\mciteBstWouldAddEndPuncttrue
\mciteSetBstMidEndSepPunct{\mcitedefaultmidpunct}
{\mcitedefaultendpunct}{\mcitedefaultseppunct}\relax
\EndOfBibitem
\end{mcitethebibliography}

\end{document}